\documentclass[trackchanges]{aastex701}

\usepackage{subcaption}
\usepackage{orcidlink}
\usepackage{amsmath}

\begin{document}

\title{A Quest for Changing-State AGNs}

\correspondingauthor{Kasra Mokhtarpour}
\email{kasra.mokhtarpour@gmail.com}

\author[orcid=0009-0007-2139-1791]{Kasra Mokhtarpour}
\affiliation{Department of Physics, Florida Atlantic University, Boca Raton, FL 33431, USA}
\email{kmokhtarpour2019@fau.edu}

\author[orcid=0000-0002-2304-0908]{Vicki Sarajedini}
\affiliation{Department of Physics, Florida Atlantic University, Boca Raton, FL 33431, USA}
\email{vsarajedini@fau.edu}

\begin{abstract}

We use Gaia Data Release 3 (DR3) light curves to identify photometrically selected changing-state active galactic nucleus (CSAGN) candidates. We employ unsupervised deep learning and anomaly detection methods to identify unusual patterns of variability in the light curves of 872,228 AGNs with known variability. We find that 8,504 light curves show anomalous behavior. Among these, 682 sources show evidence for two-state photometric behavior during the Gaia DR3 observing baseline. These 682 CSAGN candidates were selected using an automated statistical pattern recognition approach based on Gaussian mixture models (GMMs) to identify sources whose magnitude distributions are consistent with two preferred brightness levels, substantially reducing the need for human visual inspection in selecting the CSAGN candidates. The candidates tend to have fainter absolute magnitudes and lower redshifts than the non-anomalous Gaia variable AGN sample, suggesting that some may be accreting at lower rates. Our approach builds a large candidate sample with less reliance on human visual selection. Follow-up observations of these sources can help identify what drives changing-state transitions and clarify their connection to accretion-disk physics and supermassive black hole (SMBH) fueling.

\end{abstract}



\section{Introduction}
\label{sec:introduction}

Active galactic nuclei (AGNs) are variable over a wide range of timescales and wavelengths \citep{1997ARA&A..35..445U}. At optical wavelengths, ordinary quasar variability is  typically around 0.1 mag on timescales of months, increasing to a few tenth of a magnitude over years \citep{2010ApJ...721.1014M, 2012ApJ...753..106M, 2016MNRAS.457..389M}. As a first approximation, much of this behavior can be described as a stochastic process. The damped random walk (DRW) model has therefore been widely used to characterize quasar light curves \citep{2009ApJ...698..895K,2010ApJ...721.1014M}, although it does not describe all forms of AGN variability. The growth of time-domain astronomy together with longer observing baselines and data-driven searches have revealed sources whose light curves depart from their usual stochastic behavior, including objects that undergo a coherent rise or decline or spend extended intervals at different brightness levels \citep{2020MNRAS.491.4925G,2021ApJ...907...96S,2021AJ....162..206S}. Such changes may trace a more substantial reorganization of the accretion flow than the smaller fluctuations normally seen in AGNs.

The optical classification of an AGN is based largely on its emission-line spectrum. Type~1 AGNs show broad permitted emission lines from gas moving close to the central supermassive black hole, while these broad lines are absent from the direct spectrum of a Type~2 AGN. In the classical unification scheme, this difference is explained mainly by viewing angle and obscuration by circumnuclear dust \citep{1993ARA&A..31..473A,1995PASP..107..803U}. The picture becomes more complicated when the same AGN undergoes substantial changes in both flux and spectral properties on a human observing timescale. Such objects are generally referred to as changing-look AGNs (CLAGNs). Following the terminology summarized by \citet{2023NatAs...7.1282R}, CLAGNs can be divided into two broad classes according to the physical origin of the spectral transition: changing-obscuration AGNs (COAGNs) and changing-state AGNs (CSAGNs). COAGNs are associated with changes in line-of-sight column density, while CSAGNs are associated with intrinsic changes in the accretion-powered continuum radiation together with the appearance, disappearance, or pronounced weakening or strengthening of broad emission lines. For a CSAGN, the spectral change is the key evidence. Photometric variability can point to the expected continuum behavior, but it cannot by itself show that the broad-line spectrum has changed. Throughout this work, we therefore refer to sources selected only from their light curves as \emph{CSAGN candidates}, or more precisely \emph{photometrically selected CSAGN candidates}.

Several well-observed events have shown that a state change can involve the continuum, broad-line region, dust, and X-ray corona during the same episode. In SDSS~J015957.64+003310.5, the first reported changing-look quasar, the optical AGN flux fell by a factor of about six as the source changed from Type~1 to Type~1.9, while the X-ray flux also declined without evidence for increased absorption \citep{2015ApJ...800..144L}. A mid-infrared-selected changing-look quasar showed a similar pattern, with the infrared emission varying together with the optical continuum, which is difficult to explain with a simple foreground obscuration scenario \citep{2018ApJ...864...27S}. SDSS~J110057.70$-$005304.5 faded and later brightened again, and this behavior was interpreted in terms of cooling and heating fronts propagating through the disk \citep{2018MNRAS.480.4468R}. In 1ES~1927+654, the sequence was different; a blue continuum appeared first, broad Balmer lines followed after a delay, and the X-ray corona then weakened dramatically before recovering \citep{2019ApJ...883...94T,2020ApJ...898L...1R}. These examples do not necessarily point to a single physical trigger, but they show that changing-state events can connect parts of the AGN that are often studied separately.

For most optically selected changing-state events, variable obscuration alone is difficult to reconcile with all of the observations. A cloud large enough to cover the broad-line region generally has a crossing time longer than the observed transition, and many sources show neither the reddening nor the additional X-ray absorption expected from such a cloud. Correlated mid-infrared variability also points to a real change in the radiation heating the dust rather than to an obscurer moving only through our line of sight \citep{2023NatAs...7.1282R}. In a recent study of 20 nearby optically identified CLAGNs from the BAT AGN Spectroscopic Survey (BASS), every transition was accompanied by a change in Eddington ratio, while no corresponding change in obscuration was found \citep{2025A&A...693A..35J}. The physical cause is still unsettled. Proposed explanations include changes in the mass accretion rate, thermal or magnetic instabilities, heating or cooling fronts in a geometrically thicker disk, and major perturbations such as tidal disruption events. The main difficulty is the timescale; transitions observed over months to years can be much faster than the viscous timescale of a standard thin disk. Samples of changing-state quasars have nevertheless shown that heating or cooling fronts can reproduce some of the observed long-term light curve behavior \citep{2020MNRAS.491.4925G}. Different mechanisms may operate in different objects, making larger samples with clear and reproducible selection criteria especially important.

The number of known CSAGNs has increased rapidly as repeat spectroscopy has become available for larger samples. Many of the earlier systematic searches started from quasars that exhibited unusually large photometric changes and then obtained follow-up spectra. For instance, \citet{2019ApJ...874....8M} selected quasars with $|\Delta g|>1$~mag and $|\Delta r|>0.5$~mag, confirming 17 new changing-look quasars and identifying more than 200 highly variable candidates. Using Catalina Real-time Transient Survey (CRTS) light curves together with mid-infrared photometry and spectral information, \citet{2020MNRAS.491.4925G} identified 111 changing-state quasars. Repeated spectroscopy of nearly 29,000 previously known AGNs in the first year of SDSS-V later yielded 116 CSAGNs, including 107 new identifications \citep{2024ApJ...966...85Z}. A comparison of SDSS and DESI DR1 spectra has also produced a catalog of 561 confirmed CLAGNs at $z\leq0.9$ \citep{2025ApJS..278...28G}. With samples now large enough for population studies, the effects of the selection method have also become clearer. Searches that begin with known Type~1 AGNs naturally favor fading or turn-off events, while transient surveys triggered by a strong brightening are more likely to recover turn-on events. As a result, the measured occurrence rate and the relative numbers of the turn-on and turn-off events depend on the parent sample, cadence, and selection strategy.

Spectroscopy is required for confirmation, but imaging surveys provide the time coverage needed to identify promising targets and to determine when a transition may have taken place. Photometric searches have used several kinds of information, including variability amplitude, color changes, and cases where the observed variability does not match the source's earlier spectral classification. Using Zwicky Transient Facility (ZTF) light curves, \citet{2024ApJ...966..128W} identified 35 turn-on and 12 turn-off candidates with a DRW-based variability metric and confirmed 21 of the 26 candidates that received follow-up spectroscopy. A more recent ZTF study used changes in both $g$ magnitude and $g-r$ color to locate photometric transitions and estimate their durations \citep{2026ApJ..1004..121V}. These approaches produce well-defined samples, but fixed amplitude or color thresholds inevitably limit which kinds of changing-state events are recovered. This becomes important when host galaxy light dilutes the apparent nuclear variability, when only part of a transition is captured within the survey window, or when the continuum change falls below the usual large-amplitude cuts. Extreme continuum variability by itself also does not guarantee a change in the broad line state. A systematic comparison of known CLAGNs with matched samples of Type~1 AGNs, Type~2 AGNs, and extremely variable quasars found that CLAGNs show distinct ZTF variability properties, suggesting that they are not simply the most variable end of an otherwise ordinary AGN population \citep{2026ApJ...997..100W}. The shape of the light curve and the time spent at different brightness levels can therefore carry information that is missed by amplitude alone.

This problem is well suited to unsupervised learning because the unusual sources are rare and their light curves do not all follow the same pattern. Autoencoders can represent quasar variability in a low-dimensional latent space without requiring a manually defined set of variability features; for decade-long Catalina light curves, such a model reproduced and forecast quasar variability more successfully than a DRW model \citep{2020ApJ...903...54T}. \citet{2021AJ....162..206S} applied this approach in a search for changing-state candidates. They modeled 230,451 ZTF DR5 AGN light curves with a variational recurrent autoencoder (VRAE), used the reconstruction errors and an Isolation Forest in the latent space to select 8,809 anomalies, and identified 75 promising CSAGN candidates after visual inspection. Many of their candidates showed a rapid rise or decline followed by a relatively stable brightness level lasting hundreds of days, which they described as a plateau feature. Their work showed that anomaly detection can reduce a very large AGN sample to a much smaller set without assuming a single transition shape. A practical limitation remained; the anomaly sample still included artifacts and other unusual variables, so the final identification of changing-state behavior relied on visual inspection. For a much larger parent sample, a reproducible second-stage criterion is needed to connect a general photometric anomaly to the more specific signature of preferred brightness levels.

Gaia provides a particularly useful data set for this search. The Gaia Data Release~3 variability processing produced the first Gaia catalog of variable AGNs, known as GLEAN, with 872,228 sources and an estimated purity above 95\%. Sources showing artificial variability associated with Gaia’s scan angle were filtered out when constructing the GLEAN sample \citep{2023A&A...674A..25H,2023A&A...674A..24C}. Its all-sky coverage and 34-month observing window sample AGN variability differently from the longer and denser light curves available from ground-based surveys. Gaia's cadence is sparse and irregular, so some sources may be observed in both brightness regimes and during the transition, whereas in other cases Gaia may sample only part of the transition. Gaia Alerts has already demonstrated that Gaia photometry can be used to identify changing-look candidates. Spectroscopic follow-up of Gaia-selected AGN candidates reported 56 secure and 23 probable changing-look quasars, together with other nuclear transients such as tidal disruption events, supernovae, and microlensing events \citep{2026Ap&SS.371...12D}. The GLEAN light curves allow a complementary search strategy; rather than beginning with objects that crossed an alert threshold, the analysis can begin with the full Gaia DR3 variable-AGN sample and ask which light curves show evidence for two preferred brightness levels.

In this work, we carry out that search using only the Gaia DR3 G band light curves. Starting with all 872,228 GLEAN sources, we use a VRAE to learn the common variability patterns and identify 8,504 anomalous light curves using both reconstruction errors and Isolation Forest (IF) anomaly scores in the latent space. We then apply an automated bimodality analysis based on Sarle's bimodality coefficient and a comparison of one- and two-component Gaussian mixture models. This second stage selects 682 sources whose observed G magnitude distributions satisfy the combined bimodality criteria during the Gaia DR3 observing window. The selected light curves do not follow a single morphology. Some show a well-sampled change between two brightness levels, whereas others contain only part of a rise or decline, highly unbalanced modes, or flare-like behavior within the Gaia observing window. These patterns are treated as photometric evidence for preferred brightness levels, not as proof that a physical state transition occurred.
Section~\ref{sec:data} describes the Gaia light curves, preprocessing, VRAE model, anomaly selection, and bimodality criteria. Section~\ref{sec:results} presents the CSAGN candidate sample and its main properties. We discuss the amplitude change in brightness levels, the candidates' anomaly classifications, comparison with previous searches, and what the lower-luminosity candidates may tell us about accretion in Section~\ref{sec:discussion}. We summarize the main findings in Section~\ref{sec:conclusions}.

\section{Data and Methods} \label{sec:data}
\subsection{Gaia Data Release 3 \texttt{vari\_agn} table} \label{sec:DR3}
Gaia data processing follows relativistic definitions of reference systems in which the time scales are treated as their integral parts. The coordinate time of the solar system Barycentric
Celestial Reference System (BCRS) - TCB (Temps Coordonnée Barycentrique - Barycentric Coordinate Time \citet{2003AJ....126.2687S}) - is used throughout data processing and in the final Gaia catalog. Therefore, the Gaia DR3 light curves use Barycentric Julian Date (\(BJD\)) in Barycentric Coordinate Time, \(BJD_{TCB}\), to ensure high-precision timing for astronomical events, correcting for light travel time from the position of Gaia spacecraft to the barycenter of the solar system. Times are typically formatted as \(JD - 2,455,197.5\), referencing January 1, 2010, 00:00:00 UTC as the reference epoch, \(T_0\).

We use the light curves from Gaia Data Release 3 (Gaia DR3) \citep{2023A&A...674A...1G, 2016A&A...595A...1G}, which includes an extensive classification of variable sources based on Gaia time series data \citep{2023A&A...674A..14R}. In addition, Gaia DR3 provides the first Gaia catalog of variable AGNs, referred to as GLEAN in the literature \citep{2023A&A...674A..24C}, and provided in the Gaia archive as table \texttt{gaiadr3.vari\_agn}, comprising 872,228 objects. These objects were selected based on their consistency with the characteristics of QSO-like sources from the Gaia Celestial Reference Frame (Gaia-CRF3). Each source's variability is described by several parameters, and light curves for field-of-view transits are provided in the G, G\textsubscript{BP} and G\textsubscript{RP} bands (representing white, blue, and red, respectively). Photometric data in these three passbands span a period of 34 months, from July 25, 2014 to May 28, 2017 \citep{2023A&A...674A..13E}. Throughout this work, we restrict the analysis to the Gaia G band, which provides the densest and most uniform temporal coverage across the dataset. The G band is also the broadest Gaia optical passband, so it collects more light than BP or RP and provides better coverage for faint sources.

Gaia DR3 also provides low-resolution mean BP/RP spectra for a subset of sources \citep{2023A&A...674A...1G}, which can be used for source classification and, in the case of quasars, for redshift estimation \citep{2023A&A...674A..41G}.

\subsection{Pre-processing of the light curves} \label{sec:pre-lighcurve}
 The Gaia archive is accessible through TAP+, which is an extension of the Table Access Protocol (TAP). Queries through TAP are written in the Astronomical Data Query Language (ADQL), a language similar to SQL that is widely used to query astronomical databases. Before the analysis, we cleaned the epoch photometric time series retrieved from the Gaia archive. Careful cleaning is particularly important for anomaly detection, since poor measurements can otherwise be mistaken for unusual variability. As a first step, we removed G-band epochs with missing magnitude measurements. We then removed epochs with a \texttt{TRUE} value for either of the quality flags \texttt{rejected\_by\_variability} or \texttt{rejected\_by\_photometry}. The \texttt{rejected\_by\_variability} flag identifies transits rejected by the Gaia DR3 variability-processing cleaning step. The \texttt{rejected\_by\_photometry} flag refers to measurements that were unavailable or rejected by the DPAC photometric processing, or had unphysical negative flux values; although this flag was not used by the Gaia variability analysis itself, we removed these epochs to keep only reliable photometric measurements. The final cleaning step was designed to remove epochs with unusually large magnitude uncertainties without discarding genuine signals or leaving too many outliers. For each light curve, we removed epochs whose magnitude uncertainties exceeded the median uncertainty by more than $n$ times the interquartile range. An epoch was therefore removed when its magnitude uncertainty satisfied
 \begin{equation}
 \sigma_G > \mathrm{median}(\sigma_G) + n \times \mathrm{IQR}(\sigma_G)
 \end{equation}
 The choice of the tuning parameter $n$ is left to the discretion of the researcher. We found that \(n = 8\) provides a reasonable balance between removing highly noisy epochs and preserving real signals.

\begin{figure*}[t]
    \centering

    \begin{subfigure}{0.48\textwidth}
        \centering
        \includegraphics[width=\linewidth]{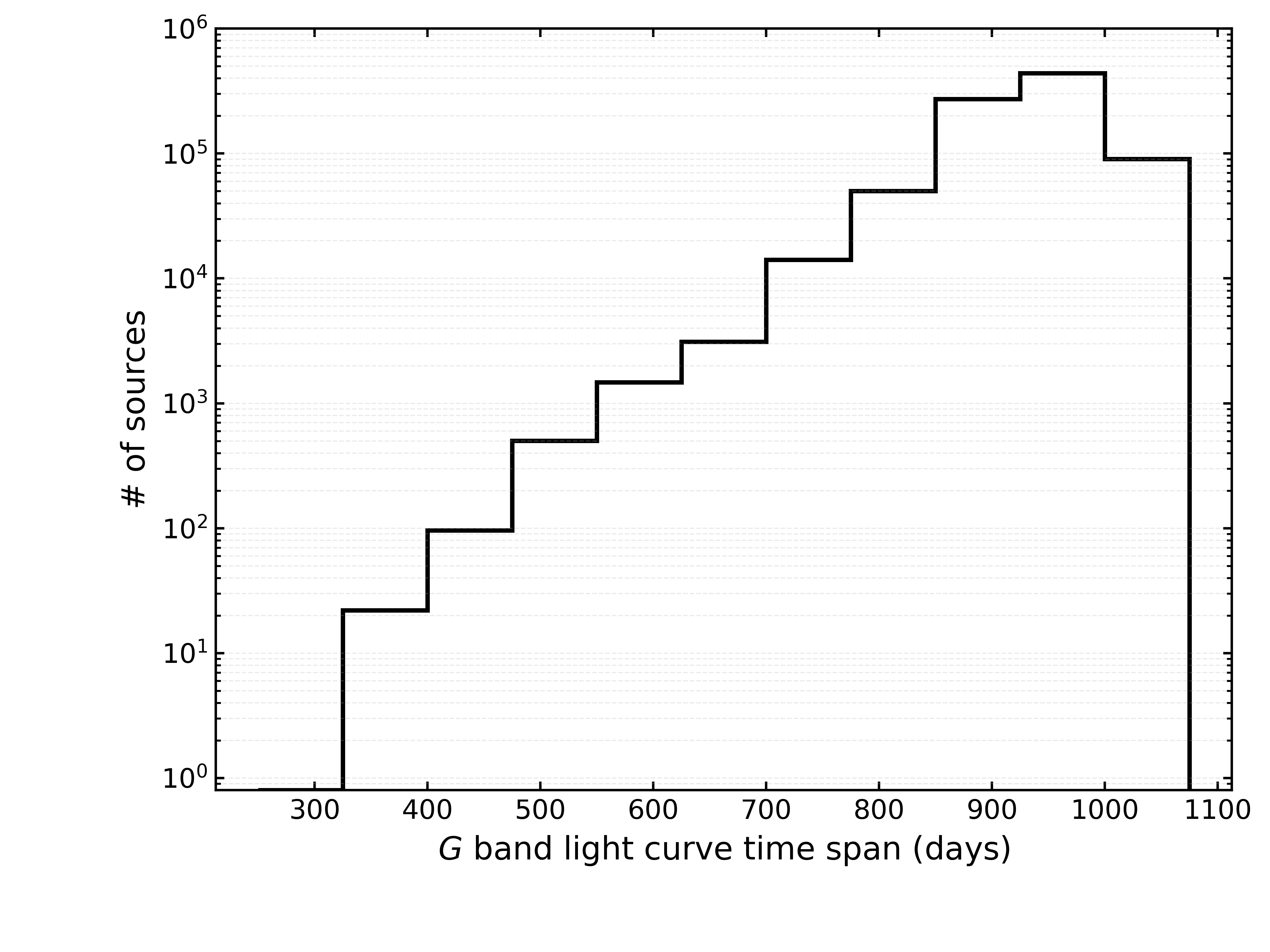}
        \caption{}
        \label{fig:time_span_distribution}
    \end{subfigure}
    \hfill
    \begin{subfigure}{0.48\textwidth}
        \centering
        \includegraphics[width=\linewidth]{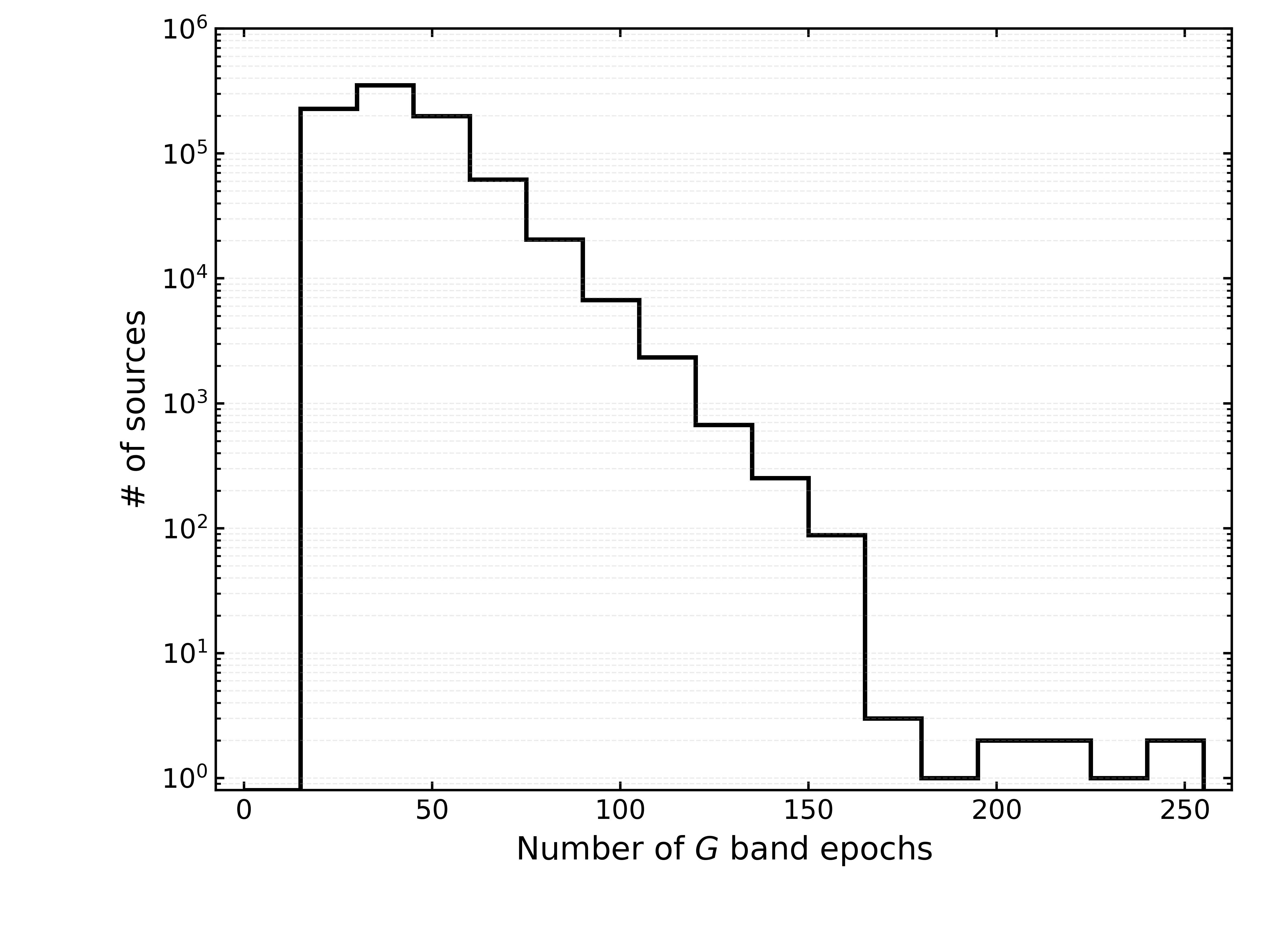}
        \caption{}
        \label{fig:epochs_per_source_distribution}
    \end{subfigure}

    \caption{Distributions of two observational properties of the Gaia DR3 G band light curves. Panel (a) shows the light curve time span in days, and panel (b) shows the number of epochs per source.}
    \label{fig:lc_time_span_epochs}
\end{figure*}

After removing epochs with poor data quality, the final dataset still consists of 872,228 sources. The number of valid G band epochs falls in the ranges of 18 to 242, with an average of approximately 40 epochs per source. The light curves span baselines from a minimum of 327 days to a maximum of 1038 days, with the average time span in the G band being approximately 933 days. Figures \ref{fig:time_span_distribution} and \ref{fig:epochs_per_source_distribution} show the distribution of the time span and the number of epochs per source, respectively.
    
\subsection{Variational Recurrent Autoencoder (VRAE)}

To characterize the variability without assuming a predefined stochastic model, we employed a Variational Recurrent Autoencoder (VRAE) on the Gaia DR3 G band light curves. An autoencoder is trained to reconstruct the input data after compressing them into a lower-dimensional latent representation. The basic idea is that an autoencoder first compresses a light curve into a lower dimensional representation and then tries to reconstruct it. Variability patterns that are common in the sample should be easier for the model to reproduce (i.e., smaller reconstruction error), while rare or unusual patterns should generally be reconstructed less accurately (i.e., larger reconstruction error). Therefore, the reconstruction error can be used as an indicator of how atypical a given light curve is relative to the overall sample.

The model consists of a probabilistic recurrent encoder--decoder architecture that maps each light curve into a 16-dimensional latent space and then reconstructs the original time series. Unlike a standard recurrent autoencoder, the VRAE is a probabilistic generative model in which the latent representation is described by a mean and variance, from which the latent vector is sampled. The encoder is based on bidirectional gated recurrent unit (GRU) layers, while the decoder uses GRU layers to reconstruct the normalized magnitude sequence. The input light curves are described by the normalized time difference between consecutive observations and the normalized Gaia G band magnitude. Following the normalization used by \citet{2021AJ....162..206S}, we normalized the time and magnitude quantities separately for each source. For a light curve with valid Gaia G band magnitudes $m_i$, the normalized magnitude was defined as
\begin{equation}
m_{{\rm norm},i}
=
\frac{m_i-\bar{m}}{s_m},
\end{equation}
where $\bar{m}$ and $s_m$ are the mean and standard deviation of the valid G band magnitudes for that source. The corresponding photometric uncertainty was scaled by the same standard deviation,
\begin{equation}
\sigma_{{\rm norm},i}
=
\frac{\sigma_i}{s_m}.
\end{equation}
The observations were ordered in time, and the interval between consecutive epochs was expressed in years as
\begin{equation}
\Delta t_{{\rm norm},i}
=
\begin{cases}
0, & i=1,\\[4pt]
\dfrac{t_i-t_{i-1}}{365}, & i>1.
\end{cases}
\end{equation}
This normalization removes the source-to-source differences in mean brightness and overall variability scale while retaining the relative shape of each light curve. The normalized photometric uncertainties were then used to weight the reconstruction term of the VRAE loss function. The reconstruction loss used inverse-variance weighting based on the normalized photometric uncertainties. For each valid epoch, the training weight was defined as
\begin{equation}
    w_t^{\rm train} =
\frac{1}{2\sigma_{{\rm norm},t}^{2}},
\end{equation}
where $\sigma_{{\rm norm},t}$ is the normalized magnitude uncertainty. Padded epochs were assigned zero weight through the sequence mask. In this way, epochs with larger photometric uncertainties contribute less to the reconstruction loss. The \(\Delta t\) input carries information about the irregular Gaia sampling, whereas the uncertainty weighting reduces the influence of noisier epochs. This allows the model to account for the irregular sampling and photometric uncertainties of the Gaia epoch photometry. We use the trained VRAE in two ways:
\begin{enumerate}
    \item \textbf{Reconstruction error:} We quantify the reconstruction quality of each source using a variance-weighted reconstruction error (hereafter \(R\)). The form of \(R\) is analogous to a reduced-\(\chi^2\) statistic. This quantity measures how well each light curve is reproduced by the VRAE while accounting for the photometric uncertainties, and therefore provides a measure of how atypical the variability behavior is relative to the overall sample. A larger value of \(R\) corresponds to a poorer reconstruction and indicates that the variability pattern is less well-represented by the behavior learned from the overall sample. For the anomaly-detection analysis, we define
    \begin{equation}
        R =
    \frac{1}{N-1}
    \sum_{i=1}^{N}
    \frac{\left(m_i - m_{{\rm pred},i}\right)^2}
    {\sigma_i^2},
    \end{equation}
    where $m_i$ and $\sigma_i$ are the observed Gaia $G$ band magnitude and its uncertainty at epoch $i$, respectively, and $N$ is the number of valid epochs in the light curve. The reconstructed magnitude $m_{{\rm pred},i}$ is the mean of $S=10$ stochastic reconstructions obtained by sampling the latent distribution.

    \item \textbf{Latent-space representation:} The latent mean vector summarizes each light curve's variability pattern in a compact form. In addition to the reconstruction error, we also use these latent vectors as the input to the Isolation Forest anomaly detection step.
\end{enumerate}

We adopted the VRAE architecture of \citet{2021AJ....162..206S}, with some modifications for the Gaia DR3 epoch photometry. Our implementation also uses two bidirectional GRU layers in the encoder and two GRU layers in the decoder, with 32 units in each recurrent layer, a dropout value of 0.25, and a 16-dimensional latent space. Following \citet{2013arXiv1312.6114K}, the latent distribution is regularized using the Kullback--Leibler (KL) divergence. We scaled the KL-divergence term by a factor of \(1/50\), since about 50 epochs is representative of the Gaia DR3 AGN light curves and keeps the reconstruction and latent terms on comparable scales.

\begin{figure}[t]
\centering

\begin{subfigure}[t]{0.49\linewidth}
  \centering
  \includegraphics[width=\linewidth]{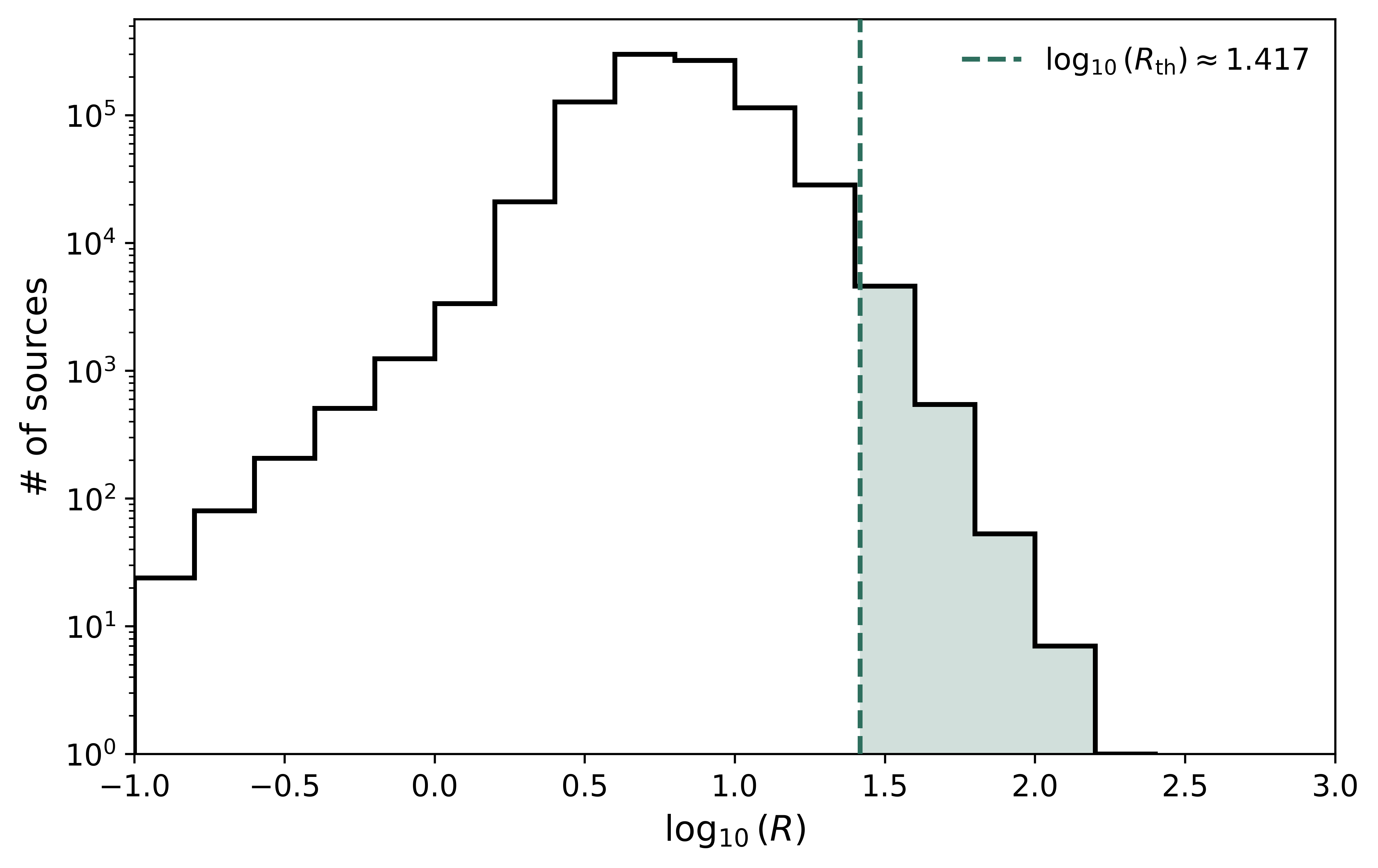}
  \caption{}
  \label{fig:R_dist}
\end{subfigure}
\hfill
\begin{subfigure}[t]{0.49\linewidth}
  \centering
  \includegraphics[width=\linewidth]{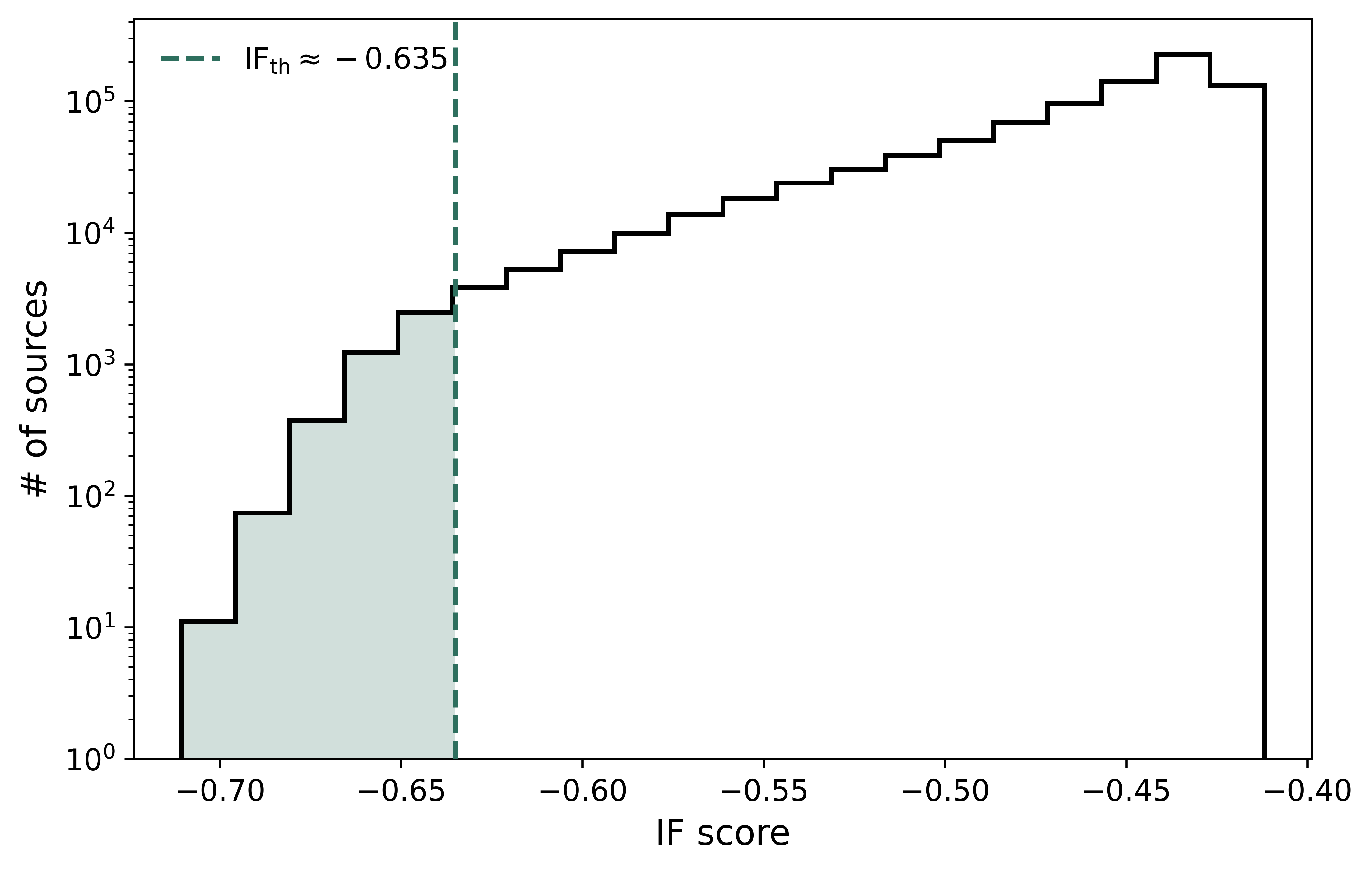}
  \caption{}
  \label{fig:IF_dist}
\end{subfigure}

\caption{%
(a) Histogram of the variance-weighted reconstruction error ($R$) per source, shown as $\log_{10}R$; the dashed line marks the threshold corresponding to the upper 0.5\% tail.
(b) Histogram of Isolation Forest score per source; the dashed line corresponds to a contamination of 0.5\%.
}
\label{fig:R_IF_panels}
\end{figure}

\subsection{Anomaly Identification}

In this study, two anomaly measures were used independently: the reconstruction error from the VRAE model and the anomaly score from the Isolation Forest applied to the VRAE latent space.

\subsubsection{Reconstruction-based anomalies}

After computing \(R\) for all sources, we flagged sources with
\begin{equation}
\log_{10}(R) \geq 1.417
\end{equation}
as reconstruction error anomalies. This threshold corresponds to the upper \(0.5\%\) tail of the reconstruction error distribution, selecting light curves that are poorly reconstructed by the VRAE model (see Figure~\ref{fig:R_dist}). This is an arbitrary threshold rather than a physical cutoff, and its exact value can be adjusted at the discretion of the researcher, depending on the desired level of selectivity. A different threshold would naturally increase or decrease the number of reconstruction-based anomalies.

\subsubsection{Latent-space anomalies: Isolation Forest}

As a second anomaly measure, we applied an Isolation Forest (IF) to the latent mean vector, \(\mu\), obtained from the VRAE model. Isolation Forest is an unsupervised method designed to detect outliers by how easily they can be separated from the rest of the sample rather than learning from labeled examples of normal and abnormal sources. It works by building an ensemble of randomly generated decision trees, where sources that require fewer random splits to be isolated receive lower scores and hence are more anomalous. In our case, this isolation is performed in the VRAE latent space, where each source is represented by its latent mean vector. Therefore, the IF score quantifies how easily a source is isolated in the latent space and provides a complementary measure to the reconstruction-based metric.

We used the implementation available in \texttt{scikit-learn}, with 500 base estimators and a contamination value of \(0.5\%\). Lower IF scores correspond to more anomalous sources. Sources with
\begin{equation}
{\rm IF}_{\rm score} \le -0.635171
\end{equation}
were flagged as latent space anomalies, corresponding to the adopted \(0.5\%\) contamination level.

A source was included in the final anomaly set if it satisfied either the reconstruction error criterion or the latent space Isolation Forest criterion. Of the 8,504 sources in the combined anomaly sample, 4,142 were selected only by the reconstruction error criterion, 4,142 were selected only by the latent space Isolation Forest criterion, and 220 satisfied both. Thus, most candidates were identified by only one of the two methods. The relatively small overlap is also useful in itself, since it shows that the two methods are sensitive to different kinds of anomalous behavior.

\subsection{Bimodality Analysis of Magnitude Distributions}
\label{sec:bimodality_analysis}

A light curve with changing-state behavior is expected to show two preferred brightness states. In the ideal case, both states are sampled for a significant part of the observed baseline. In practice, however, the observed light curve is limited by the survey time window and cadence. As a result, a Gaia DR3 light curve, during its 34-month baseline, may capture a well-sampled transition between two states, but it may also record only part of the event, such as one state followed by a transition, a transition into a new state, or a plateau-like segment of a longer change.

This type of behavior is closely related to the ``plateau feature'' discussed by \citet{2021AJ....162..206S}, in which changing-state candidates show a fast rise or decay followed by a relatively stable brightness level over hundreds of days. The states do not have to be perfectly constant, since AGNs are intrinsically variable even within a given state. However, for a source to be considered a strong changing-state candidate, the variability within each state should be small compared with the overall magnitude change across the light curve.

Such behavior should leave an imprint on the magnitude distribution of the source. If the light curve spends a significant amount of time in two different brightness states, its magnitude distribution is expected to show two preferred regions rather than a single broad distribution. Therefore, after identifying the anomalous sources, we analyzed the magnitude distribution of each anomalous light curve to search for evidence of bimodality. This allowed us to separate light curves with possible changing-state behavior from other types of anomalies.

\subsubsection{Gaussian Mixture Modeling (GMM)}
\label{sec:GMM}

For each anomalous light curve, we fit both one- and two-component Gaussian Mixture Models (GMMs) to the distribution of Gaia G band magnitudes. A GMM represents the observed distribution as a weighted combination of Gaussian components. In this context, a one-component model describes a light curve with one dominant brightness state, while a two-component model allows for two preferred brightness states.

The two models were compared using Schwarz's Bayesian Information Criterion (BIC; \citealt{1978AnSta...6..461S}),
\begin{equation}
{\rm BIC} = -2\ln(\mathcal{L}) + d\ln(N),
\end{equation}
where \(\mathcal{L}\) is the likelihood of the model, \(d\) is the number of free parameters, and \(N\) is the number of data points. The BIC penalizes models with more free parameters, thereby reducing the likelihood that the two-component model is preferred simply because it is more flexible. Lower BIC values indicate a better-supported model. We therefore defined
\begin{equation}
  \Delta{\rm BIC} = {\rm BIC}_{1} - {\rm BIC}_{2},
\end{equation}
where \({\rm BIC}_{1}\) and \({\rm BIC}_{2}\) are the BIC values obtained from the one- and two-component models, respectively. With this definition, a positive value of \(\Delta{\rm BIC}\) indicates that the two-component model is preferred over the one-component model, while increasingly large positive values provide stronger support for the two-component model.

We adopted a threshold of \(\Delta{\rm BIC}>40\) after testing a range of values and visually examining the resulting magnitude distributions and light curves. Although any positive value of \(\Delta{\rm BIC}\) indicates that the two-component model is preferred over the one-component model, small positive values often corresponded to weakly separated or ambiguous distributions. Increasing the threshold progressively favored more pronounced bimodal distributions, and we found that a value of 40 provided a suitable empirical cutoff for retaining the more convincing cases. This threshold is therefore used as a selection criterion rather than as a physical boundary or a uniquely defined statistical cutoff and was then applied uniformly to the full anomalous sample, without source-by-source visual inspection.

In addition to \(\Delta{\rm BIC}\), we computed Sarle's bimodality coefficient (BC), which is based on the skewness and kurtosis of the magnitude distribution. The BC ranges between 0 and 1, with larger values indicating stronger evidence that the distribution is not described by a single centrally concentrated peak. A commonly used reference value is \(5/9 \simeq 0.555\), corresponding to the value obtained for a uniform distribution; values above this level may indicate bimodal or multimodal structure \citep{10.3389/fpsyg.2013.00700}. In this work, we adopted this reference value as the BC threshold.

We therefore required both indicators to support bimodality, adopting the following combined criteria:
\begin{itemize}
    \item \(\mathrm{BC} > 5/9\), and
    \item \(\Delta{\rm BIC} > 40\).
\end{itemize}

Applying the combined bimodality criteria to the anomalous subset yielded 682 photometrically selected CSAGN candidates, corresponding to about \(8\%\) of the 8,504 anomalous light curves.

\begin{figure*}[t]
\centering

\includegraphics[width=0.24\textwidth]{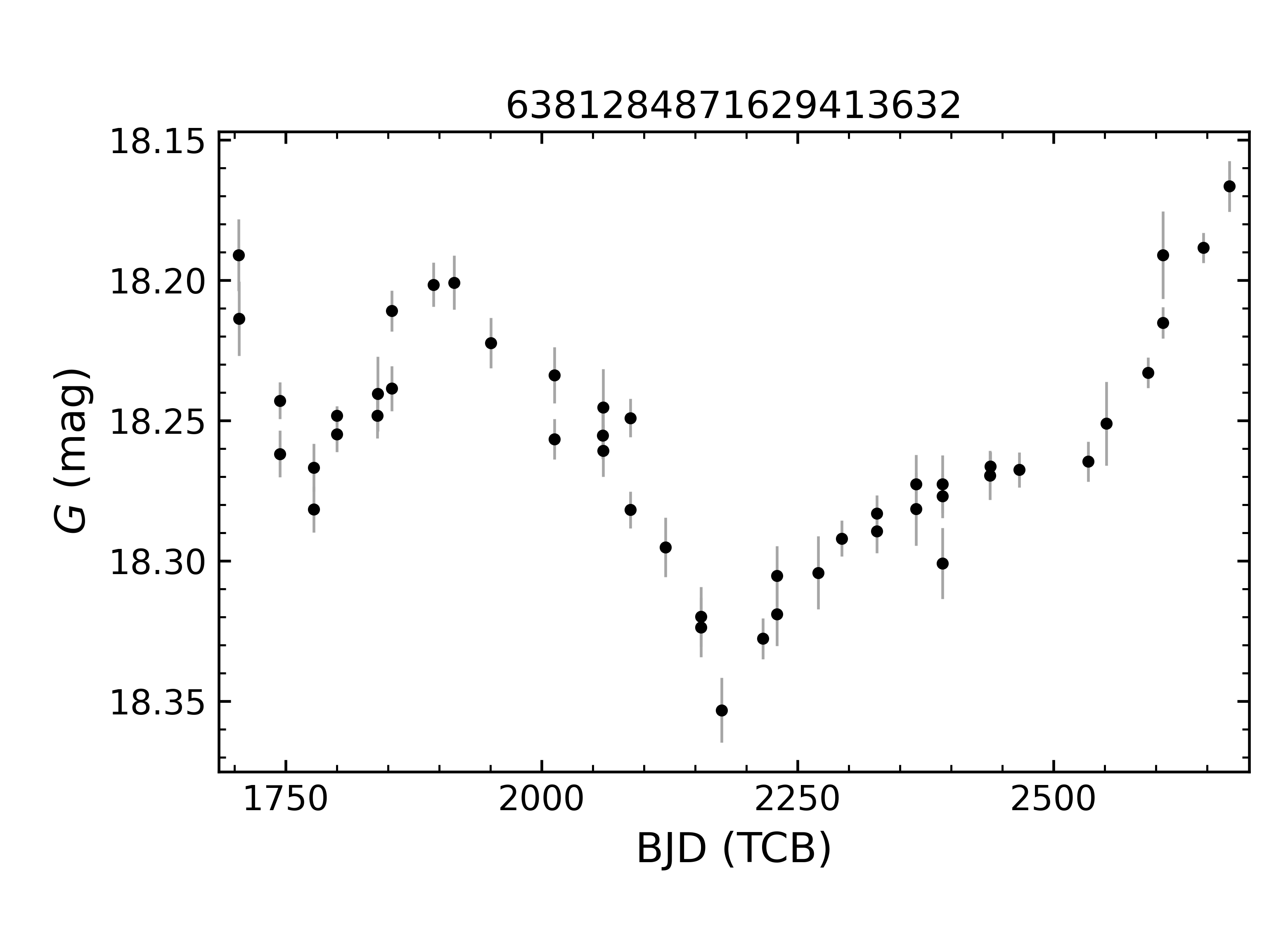}\hfill
\includegraphics[width=0.24\textwidth]{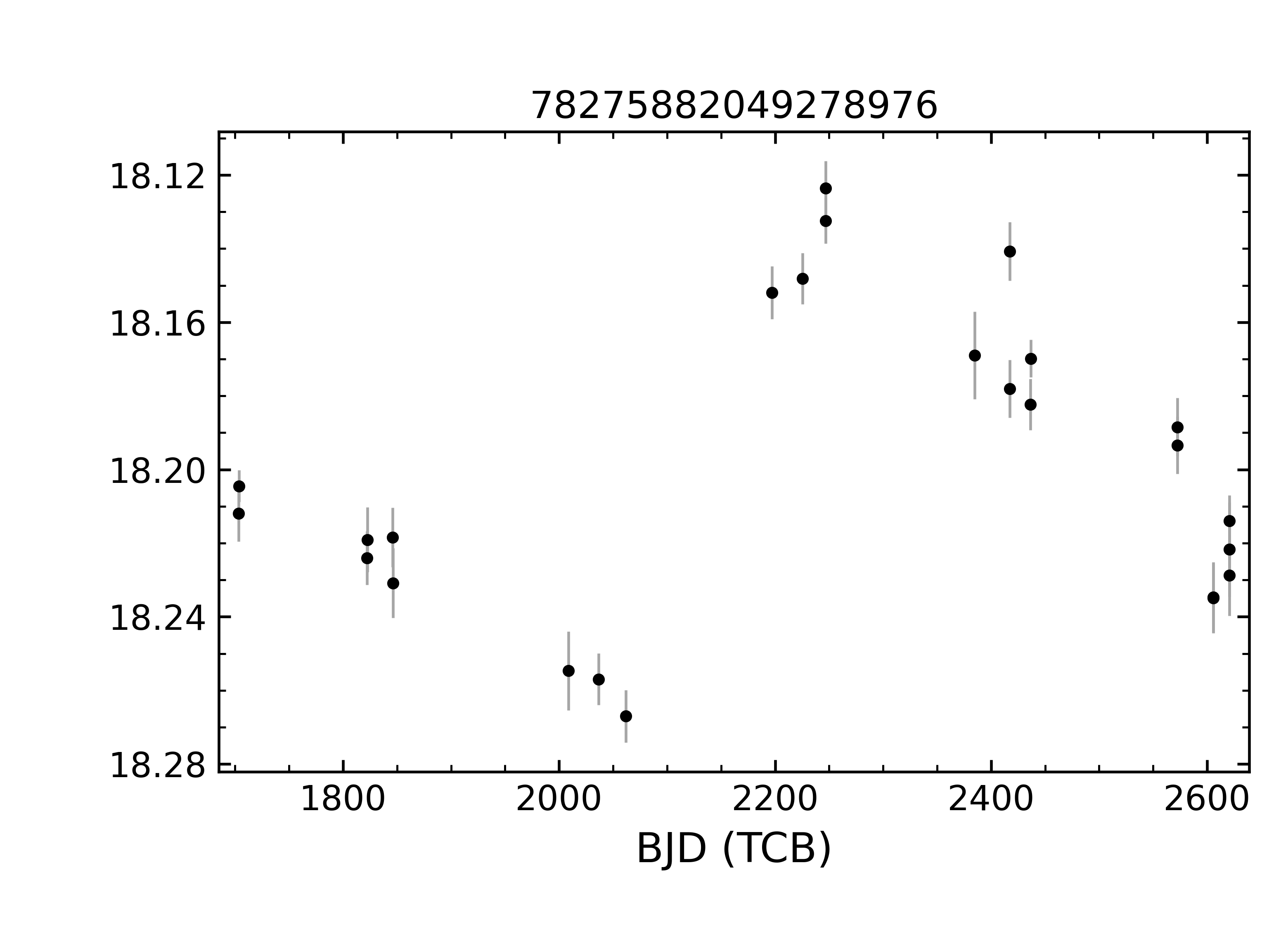}\hfill
\includegraphics[width=0.24\textwidth]{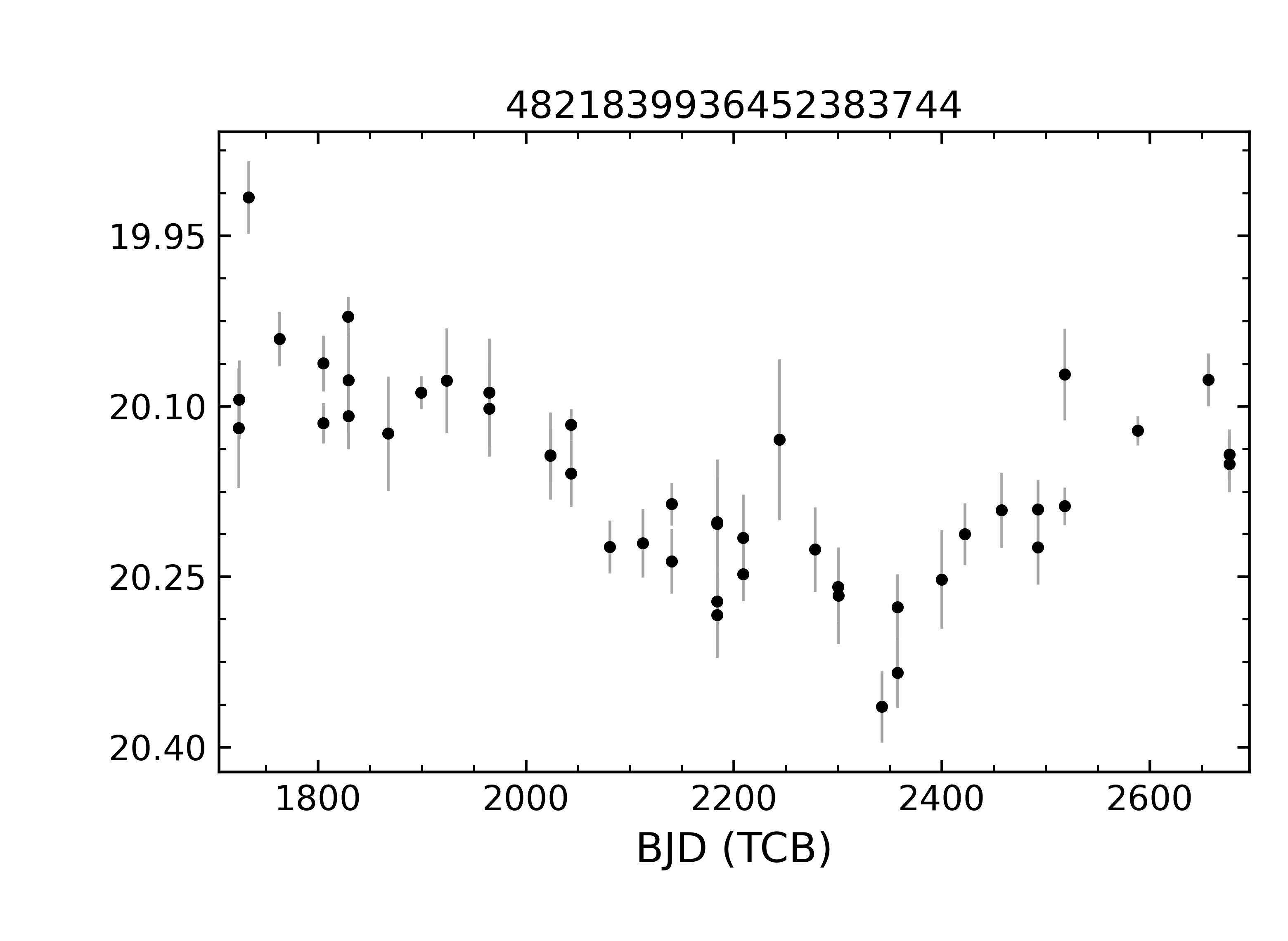}\hfill
\includegraphics[width=0.24\textwidth]{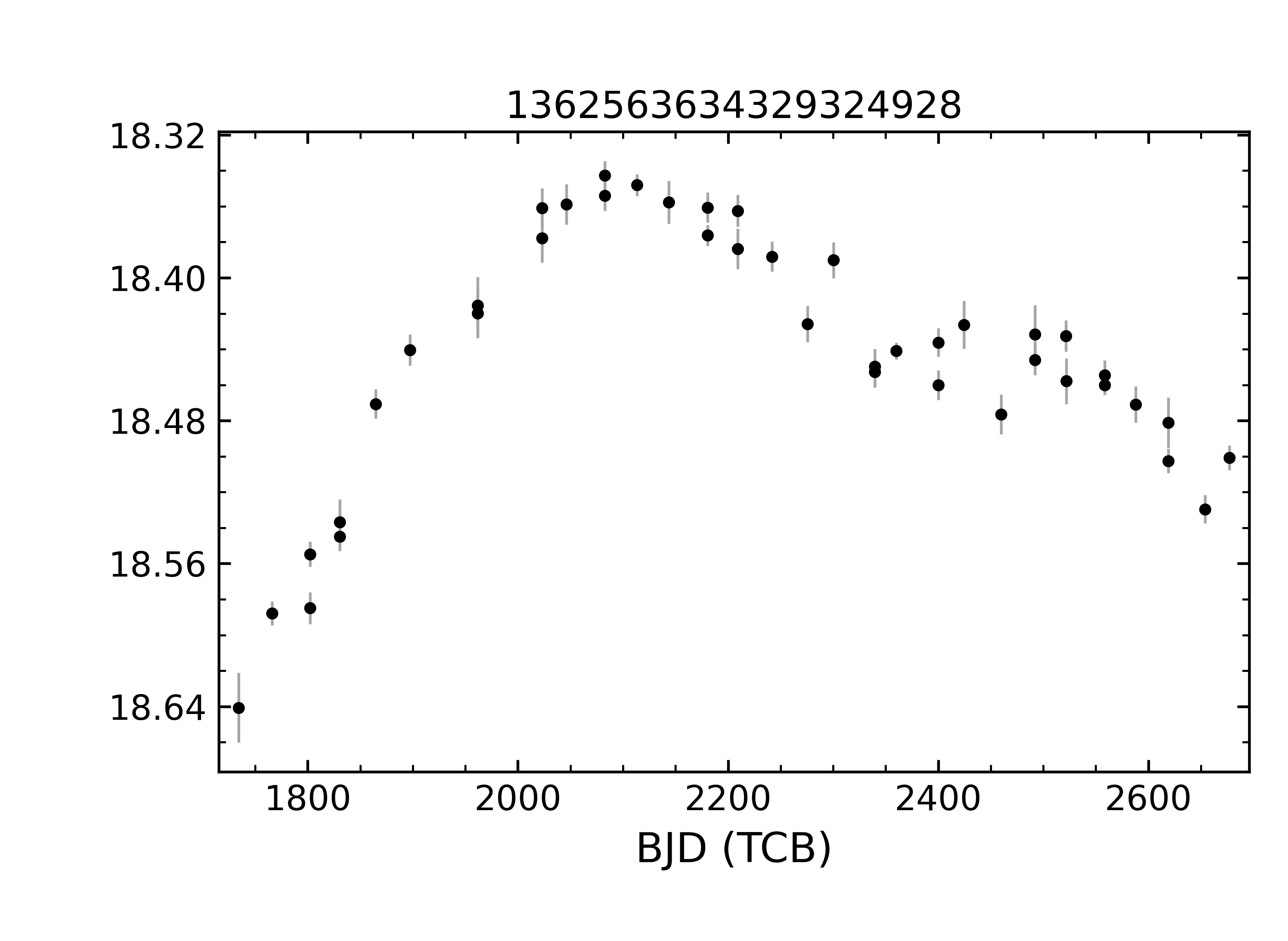}\\[-1.5mm]

\includegraphics[width=0.24\textwidth]{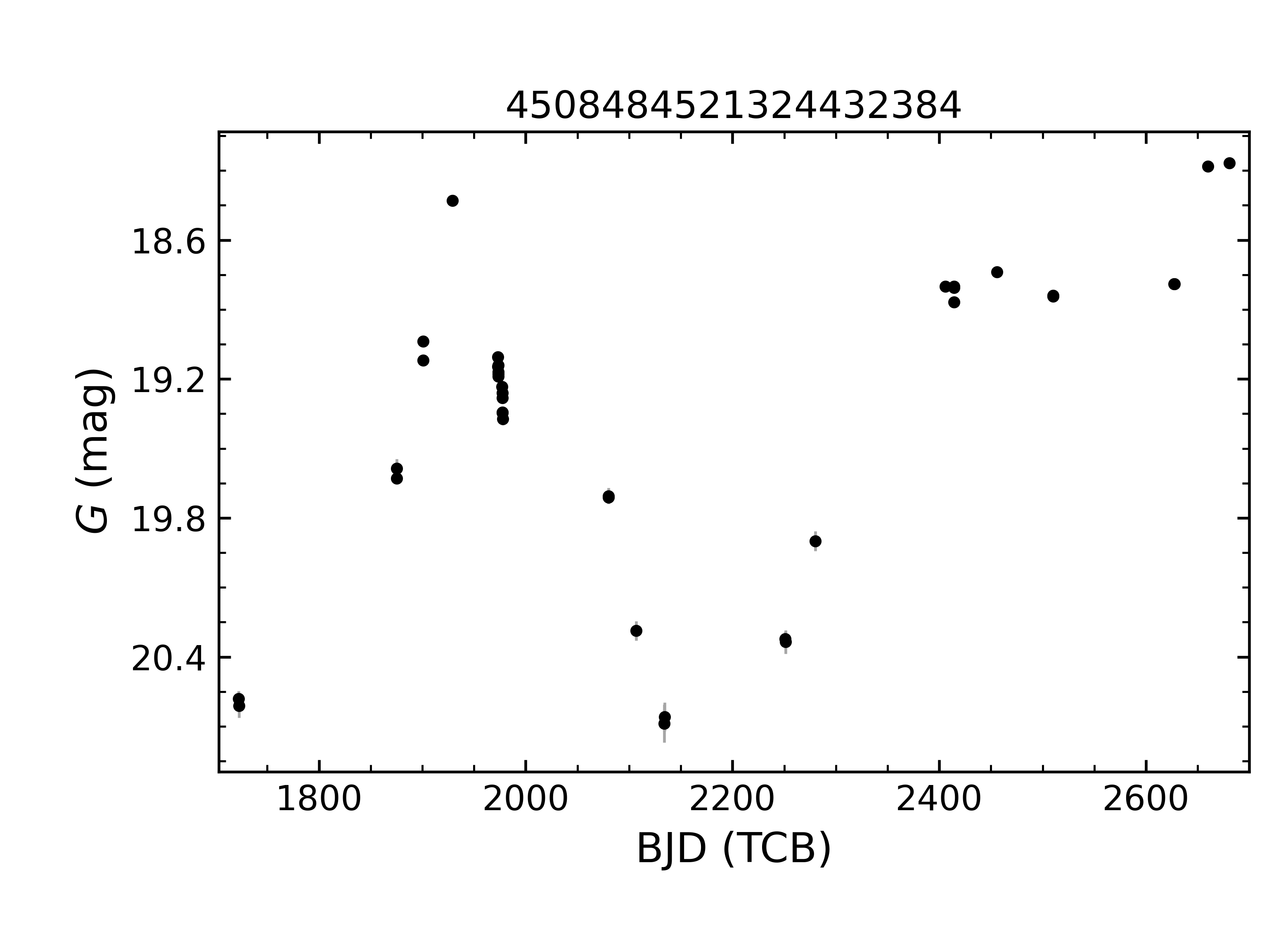}\hfill
\includegraphics[width=0.24\textwidth]{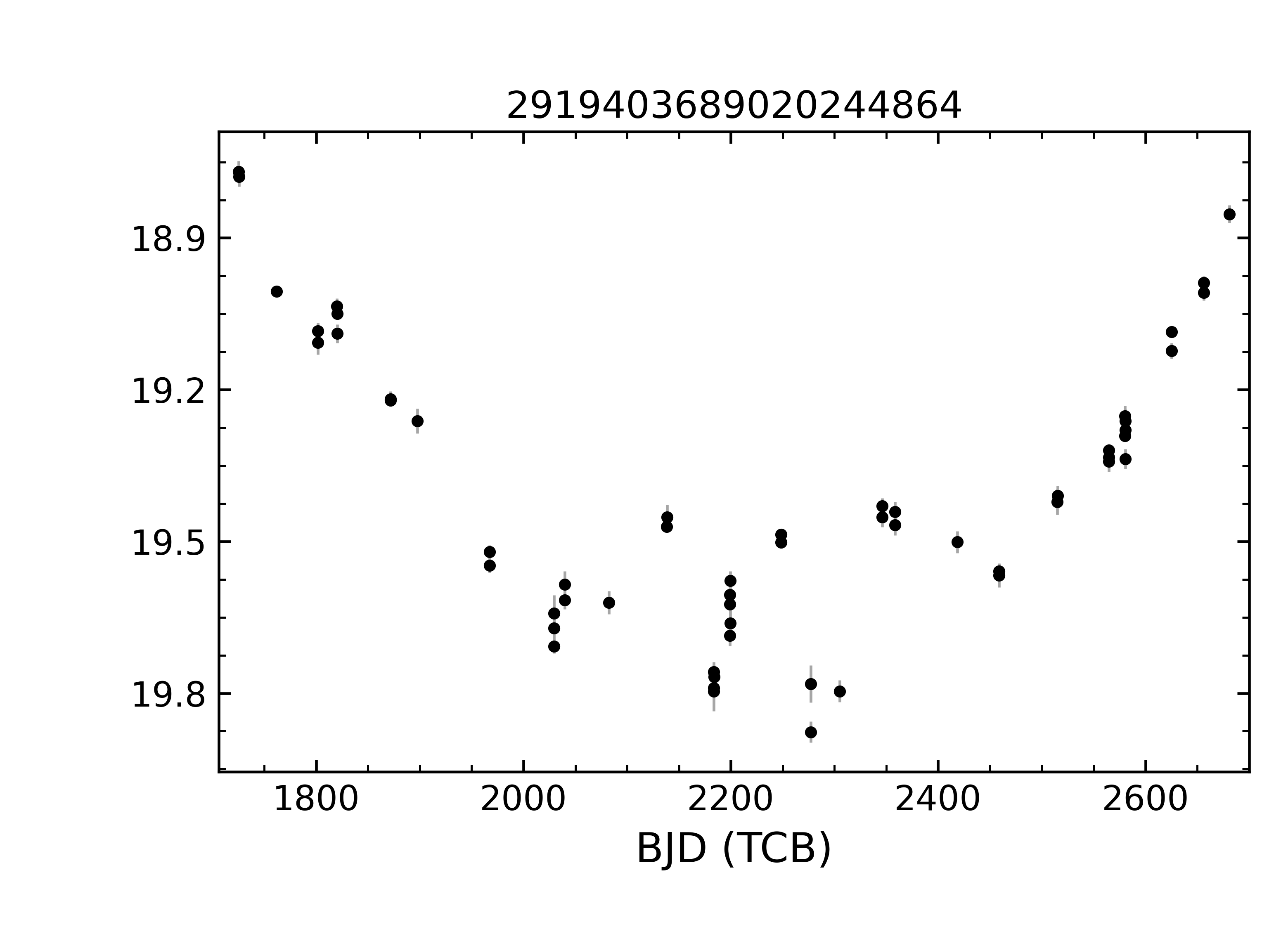}\hfill
\includegraphics[width=0.24\textwidth]{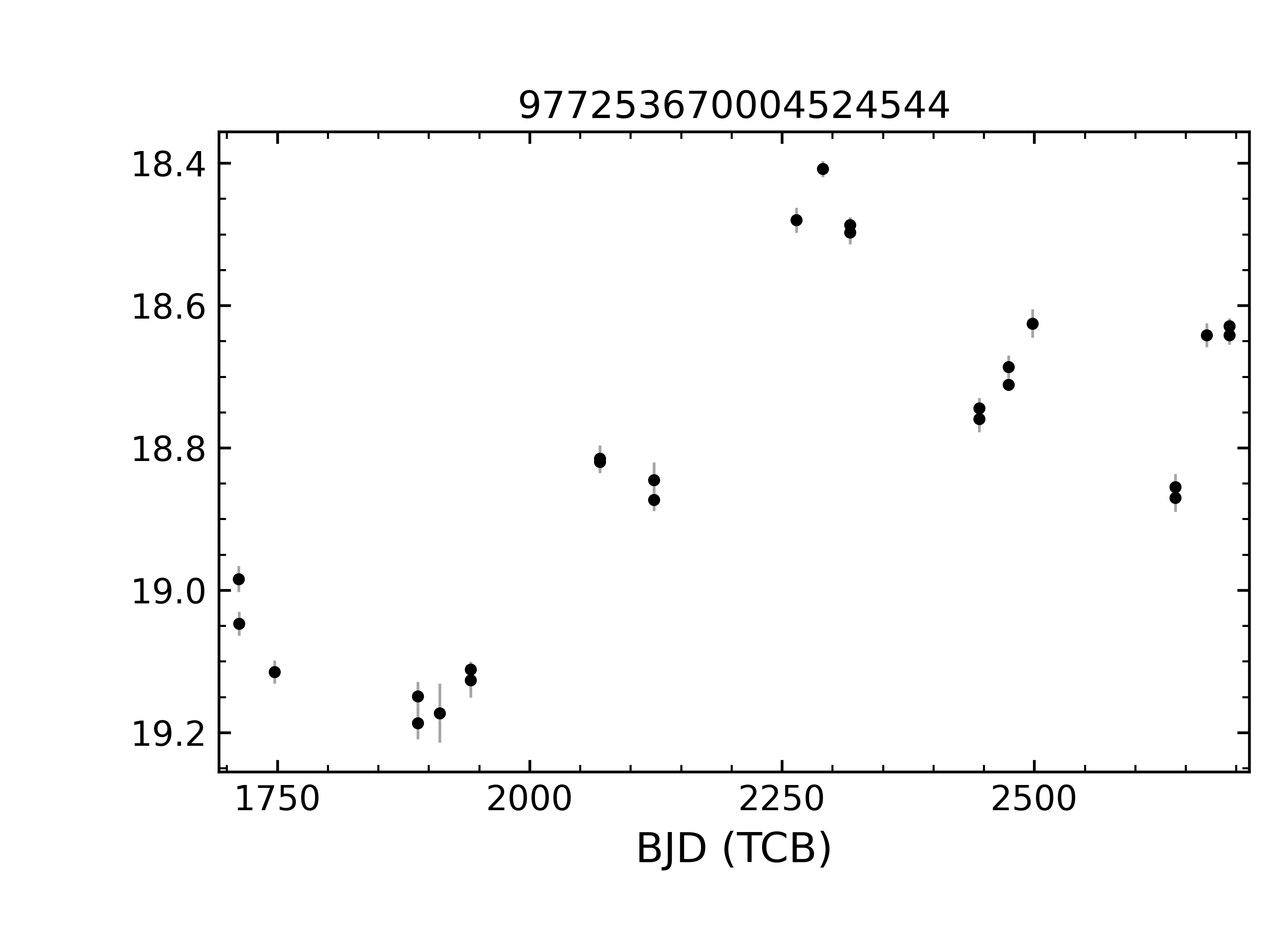}\hfill
\includegraphics[width=0.24\textwidth]{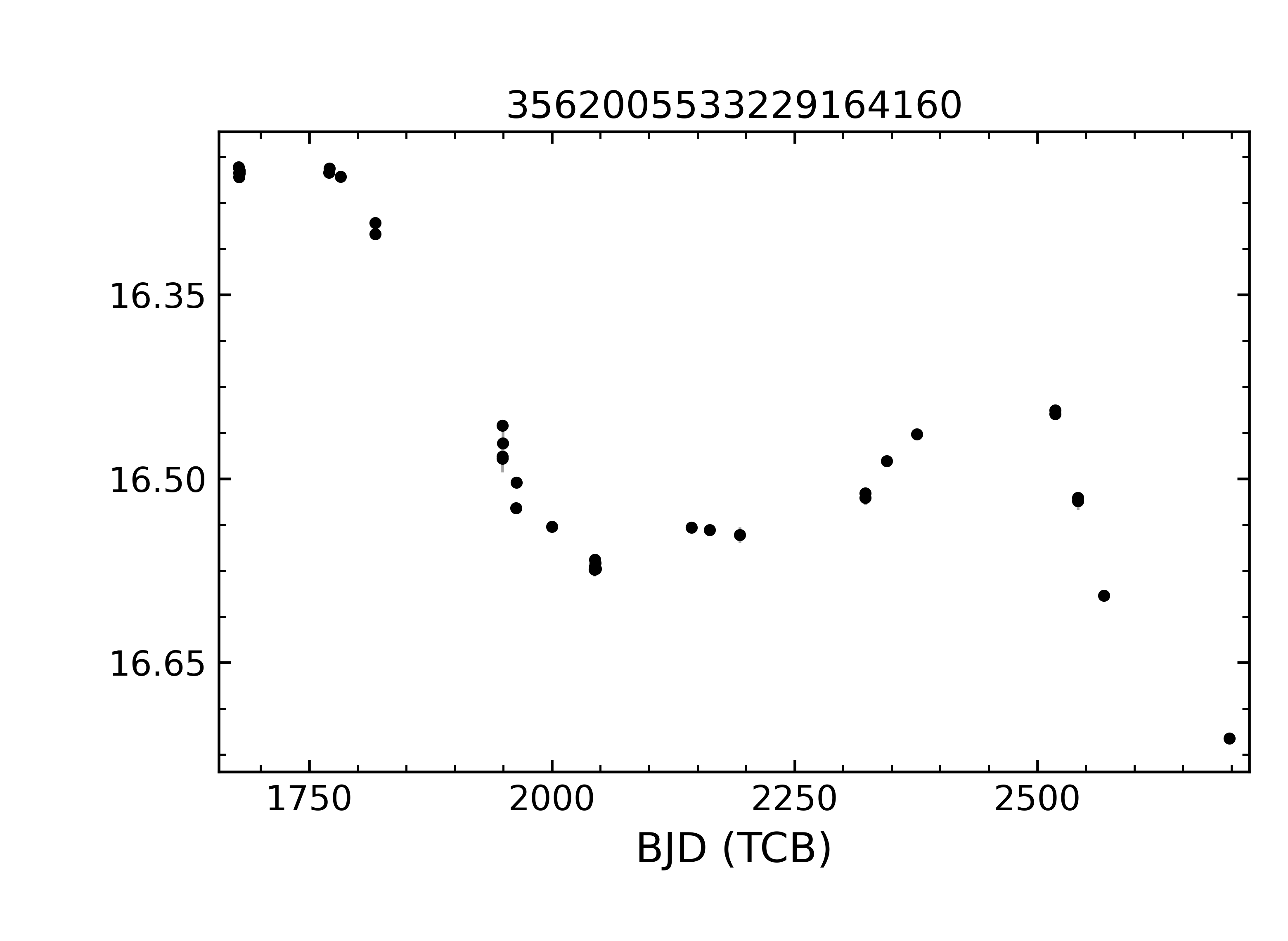}\\[-1.5mm]

\includegraphics[width=0.24\textwidth]{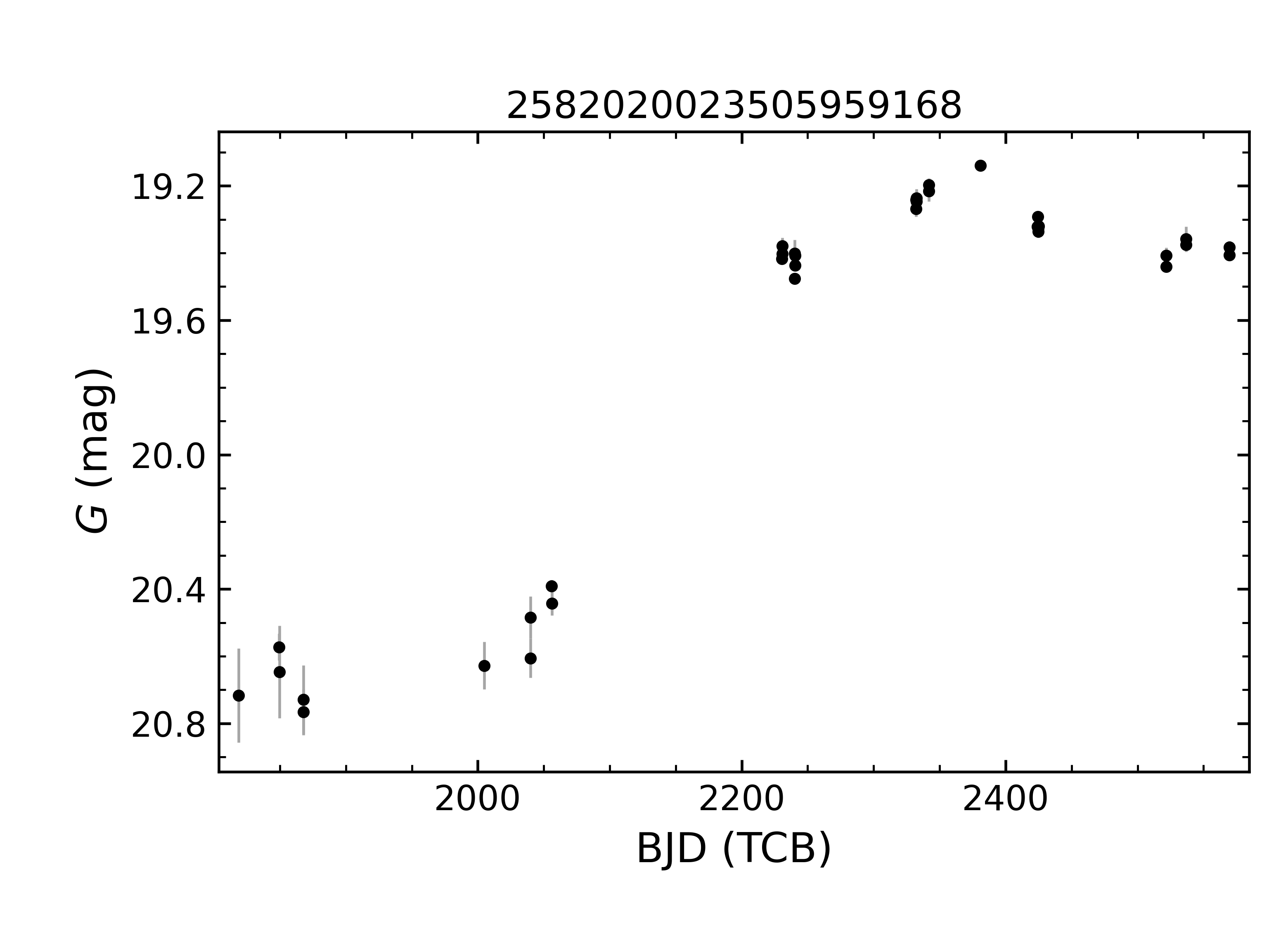}\hfill
\includegraphics[width=0.24\textwidth]{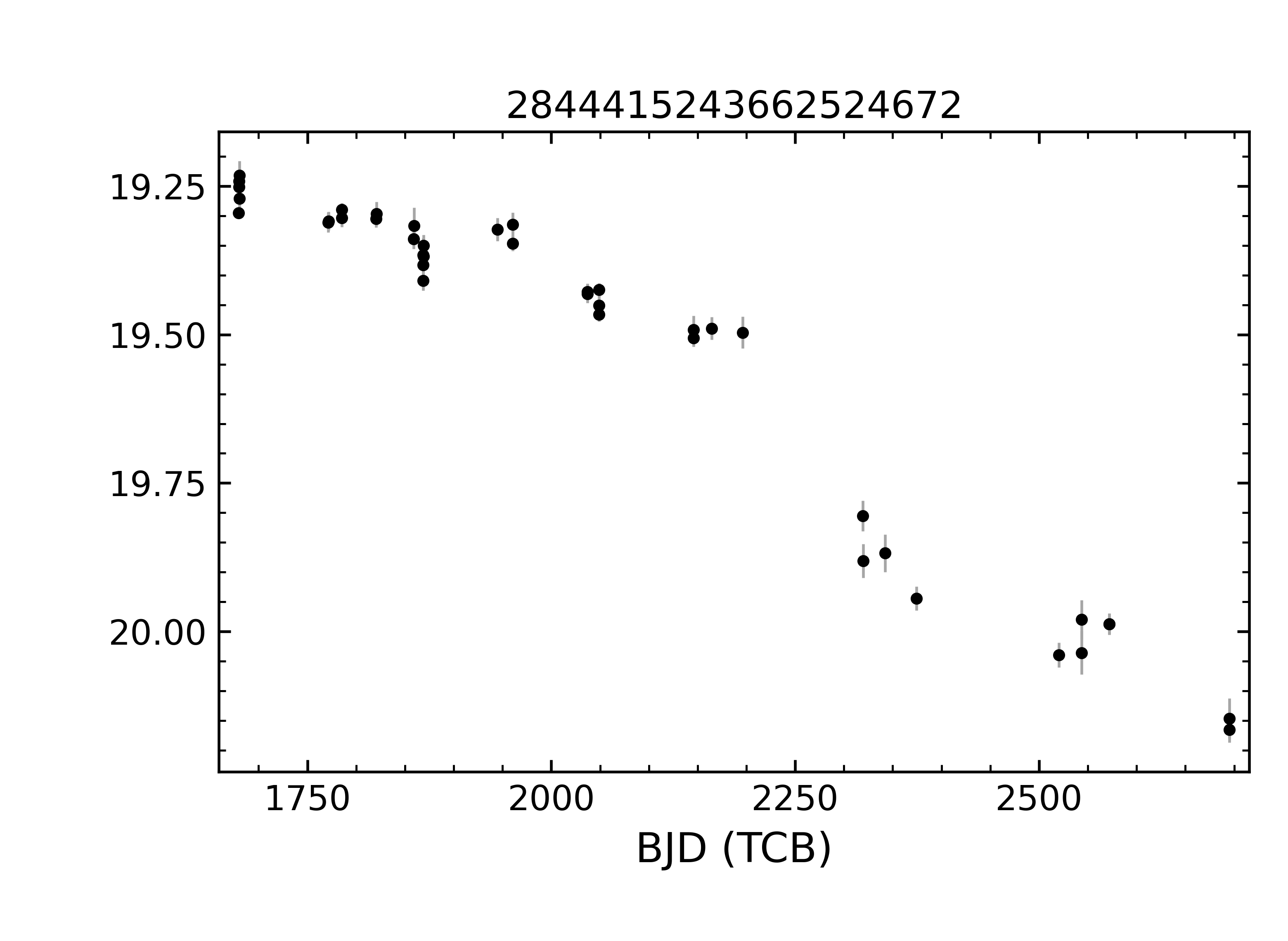}\hfill
\includegraphics[width=0.24\textwidth]{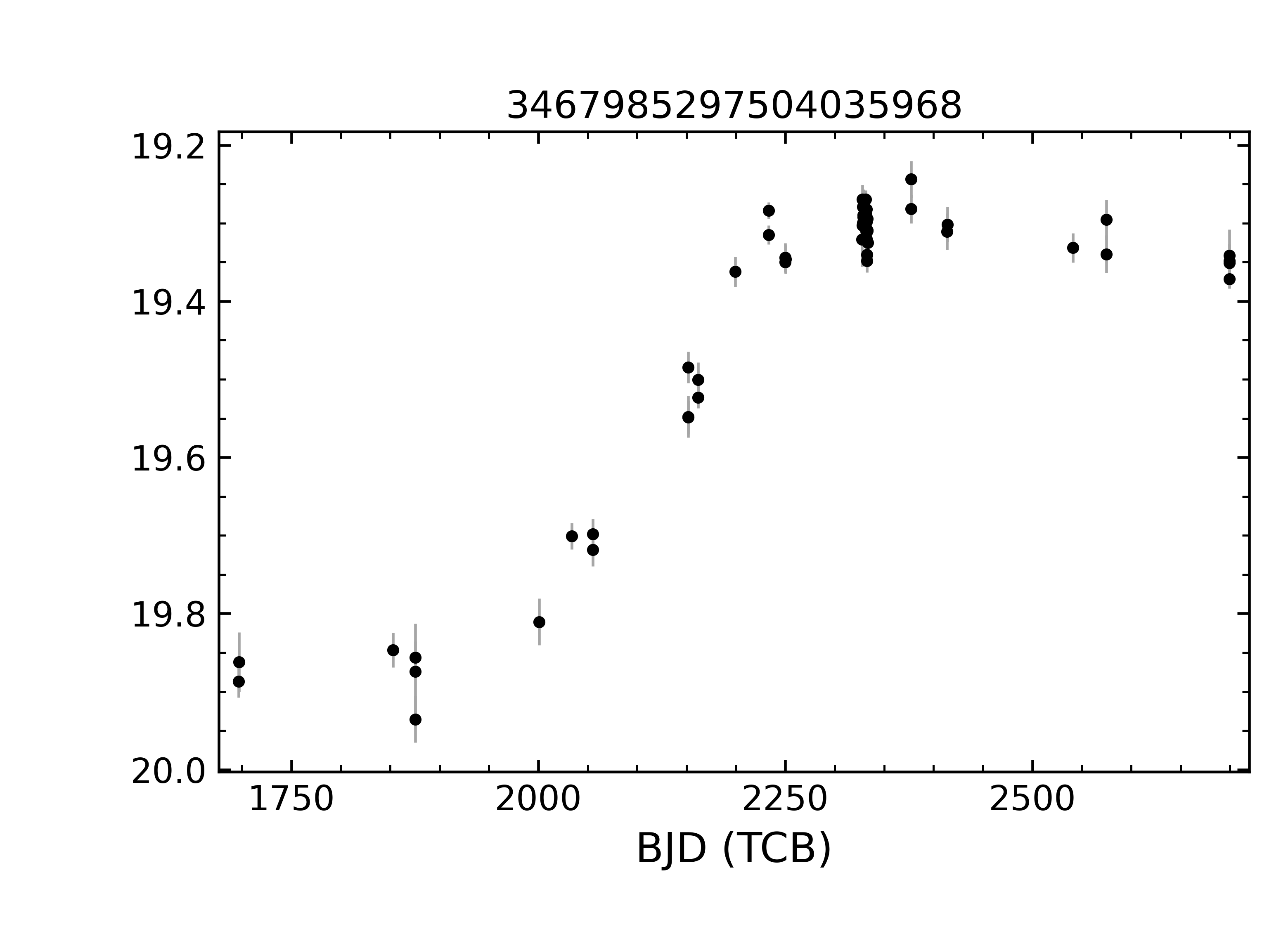}\hfill
\includegraphics[width=0.24\textwidth]{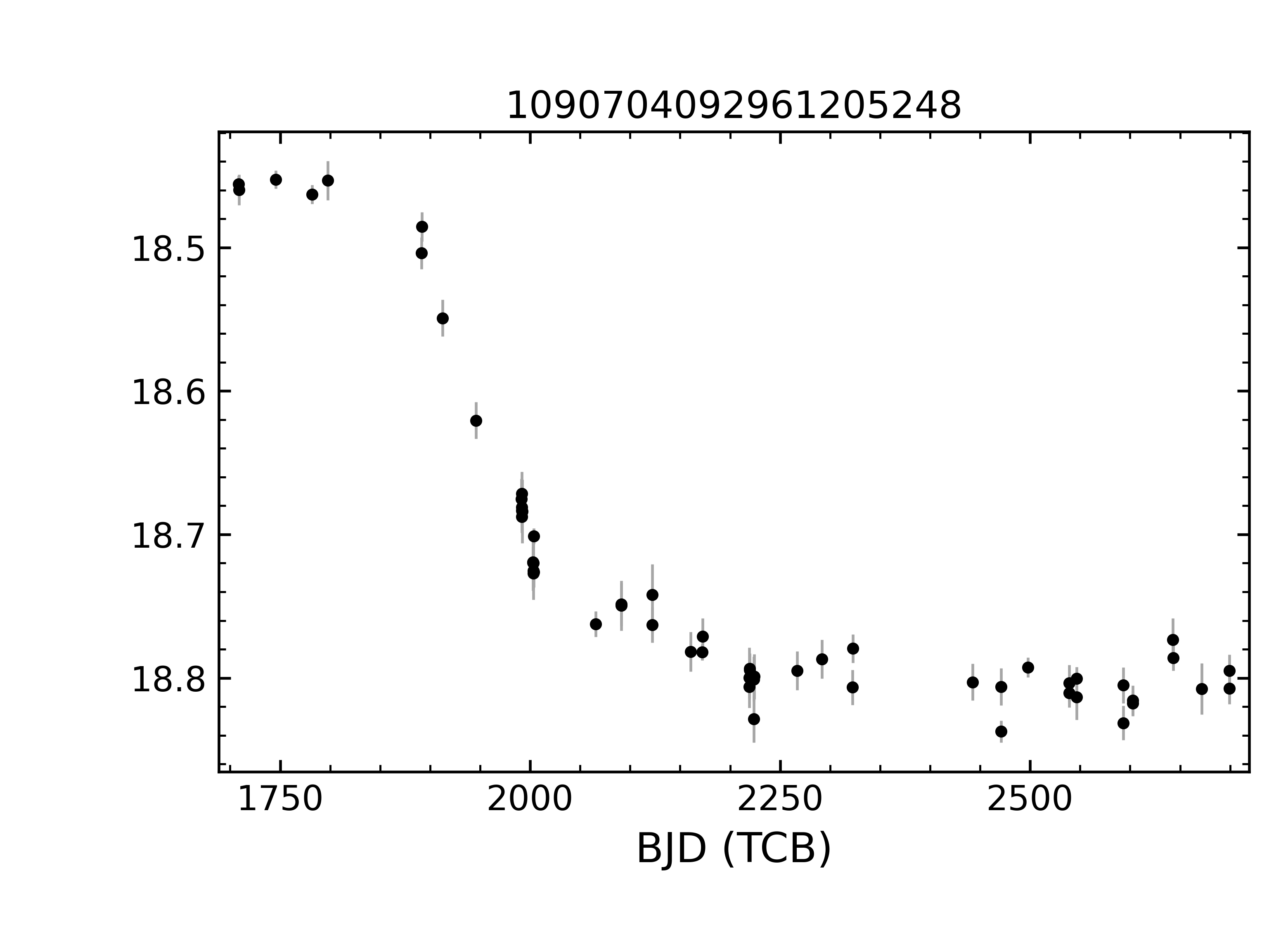}

\caption{Representative Gaia DR3 $G$-band light curves for three samples. 
Top row: non-anomalous control sources. Middle row: anomalous sources that did not satisfy the bimodality criteria. Bottom row: CSAGN candidates. 
Times are shown as BJD (TCB) and magnitudes are in $G$ (mag); error bars are shown in gray.}
\label{fig:rep_lightcurves}
\end{figure*}

To illustrate the outcome of the selection procedure, Figure \ref{fig:rep_lightcurves} presents representative Gaia G band light curves from different stages of the pipeline. The top row shows examples of non-anomalous Gaia DR3 AGNs, the middle row shows anomalous sources that passed the anomaly criteria but did not satisfy the bimodality selection criteria and the bottom row shows representative photometrically selected CSAGN candidates.

\subsubsection{GMM Separation and Peak Balance}

For light curves satisfying the bimodality criteria, we further examined the properties of the two GMM components. This step was used to check whether the two fitted components correspond to two meaningful brightness states, rather than a broad distribution with a weak secondary component.

We measured the GMM separation as
\begin{equation}
    S_{\rm GMM} =
\frac{|\mu_1-\mu_2|}
{\sqrt{\sigma_1^2+\sigma_2^2}},
\end{equation}
where \(\mu_1\) and \(\mu_2\) are the means of the two Gaussian components, and \(\sigma_1\) and \(\sigma_2\) are their standard deviations. This quantity measures how clearly separated the two fitted components are in magnitude space.

We also measured the peak balance as
\begin{equation}
    P_{\rm balance} =
\frac{\min(w_1,w_2)}{\max(w_1,w_2)},
\end{equation}
where \(w_1\) and \(w_2\) are the GMM mixing weights. This quantity measures how evenly the two fitted components are populated. Values close to unity indicate that the two components contain comparable fractions of the light curve data points, while very small values indicate that one component contains only a small fraction of the epochs.

Together, these quantities help distinguish clearer two-component light curves from cases where the magnitude distribution is formally better described by two components, but one component is weak, poorly separated, or represents only a small number of points. These quantities were not used as additional hard selection criteria, but were later incorporated into the candidate-ranking scheme.

\subsection{CSAGN Scoring}

To rank the sources that satisfy the bimodality criteria, we constructed a composite CSAGN score using the following five quantities:
\begin{itemize}
    \item \(\Delta{\rm BIC}\),
    \item GMM separation,
    \item peak balance,
    \item bimodality coefficient (BC),
    \item number of valid epochs.
\end{itemize}

Before combining these quantities, each metric was replaced by its percentile rank within the selected bimodal sample, placing all five diagnostics on a common 0--1 scale. The final CSAGN score was then defined as a weighted linear combination:
\begin{equation}
    S_{\rm CSAGN}
=
4p_{\Delta{\rm BIC}}
+
3p_{\rm sep}
+
3p_{\rm balance}
+
p_{\rm BC}
+
p_{\rm epochs},
\end{equation}
where each \(p\) denotes the percentile-rank value of the corresponding quantity.

The largest weight was assigned to \(\Delta{\rm BIC}\), since it measures how strongly the two-component GMM is favored over the one-component model. The GMM separation and peak balance were also given relatively large weights, because stronger candidates are expected to show two well-separated magnitude states and both states should be meaningfully populated. The BC and the number of epochs were included with smaller weights as supporting diagnostics. The adopted weights are heuristic and reflect our prioritization of the different diagnostics; they should not be interpreted as a uniquely defined prescription. The final score was used only to rank the candidates for inspection and presentation, not as an additional selection cut.

\section{Results} \label{sec:results}

\subsection{Prevalence of Bimodality}

\begin{figure}[t]
    \centering
    \includegraphics[width=0.55\linewidth]{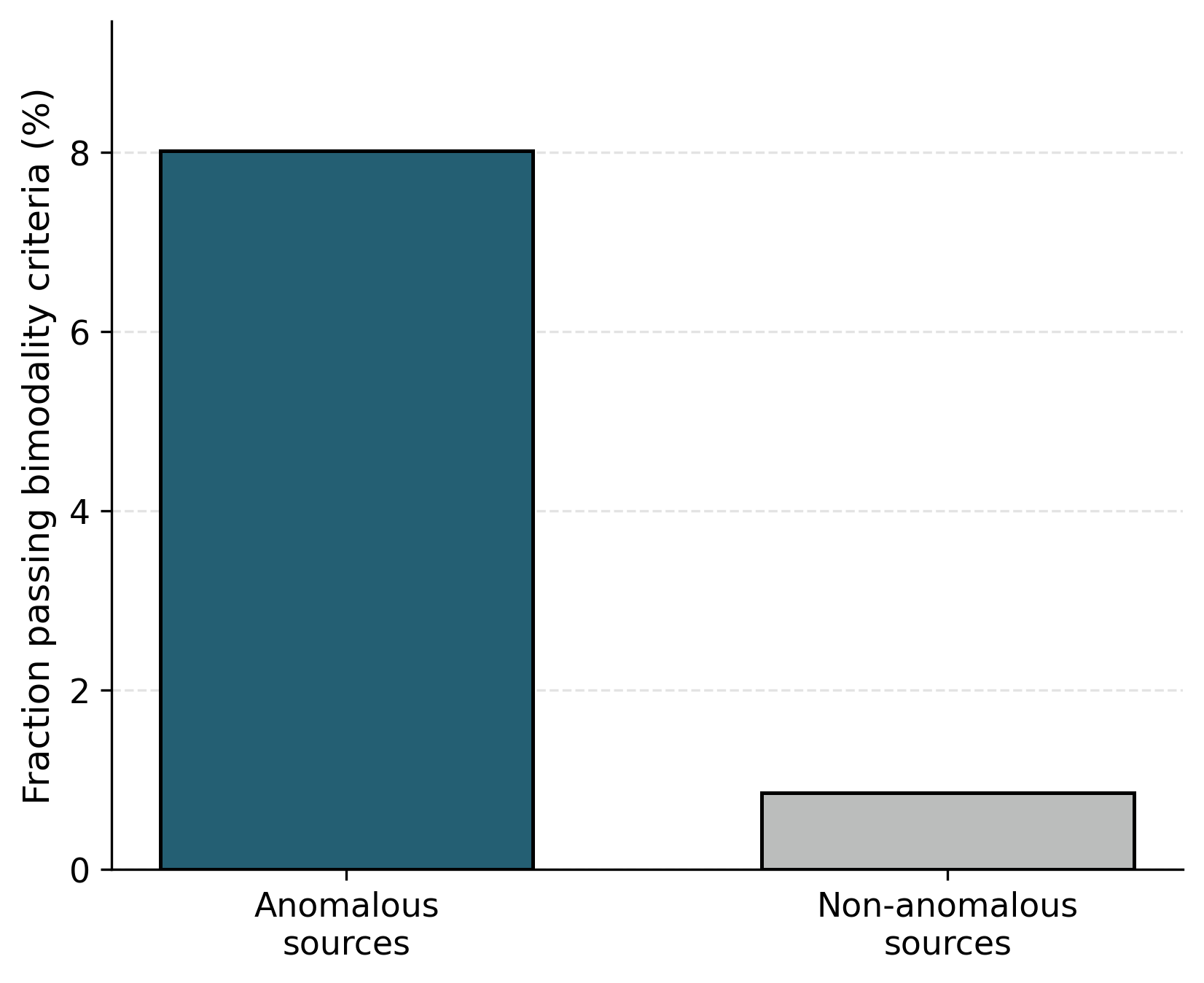}
    \caption{Fraction of sources satisfying the combined bimodality criteria in the anomalous and non-anomalous samples. Among the 8,504 anomalous light curves, 682 satisfy the BC and \(\Delta{\rm BIC}\) criteria, corresponding to \(8.02\%\). In comparison, 7,346 out of the 863,724 non-anomalous sources satisfy the same criteria, corresponding to \(0.85\%\). The anomalous sample is therefore enriched in bimodal light curves by a factor of about \(9.4\), supporting the use of anomaly selection before applying the bimodality analysis.}
    \label{fig:bimodal_fraction}
\end{figure}

Of the 8,504 anomalous sources, 682 satisfy the combined bimodality criteria, corresponding to \(8.02\%\) of the anomalous sample. To test whether this bimodal behavior is preferentially associated with anomalous variability, we applied the same bimodality analysis to the full set of 863,724 non-anomalous sources. Among these, 7,346 passed the same BC and \(\Delta{\rm BIC}\) criteria, corresponding to only \(0.85\%\) (Figure~\ref{fig:bimodal_fraction}).
Thus, the bimodal fraction among anomalous sources is larger than that of the non-anomalous set by a factor of approximately \(9.4\).
This supports the use of anomaly selection before applying the bimodality criteria, as it strongly enriches the sample in light curves with two-state magnitude distributions. At the same time, only a small fraction of the anomalous sources pass the bimodality cuts, which shows that anomalous variability alone is not sufficient to define a CSAGN candidate. The CSAGN candidate sample is therefore selected through both stages; first by identifying unusual variability, and then by isolating the subset whose magnitude distribution is consistent with two distinct brightness states.

\subsection{Dependence on Anomaly Type}

\begin{figure}[t]
    \centering
    \includegraphics[width=0.58\linewidth]{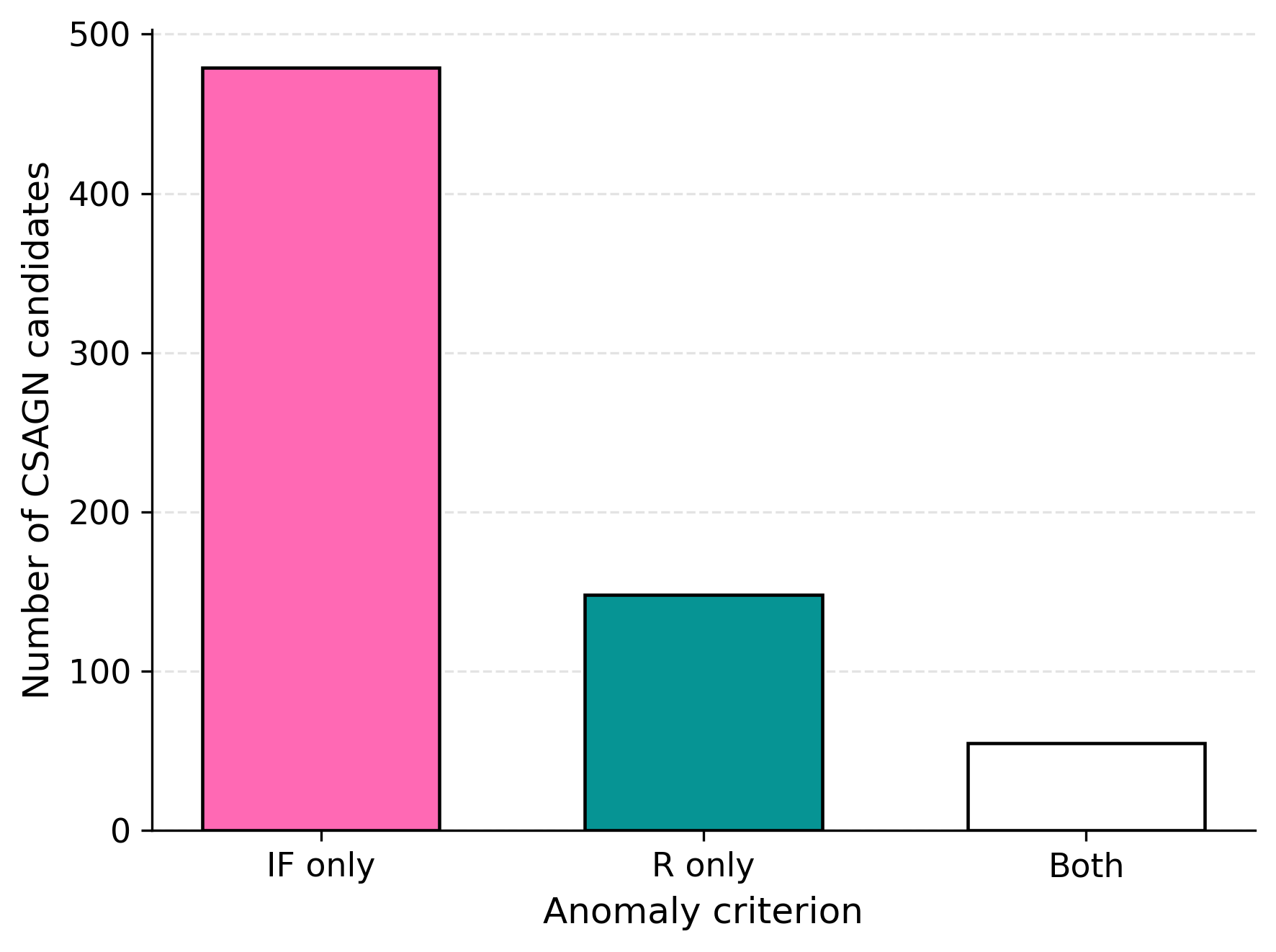}
    \caption{Distribution of the CSAGN candidates by anomaly selection channel. The dominance of IF-selected candidates shows that a large fraction of the final two-state candidates are selected through the Isolation Forest rather than through reconstruction error alone.}
    \label{fig:csagn_anomaly_type}
\end{figure}

We next examined which anomaly selection channel contributes most to the CSAGN candidate sample. Each candidate was classified according to whether it satisfied the Isolation Forest (IF) criterion in the VRAE latent space, the reconstruction error (R) criterion, or both. The CSAGN candidate sample is dominated by latent space anomalies. Out of the 682 CSAGN candidates, 479 sources are selected only by the Isolation Forest criterion, corresponding to \(70.2\%\) of the sample. In comparison, 148 sources are selected only by the reconstruction error criterion, corresponding to \(21.7\%\). Only 55 sources, or \(8.1\%\), satisfy both anomaly criteria (Figure~\ref{fig:csagn_anomaly_type}).

This result shows that most of the final two-state candidates are not simply the light curves with the largest reconstruction errors. Instead, many of them are better identified through their position in the VRAE latent space, where the Isolation Forest separates sources with unusual patterns in the latent representation. The relatively small overlap between the IF and R selections also suggests that the two anomaly measures are sensitive to different aspects of the light curves, although the fraction satisfying both criteria increases from \(2.6\%\) in the full anomalous sample to \(8.1\%\) among the CSAGN candidates. The R criterion highlights sources that are poorly reconstructed by the VRAE, while the IF criterion highlights sources whose latent space representation is isolated from the main population. Using both criteria therefore allows the anomaly selection step to recover a broader range of unusual variability behaviors before the bimodality analysis is applied.

\subsection{Distribution of CSAGN Scores}

The CSAGN score was used to rank the 682 CSAGN candidates after the bimodality selection. As described above, this score combines the percentile-rank values of \(\Delta{\rm BIC}\), GMM separation, peak balance, BC, and the number of valid epochs. Therefore, higher scores generally correspond to sources with stronger evidence for bimodality, better-separated GMM components, and a more balanced occupation of the two brightness states.

\begin{figure*}[t]
        \centering
        \includegraphics[width=\linewidth]{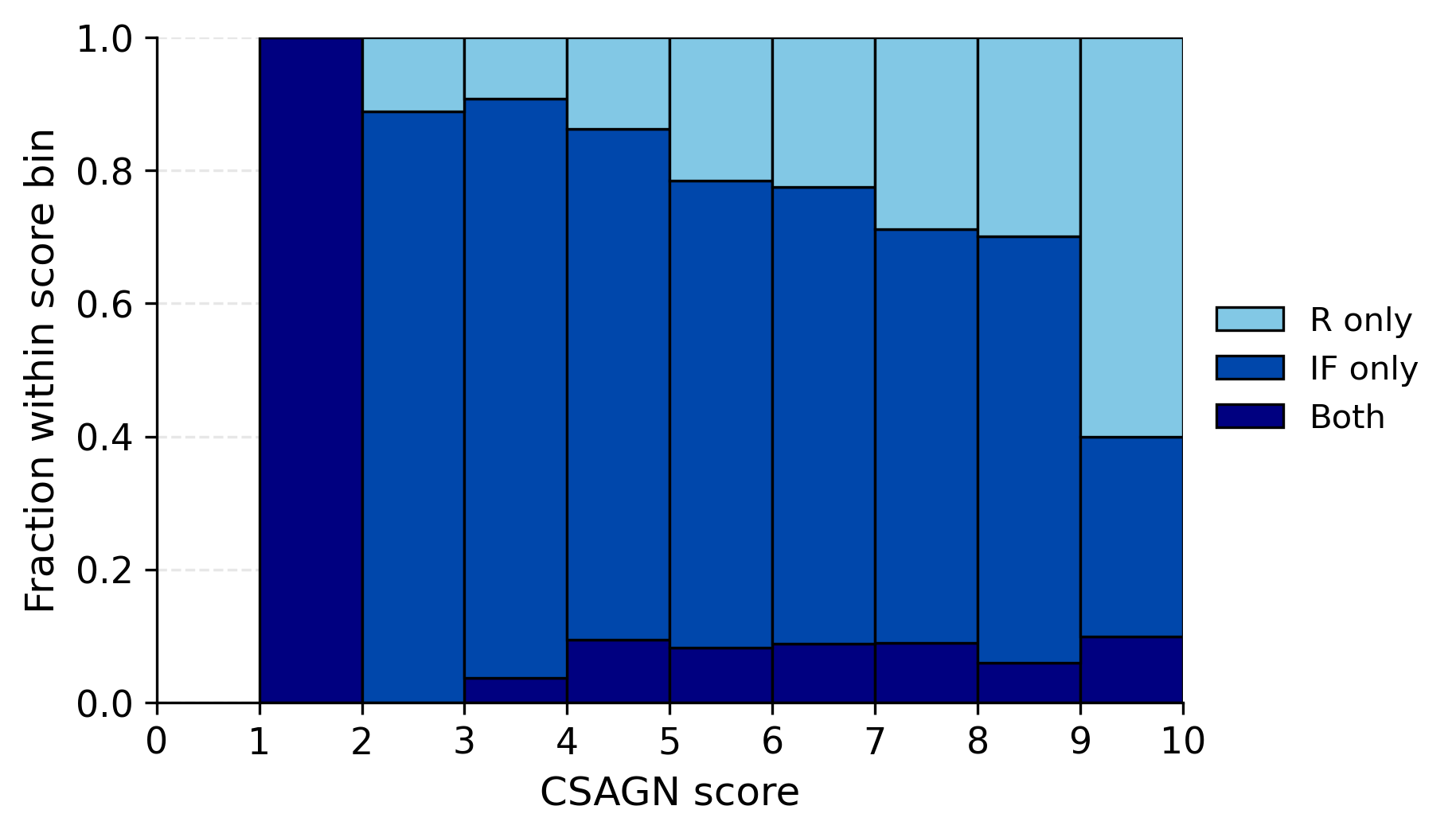}
        \caption{Relation of CSAGN score to the initial anomaly-selection criteria. Each bin is normalized by the number of candidates in that score interval. IF-selected candidates dominate most of the score ranges, though R selected sources increase with increasing CSAGN score.}
        \label{fig:csagn_score_type_fraction}
\end{figure*}

We examined how the composition of the initial anomaly selection channels changes across the CSAGN-score range. Figure~\ref{fig:csagn_score_type_fraction} shows the fractional contribution of R-only, IF-only, and jointly selected candidates in each CSAGN-score bin, and the same values are listed in Table~\ref{tab:csagn_score_anomaly_type}. The IF-only candidates dominate most of the score range, with the largest IF-only fraction occurring in the \([2,3)\) score bin. From there, the IF-only fraction generally decreases toward higher CSAGN scores. In contrast, the R-only fraction shows a small dip between the \([2,3)\) and \([3,4)\) bins, but then increases steadily up to the \([8,9)\) bin. In the highest score bin, \([9,10)\), the R-only fraction becomes the largest contribution.

The sources selected by both criteria make up a small fraction of most score bins. The only exception is the lowest occupied score bin, \([1,2)\), which contains a single source selected by both IF and R. Because this bin contains only one object, and the highest score bin contains only 10 objects, the fractions at the two ends of the score range should be interpreted with some caution. Overall, the trend is consistent with the result shown in Figure~\ref{fig:csagn_anomaly_type}. In other words, most of the CSAGN candidates are selected by the Isolation Forest in the latent space, while the highest-ranking candidates include a larger contribution from sources with large reconstruction errors. This behavior may reflect a limitation of the latent representation for the most extreme light curves. \citet{2021AJ....162..206S} note that, for sources with large reconstruction errors, the latent space vectors may no longer compress the true variability signal reliably, and the corresponding Isolation Forest scores therefore become less trustworthy. In this sense, the increasing contribution of reconstruction-selected sources toward the highest CSAGN scoring light curves may indicate that the highest-ranking bimodal candidates depart sufficiently from the variability patterns learned by the model and that the reconstruction error provides a clearer indication of their anomalous behavior than the latent space IF score.

\begin{table}[t]
\centering
\caption{Composition of the CSAGN candidate sample in bins of CSAGN score. The percentages are computed within each score bin and show the fraction of candidates selected only by the R criterion, only by the IF criterion, or by both criteria.}
\label{tab:csagn_score_anomaly_type}
\begin{tabular}{ccccc}
\hline
CSAGN score bin & $N$ & R only (\%) & IF only (\%) & Both (\%) \\
\hline
$[0,1)$  & 0   & 0.0  & 0.0  & 0.0 \\
$[1,2)$  & 1   & 0.0  & 0.0  & 100.0 \\
$[2,3)$  & 18  & 11.1 & 88.9 & 0.0   \\
$[3,4)$  & 54  & 9.3  & 87.0 & 3.7   \\
$[4,5)$  & 116 & 13.8 & 76.7 & 9.5   \\
$[5,6)$  & 158 & 21.5 & 70.3 & 8.2   \\
$[6,7)$  & 147 & 22.4 & 68.7 & 8.8   \\
$[7,8)$  & 111 & 28.8 & 62.2 & 9.0   \\
$[8,9)$  & 67  & 29.9 & 64.2 & 6.0   \\
$[9,10)$ & 10  & 60.0 & 30.0 & 10.0  \\
\hline
\end{tabular}
\end{table}

\subsection{Redshift, Source Classification, Absolute Magnitude, and Color Distributions}
\label{sec:z_class_mag_color_distributions}

We next explored the basic properties of the CSAGN candidate sample using Gaia photometry, classifications, and redshifts to compute absolute magnitudes and colors. We compiled redshift information for the CSAGN candidates by crossmatching their coordinates with several external catalogs and by using Gaia DR3 Quasar Classifier (QSOC) redshifts. The QSOC redshifts are estimated from the low-resolution Gaia BP/RP spectra \citep{2023A&A...674A...2D}, and the pipeline also provides QSOC processing flags, \texttt{flags\_qsoc}, that mark possible issues or degeneracies in the redshift estimate \citep{2023A&A...674A..41G}. When more than one redshift was available for a source, we adopted a single redshift according to the following priority order. External spectroscopic redshifts from SDSS DR16Q \citep{2020ApJS..250....8L}, LAMOST DR6 -- DR9 \citep{2023ApJS..265...25J} and DESI DR1 \citep{2026AJ....171..285D}, together with spectroscopic redshifts compiled in MILLIQUAS (version 8) \citep{2023OJAp....6E..49F}, were given highest priority. Gaia QSOC redshifts were then used when no external spectroscopic redshift was available, with \texttt{flags\_qsoc}=0 treated as the cleanest QSOC category. For two sources with no available spectroscopic or Gaia QSOC redshift, we adopted photometric redshift estimates from the DESI Legacy Imaging Surveys Data Release 8 (LS DR8; \citealt{2022MNRAS.512.3662D}). 

Redshifts were found for 662 of the 682 candidates (See Table~\ref{tab:redshift_provenance}). The largest contributions come from Gaia QSOC redshifts with \texttt{flags\_qsoc}=0, SDSS DR16Q, and Gaia QSOC redshifts with other nonzero flags. A smaller number of redshifts were obtained from LAMOST DR6--DR9, LS DR8 photo-\(z\), DESI DR1, and MILLIQUAS v8. The adopted redshifts span \(0.085\leq z \leq 5.754\), with a median redshift of \(z=1.11\) and a mean redshift of \(z=1.25\). The remaining 20 candidates do not have an adopted redshift.

We also examined the source classifications reported by the Discrete Source Classifier (DSC) \citep{2023A&A...674A..31D}, a module of Coordination Unit 8 (CU8) within the Gaia Data Processing and Analysis Consortium (DPAC). Based on the DSC classifications, 641 of the 682 candidates are classified as quasars, 32 as stars, and 9 as galaxies. Of the 20 candidates without an adopted redshift, 19 are classified by the DSC as stars and one as a galaxy. Therefore, the CSAGN candidate sample is strongly dominated by quasar-classified sources, although a small number of candidates may be contaminants or may have ambiguous classifications.

\begin{table}[t]
\centering
\caption{Adopted redshift provenance for the 682 CSAGN candidates used in the source-property analysis.}
\label{tab:redshift_provenance}
\begin{tabular}{lc}
\hline
Redshift source & Number of candidates \\
\hline
Gaia QSOC, \texttt{flags\_qsoc}=0 & 255 \\
SDSS DR16Q & 204 \\
Gaia QSOC, \texttt{flags\_qsoc}\(\neq 0\) & 196 \\
LAMOST DR6--DR9 & 3 \\
LS DR8 photo-\(z\) & 2 \\
DESI DR1 & 1 \\
MILLIQUAS v8 & 1 \\
No redshift data & 20 \\
\hline
Total & 682 \\
\hline
\end{tabular}
\end{table}

\begin{figure*}[t]
    \centering

    \begin{subfigure}{0.47\textwidth}
        \centering
        \includegraphics[width=\linewidth]{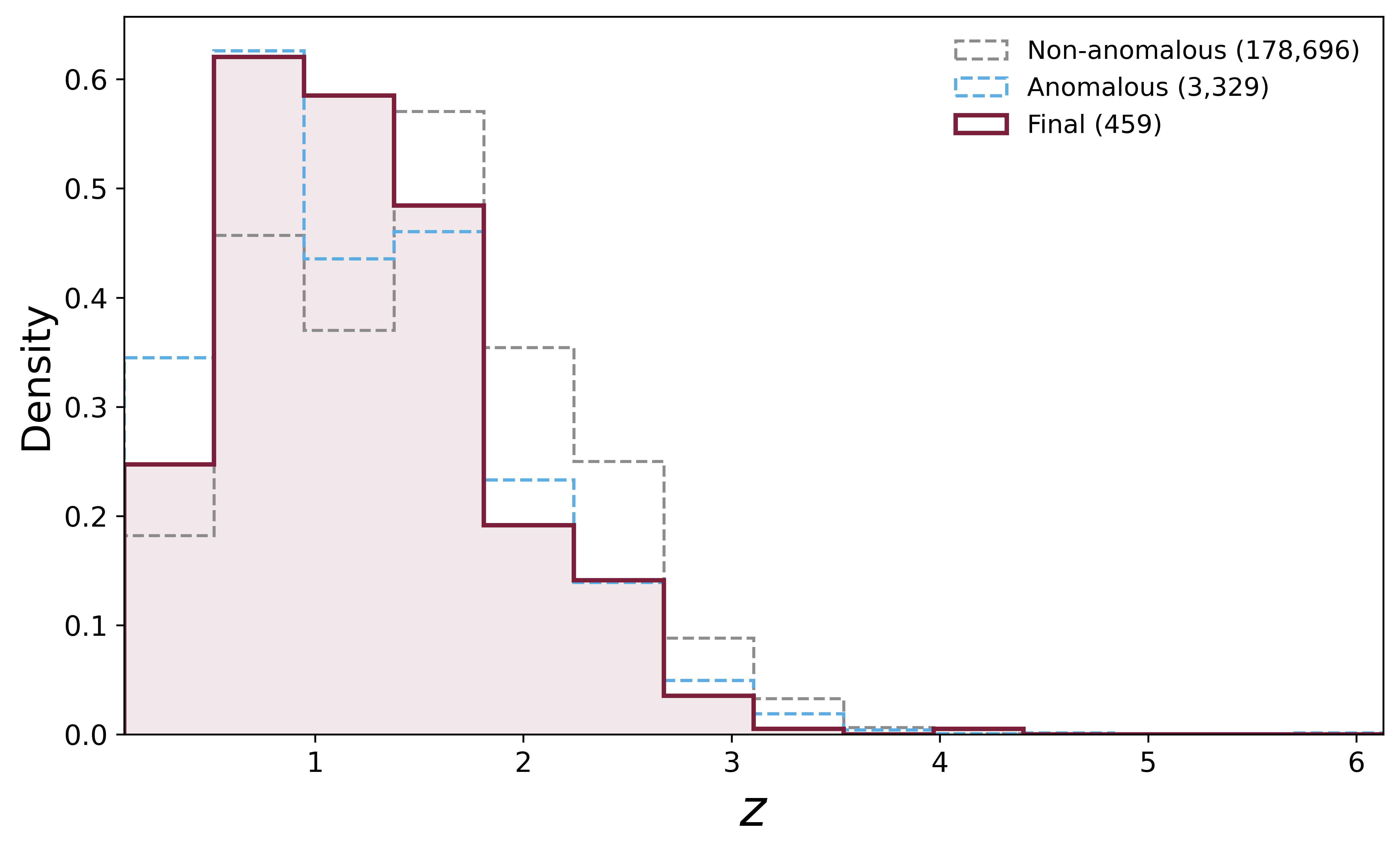}
        \caption{}
        \label{fig:z_dist_high_quality}
    \end{subfigure}
    \hfill
    \begin{subfigure}{0.47\textwidth}
        \centering
        \includegraphics[width=\linewidth]{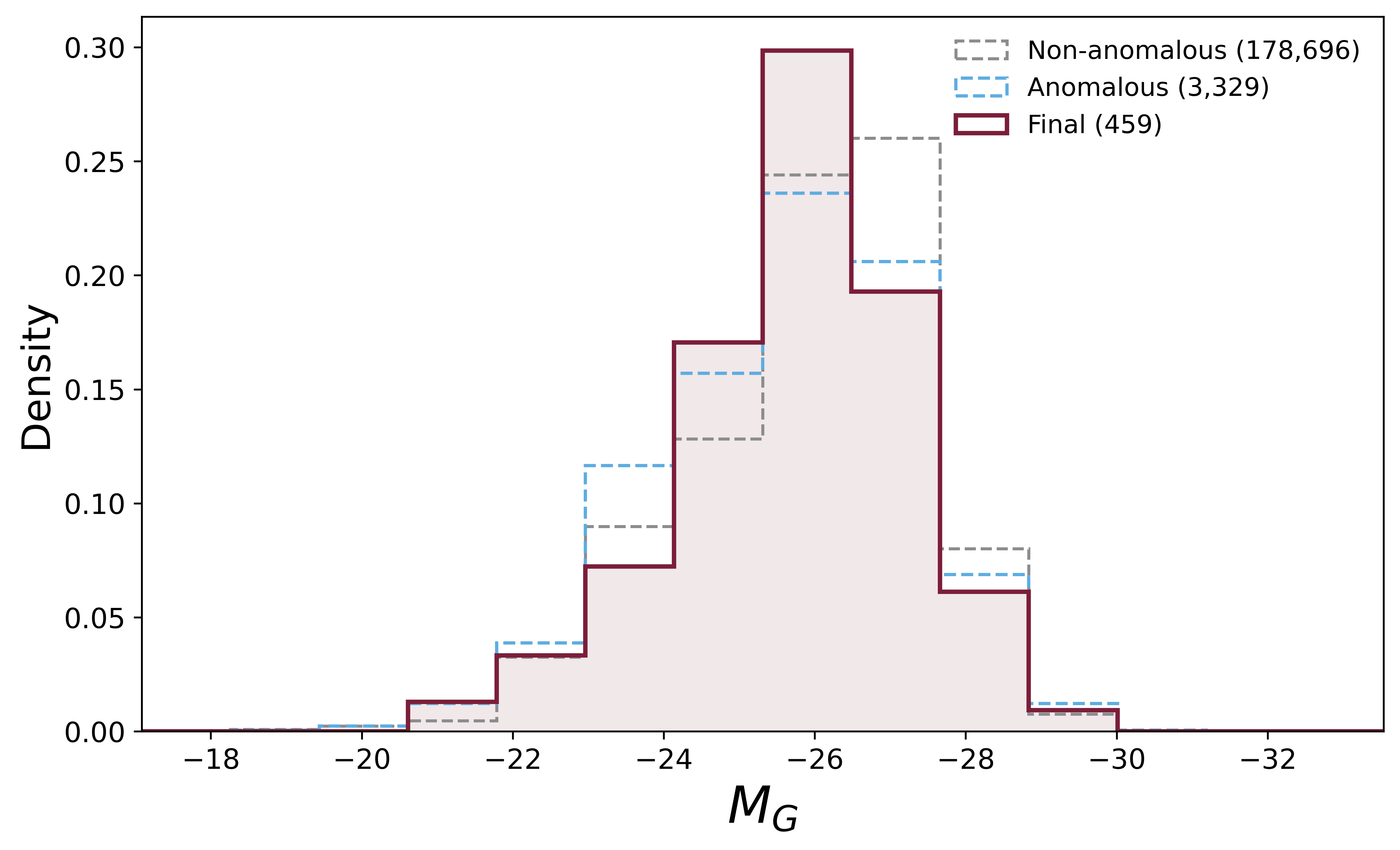}
        \caption{}
        \label{fig:MG_dist_high_quality}
    \end{subfigure}

    \vspace{0.12in}

    \begin{subfigure}{0.47\textwidth}
        \centering
        \includegraphics[width=\linewidth]{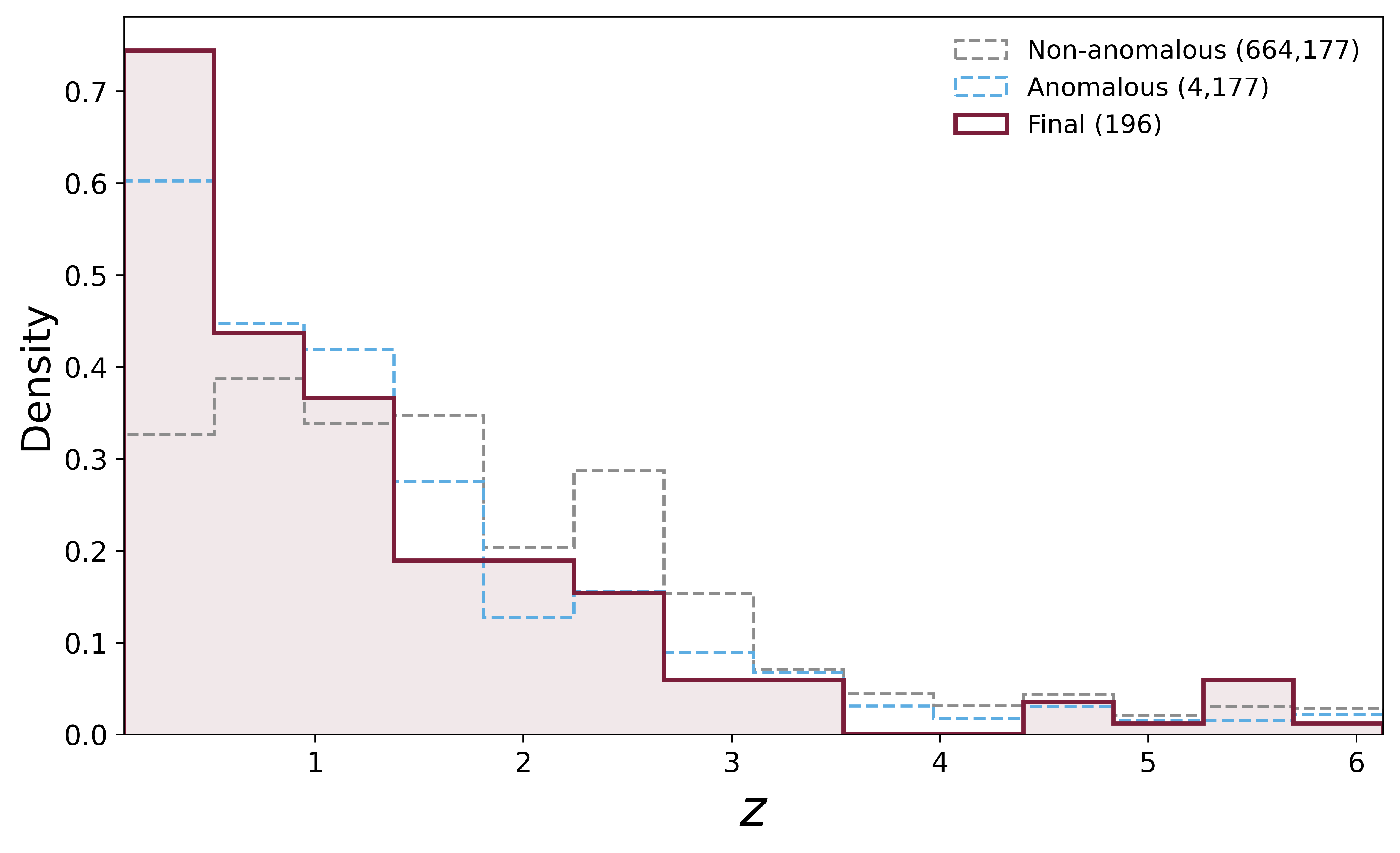}
        \caption{}
        \label{fig:z_dist_flag_nonzero}
    \end{subfigure}
    \hfill
    \begin{subfigure}{0.47\textwidth}
        \centering
        \includegraphics[width=\linewidth]{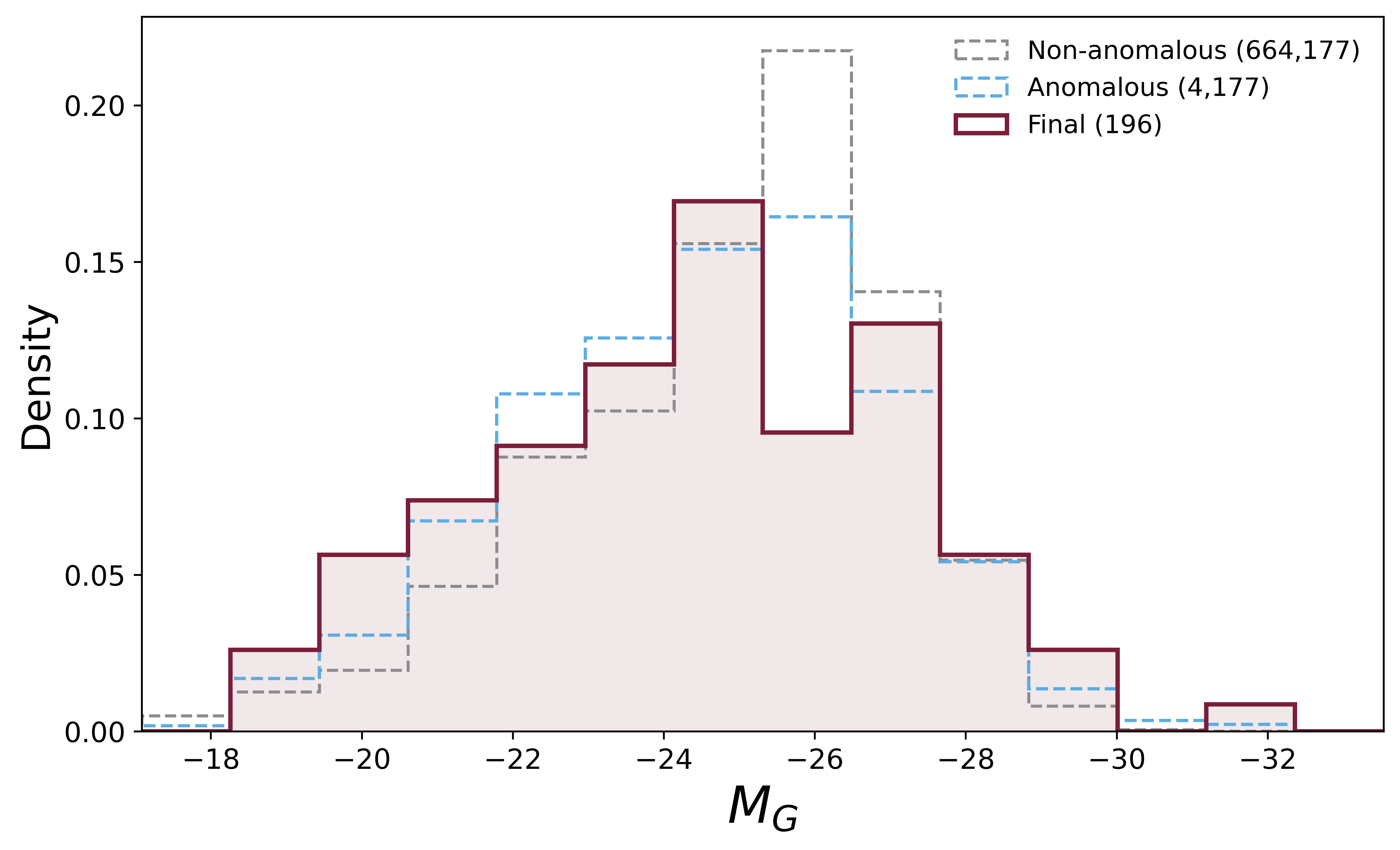}
        \caption{}
        \label{fig:MG_dist_flag_nonzero}
    \end{subfigure}

    \vspace{0.05in}

    \caption{Normalized distributions of redshift and G band absolute magnitude derived from Gaia DR3 photometry. The comparison is separated according to the Gaia QSOC redshift-flag grouping used to construct the reference samples. The top row shows the higher-quality comparison, where the CSAGN candidates have either SDSS DR16Q redshifts or Gaia QSOC redshifts with \texttt{flags\_qsoc}=0. The corresponding reference samples are drawn from sources with Gaia QSOC \texttt{flags\_qsoc}=0. The bottom row shows the nonzero-flag Gaia QSOC comparison, where the CSAGN candidates have adopted Gaia QSOC redshifts with nonzero \texttt{flags\_qsoc} and the reference samples are drawn from the same nonzero-flag QSOC category. Panels (a) and (c) show the redshift distributions, panels (b) and (d) show the corresponding \(M_G\) distributions. All histograms are normalized to unit area to allow comparison between samples of different sizes.}
    \label{fig:z_MG_color_distributions}
\end{figure*}

Figure~\ref{fig:z_MG_color_distributions} compares the redshift and absolute magnitude distributions of the CSAGN candidates with those of the non-anomalous and anomalous light curves with available Gaia QSOC redshifts. The top row compares candidates with higher quality redshifts, either from Gaia QSOC \texttt{flags\_qsoc}=0 redshifts or from ground-based spectroscopy, primarily SDSS DR16Q. The bottom row shows the corresponding comparison for the nonzero-flag Gaia QSOC sources with lower quality redshifts. All histograms are normalized to the unit area. The three samples show similar Gaia \(BP-RP\) color distributions in both redshift-quality groups.

For the higher-quality redshift comparisons, the reference samples contain 3,329 anomalous sources and 178,696 non-anomalous sources. Figure~\ref{fig:z_MG_color_distributions}(a) shows the corresponding redshift distributions. The CSAGN candidates have a median redshift of \(z=1.15\), which is similar to the anomalous reference sample with a median \(z=1.12\). Both are lower than the non-anomalous reference sample, which has a median redshift of \(z=1.48\). For the lower quality redshift comparison, the reference samples contain 4,177 anomalous sources and 664,177 non-anomalous sources. Figure~\ref{fig:z_MG_color_distributions}(c) shows the redshift distributions for these sources. Again, the CSAGN candidates have a redshift distribution similar to the anomalous sources with a median \(z=0.92\), compared with \(z=1.08\) for the anomalous reference sample and \(z=1.50\) for the non-anomalous reference sample.

We next compared the Gaia \(G\)-band absolute magnitude distributions. For each source with an adopted redshift, we computed \(M_G\) from the median observed Gaia \(G\)-band magnitude and the luminosity distance, assuming a flat \(\Lambda\)CDM cosmology with \(H_0=70~{\rm km~s^{-1}~Mpc^{-1}}\) and \(\Omega_m=0.3\). No rest-frame K-correction was applied to the Gaia \(G\)-band magnitudes.

Figure~\ref{fig:z_MG_color_distributions}(b) shows the \(M_G\) distributions for the higher-quality redshift subset. The CSAGN candidates have a median absolute magnitude of \(M_G=-25.88\), close to the anomalous reference sample with median \(M_G=-25.79\), and slightly fainter than the non-anomalous reference sample with median \(M_G=-26.17\). Figure~\ref{fig:z_MG_color_distributions}(d) shows the \(M_G\) distributions for the nonzero-flag Gaia QSOC subset. The CSAGN candidates again overlap with the corresponding reference samples, but their median absolute magnitude, \(M_G=-24.51\), is fainter than both the anomalous reference sample, with median \(M_G=-24.80\), and the non-anomalous reference sample, with median \(M_G=-25.29\). The lower luminosities observed among sources with lower-quality spectra may partly reflect increased spectral noise in intrinsically fainter objects. Although the absolute magnitude distributions are broad and overlapping, the median offsets for CSAGN candidates relative to the corresponding subset of the full Gaia AGN catalog (\(\Delta{\rm M_G} \approx 0.3\) fainter for high quality spectra and  \(\Delta{\rm M_G} \approx 0.8\) fainter for poor quality spectra) suggest that the selected CSAGN candidates are, on average, intrinsically fainter.

\subsection{Color--Magnitude Diagrams}
\label{sec:cmd}

We next examined the the CSAGN candidates in color--magnitude space. For this comparison, we used the median Gaia DR3 \(BP-RP\) color and the quantity \(M_G-5\log_{10}h\), where \(M_G\) is the Gaia G band absolute magnitude computed from the median observed G band magnitude and the adopted redshift, and \(h\equiv H_0/(100~{\rm km~s^{-1}~Mpc^{-1}})=0.7\).

\begin{figure*}[t]
    \centering

    \begin{subfigure}{0.32\textwidth}
        \centering
        \includegraphics[width=\linewidth]{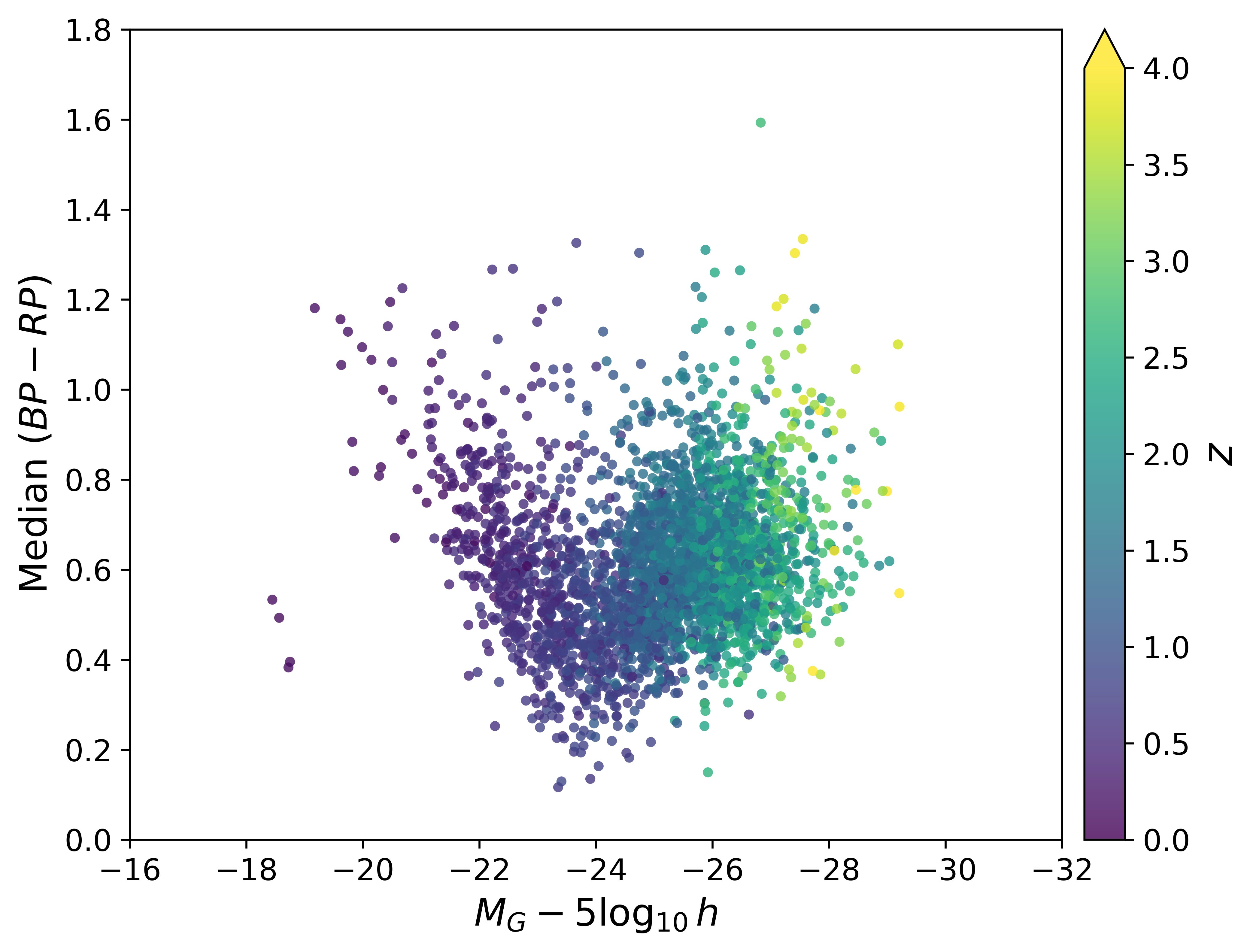}
        \caption{Non-anomalous, higher-quality}
        \label{fig:cmd_highqual_non}
    \end{subfigure}
    \hfill
    \begin{subfigure}{0.32\textwidth}
        \centering
        \includegraphics[width=\linewidth]{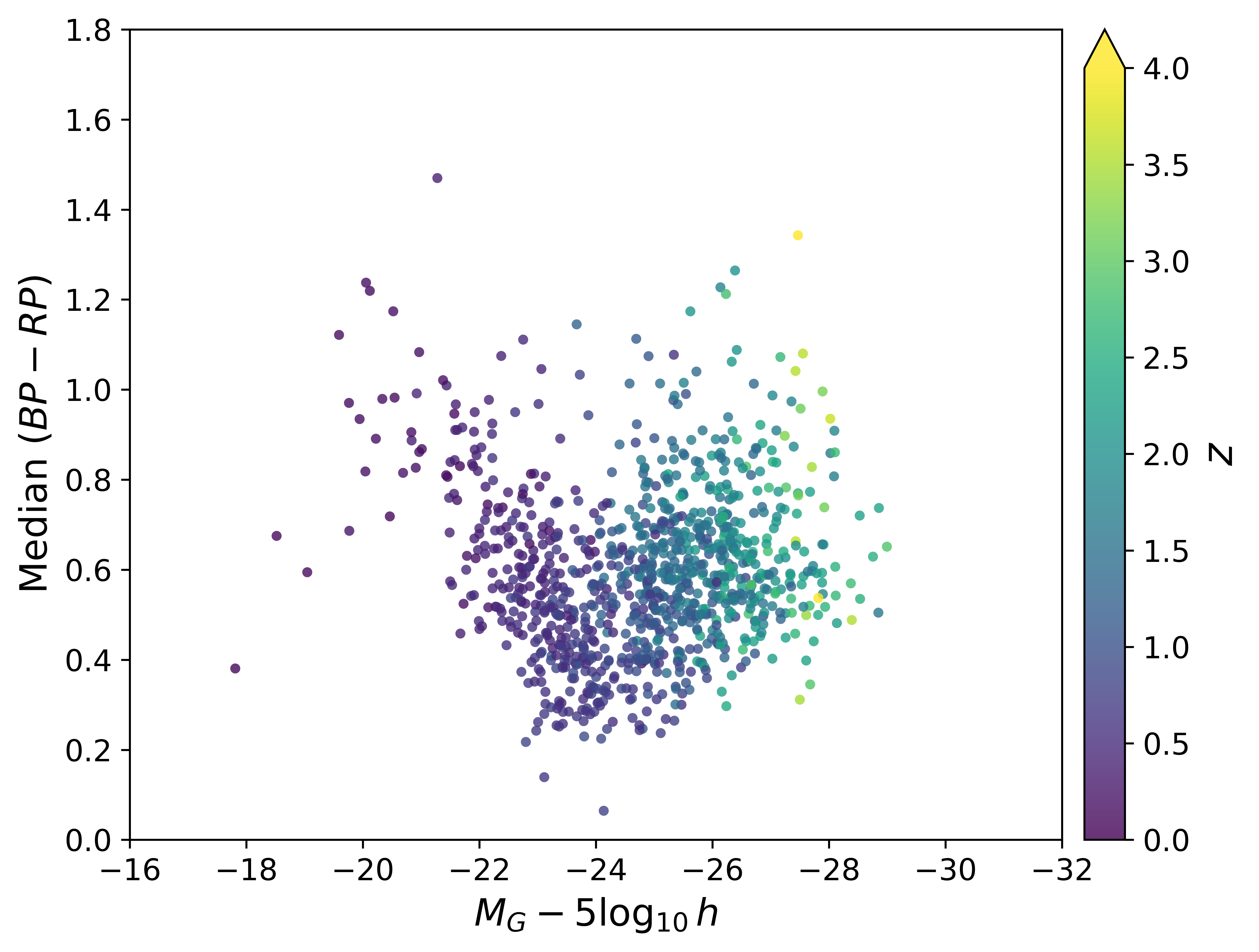}
        \caption{Anomalous, higher-quality}
        \label{fig:cmd_highqual_anom}
    \end{subfigure}
    \hfill
    \begin{subfigure}{0.32\textwidth}
        \centering
        \includegraphics[width=\linewidth]{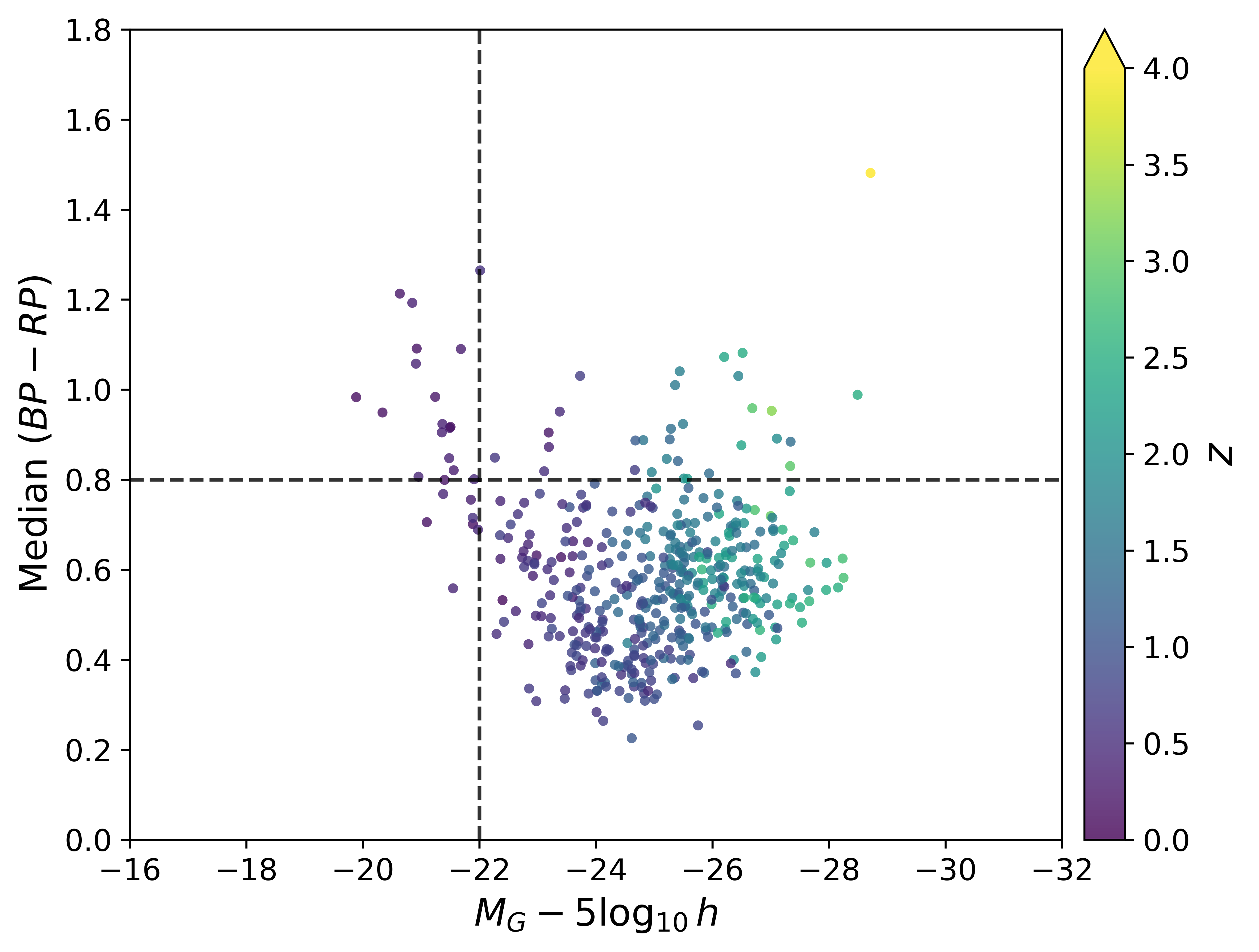}
        \caption{CSAGN candidates, higher-quality}
        \label{fig:cmd_highqual_final}
    \end{subfigure}

    \vspace{0.5em}

    \begin{subfigure}{0.32\textwidth}
        \centering
        \includegraphics[width=\linewidth]{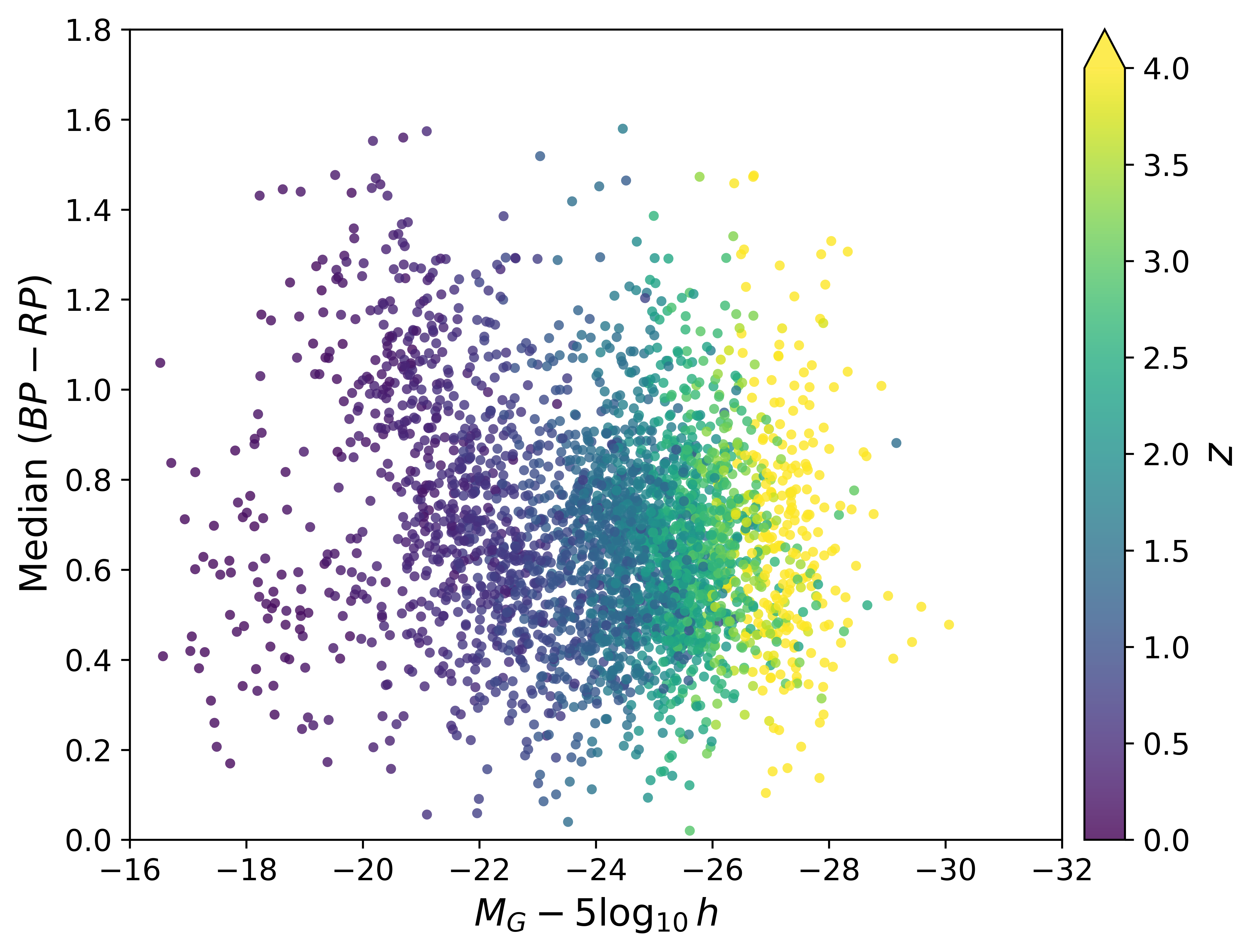}
        \caption{Non-anomalous, lower-quality}
        \label{fig:cmd_nonzero_non}
    \end{subfigure}
    \hfill
    \begin{subfigure}{0.32\textwidth}
        \centering
        \includegraphics[width=\linewidth]{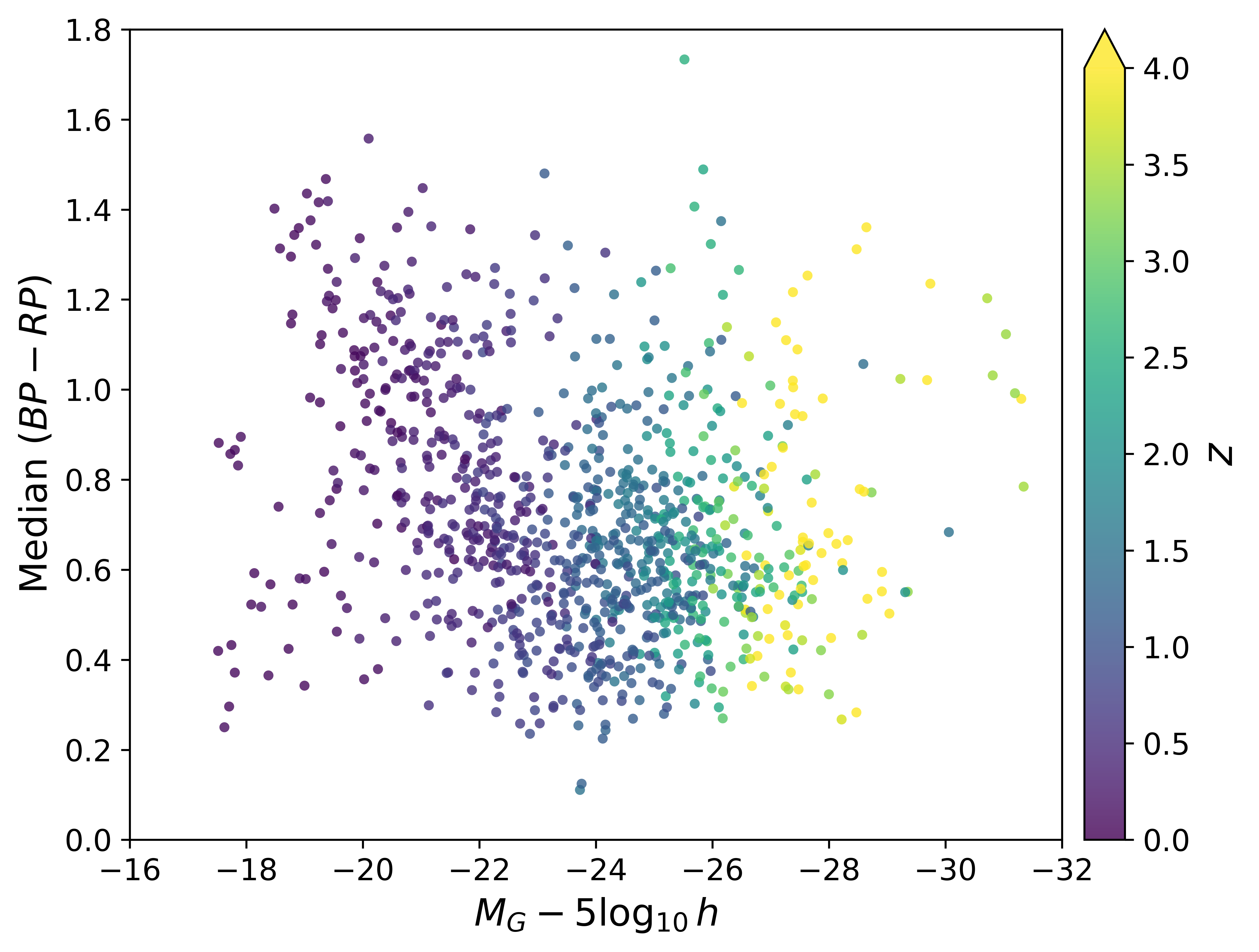}
        \caption{Anomalous, lower-quality}
        \label{fig:cmd_nonzero_anom}
    \end{subfigure}
    \hfill
    \begin{subfigure}{0.32\textwidth}
        \centering
        \includegraphics[width=\linewidth]{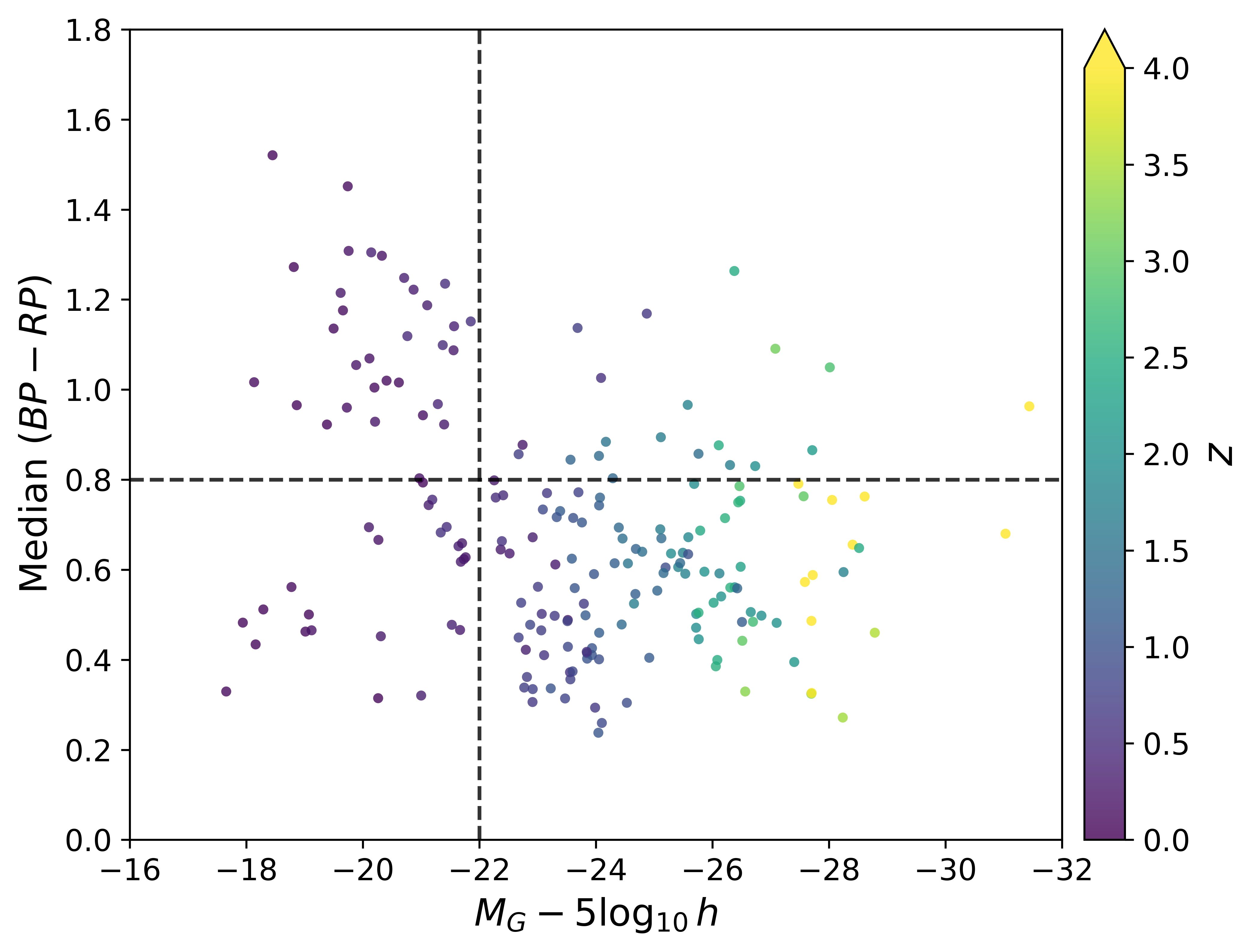}
        \caption{CSAGN candidates, lower-quality}
        \label{fig:cmd_nonzero_final}
    \end{subfigure}

    \caption{Color--magnitude diagrams of median Gaia DR3 \(BP-RP\) color versus \(M_G-5\log_{10}h\) for the non-anomalous reference sample, anomalous reference sample, and CSAGN candidate sample, separated by redshift quality. The upper row shows the higher-quality redshift subset where the comparison samples have finite Gaia QSOC redshifts with \texttt{flags\_qsoc}=0, while the CSAGN subset uses SDSS DR16Q redshifts where available together with Gaia QSOC redshifts having \texttt{flags\_qsoc}=0. The lower row shows the corresponding comparison for sources with Gaia QSOC redshifts having nonzero \texttt{flags\_qsoc}. Points are colored by redshift.}
    \label{fig:cmd_comparison}
\end{figure*}

Because the inferred absolute magnitude depends directly on redshift, we separated the CMD comparison according to redshift quality. Figure~\ref{fig:cmd_comparison} compares the CMDs of the higher-quality redshift subset in the upper row and the lower-quality subset in the lower row. The lower-quality redshifts generally result in more uncertain absolute magnitudes and produce CMDs with sources more dispersed along the x-axis. The CMDs colored by redshift show a clear redshift-dependence across the diagram. Lower-redshift candidates are more common on the lower-luminosity side of the CMD, while higher-redshift candidates are preferentially found toward the more luminous part of the diagram. This is consistent with the behavior expected in a flux-limited sample, where less luminous quasars tend to be observed at lower redshifts.

To describe the different parts of the CMDs, we use two dashed guide lines. The vertical line at \(M_G - 5\log_{10}h=-22\) separates brighter from fainter sources, while the horizontal line at \(\mathrm{median}(BP-RP)=0.80\) separates bluer from redder sources. In all panels, most sources lie on the brighter side of the line \(M_G - 5\log_{10}h=-22\), particularly in the higher-quality redshift subsets. Nevertheless, we noticed an extension of the CSAGN sample toward the fainter and redder quadrant of the CMD. To evaluate the significance of this difference, we performed a Monte Carlo test using the 655 CSAGN candidates in the two redshift-quality groups shown in Figure~\ref{fig:cmd_comparison}, 459 in higher-quality group (top row) and 196 in the lower-quality group (bottom row). In the combined CSAGN sample, 48 sources (\(7.33\%\)) lie in the upper-left quadrant of the CMD. We generated $10^{5}$ random samples from the non-anomalous population, each with the same size and composition of redshift-quality, drawing 459 from the \texttt{flags\_qsoc}=0 and 196 from the \texttt{flags\_qsoc}\(\ne 0\). These samples contained an average of 25.24 sources in the upper-left quadrant of the CMD of the non-anomalous set (\(3.85\%\)), nearly half the fraction found among the CSAGN candidates. Only four random samples contained 48 or more sources in this region, giving a corrected one-sided Monte Carlo probability of \(p\simeq5\times10^{-5}\). The fraction of CSAGN candidates in the upper-left region is therefore significantly higher than expected from random samples of the same size and redshift-quality composition drawn from the corresponding non-anomalous subsets. 

\begin{figure}[t]
    \centering
    \includegraphics[width=0.72\linewidth]{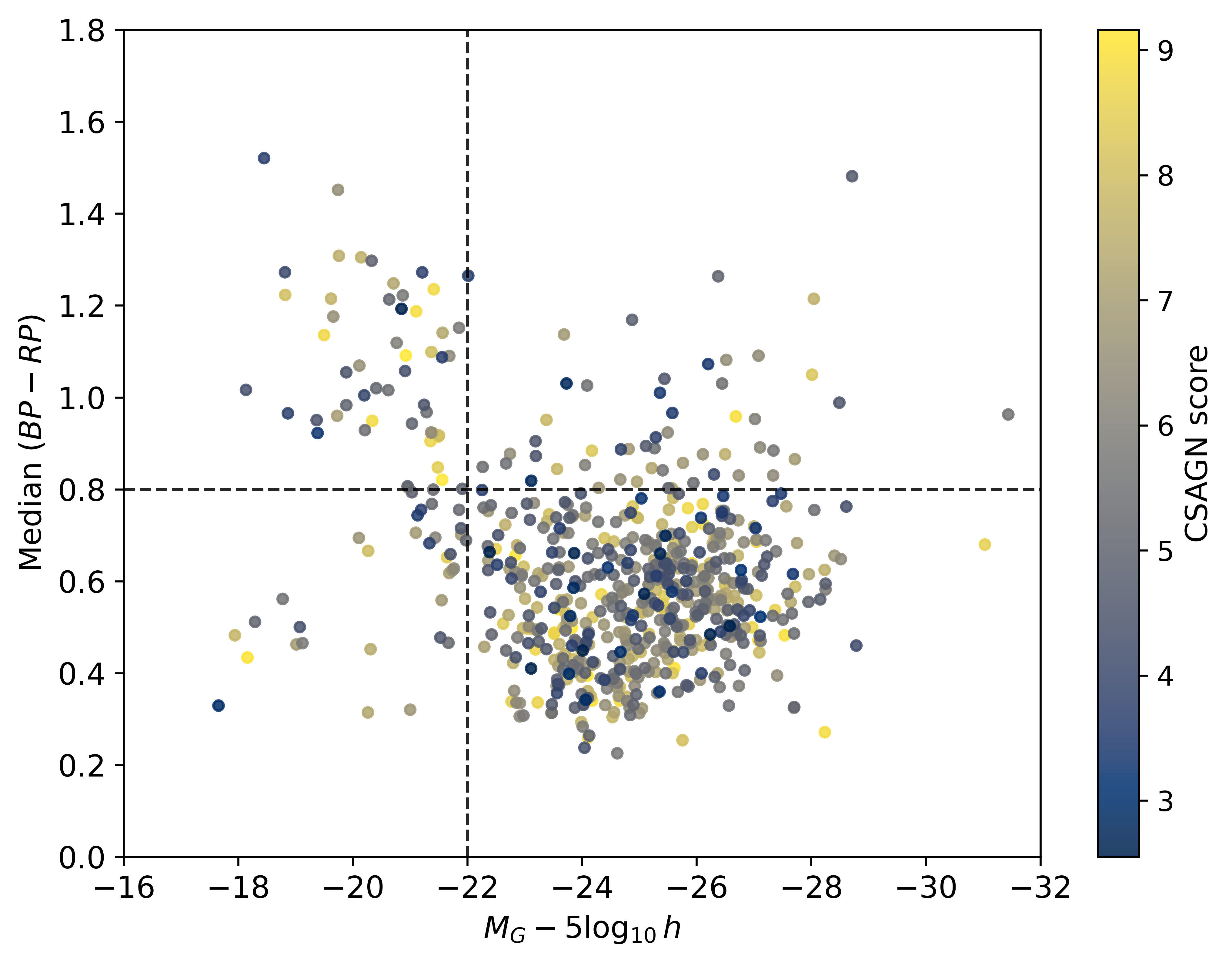}
    \caption{Color--magnitude diagram of the CSAGN candidate sample, showing median Gaia DR3 \(BP-RP\) color versus \(M_G-5\log_{10}h\), colored by CSAGN score. The dashed lines at \(M_G - 5\log_{10}h=-22\) and \(\mathrm{median}(BP-RP)=0.80\) are used only as visual guides to describe the CMD regions. The upper-left region, corresponding to fainter absolute magnitudes and redder colors, contains all galaxy-classified sources in the CMD subset and several of the higher-score candidates.}
    \label{fig:cmd_final_score}
\end{figure}

To examine this CMD feature more closely, we turn to the full CSAGN candidate sample in Figure~\ref{fig:cmd_final_score}, using the same guide lines. As stated previously, among the 662 CSAGN candidates displayed on the CMD, 641 are classified as quasars, 13 as stars, and 8 as galaxies. We find that the sources on the right side of the vertical dashed line in Figure~\ref{fig:cmd_final_score} are strongly dominated by quasars (\(98\%\)) with the other \(2\%\) classified as stars and no galaxies. The upper-left quadrant contains all 8 of the CSAGN candidates with a galaxy classification, resulting in \(16\%\) galaxies with the remainder classified as quasars with 1 star classification. The lower-left quadrant contains \(100\%\) quasar classified sources. Among the non-anomalous sources, galaxies make up \(11\%\) of the upper-left CMD region and \(13.6\%\) in the anomalous sample. On the brighter side, both samples are dominated by quasars (\(96\%\) and \(97\%\), respectively), with galaxy fractions of only \(0.6\%\) and \(0.7\%\). This suggests that the redder, lower-luminosity part of the CMD contain sources with a stronger host-galaxy contribution which is even more pronounced in the CSAGN candidate sample.

Furthermore, in Figure~\ref{fig:cmd_final_score} we examined how the CSAGN score varies across the CMD. We find that the upper-left region has a larger fraction of higher-score candidates than the other CMD regions (Table~\ref{tab:csagn_score_regions}). In this region, $37.3\%$ of the sources have CSAGN scores greater than 7 (19/51), compared with $21.2\%$ (7/33) in the lower-left region and $27.0\%$ (156/578) in the main CMD region. The same trend is more pronounced at higher scores: $23.5\%$ (12/51) of the upper-left sources have CSAGN scores greater than 8, compared with $6.1\%$ (2/33) in the lower-left region and $10.7\%$ (62/578) in the main CMD region. The corresponding difference in the fraction of sources with CSAGN score greater than 8 between the upper-left and main CMD regions is $\Delta f = 0.128 \pm 0.061$\footnote{For a fixed score threshold, each source either passes or fails, so the count $N_{\rm pass}$ in a region can be modeled as binomial, $N_{\rm pass}\sim{\rm Binomial}(N,p)$. The sample fraction $f=N_{\rm pass}/N$ then has ${\rm Var}(f)=p(1-p)/N$, approximated by substituting $p\simeq f$, which yields $\sigma_f \simeq \sqrt{f(1-f)/N}$. For the difference between two independent fractions, \(\Delta f=f_1-f_2\), the uncertainty was obtained by adding the two fraction uncertainties in quadrature, \(\sigma_{\Delta f}=\sqrt{\sigma_{f_1}^2+\sigma_{f_2}^2}\).}. 

\begin{deluxetable}{lccccc}
\tablecaption{CSAGN-score enrichment across CMD regions\label{tab:csagn_score_regions}}
\tablehead{
\colhead{Region} &
\colhead{$N$} &
\colhead{$N(S>7)$} &
\colhead{$f(S>7)$} &
\colhead{$N(S>8)$} &
\colhead{$f(S>8)$}
}
\startdata
Upper left          &  51 &  19 & 0.3725 & 12 & 0.2353 \\
Lower left          &  33 &   7 & 0.2121 &  2 & 0.0606 \\
Left part           &  84 &  26 & 0.3095 & 14 & 0.1667 \\
Main cluster / rest & 578 & 156 & 0.2699 & 62 & 0.1073 \\
\hline
Whole CSAGN CMD     & 662 & 182 & 0.2749 & 76 & 0.1148 \\
\enddata
\tablenotetext{}{Fractions are computed as $f=N_{\rm pass}/N$. Here $N$ is the total number of candidates in each CMD region and $S$ is the CSAGN score. For each threshold ($S>7$ or $S>8$), $N_{\rm pass}$ is the number of candidates above that threshold, and $f=N_{\rm pass}/N$ is the corresponding fraction.}
\end{deluxetable}

Therefore, although high-score candidates are not restricted to a single part of the CMD, the redder and lower-luminosity region contains a larger fraction of higher-ranked two-state candidates. The result suggests that the lower-luminosity, redder part of the CMD may be an important region for finding strong photometric CSAGN candidates, possibly including sources with stronger host-galaxy contribution.

\section{Discussion} \label{sec:discussion}

\subsection{Magnitude Differences Between the Two Photometric States}
\label{sec:state_magnitude_differences}

To examine the typical magnitude change among our CSAGN candidates, we measured the magnitude difference between the two photometric states. The observed epochs were assigned to one of the two GMM components, and the median magnitude of each component was used to represent the corresponding state. We then defined
\begin{equation}
    \Delta m_{\rm state} =
\left|\tilde{m}_{\rm faint}-\tilde{m}_{\rm bright}\right|,
\end{equation}
where \(\tilde{m}_{\rm bright}\) and \(\tilde{m}_{\rm faint}\) are the median observed magnitudes of the brighter and fainter states, respectively.

\begin{figure}[t]
    \centering
    \includegraphics[width=0.62\linewidth]{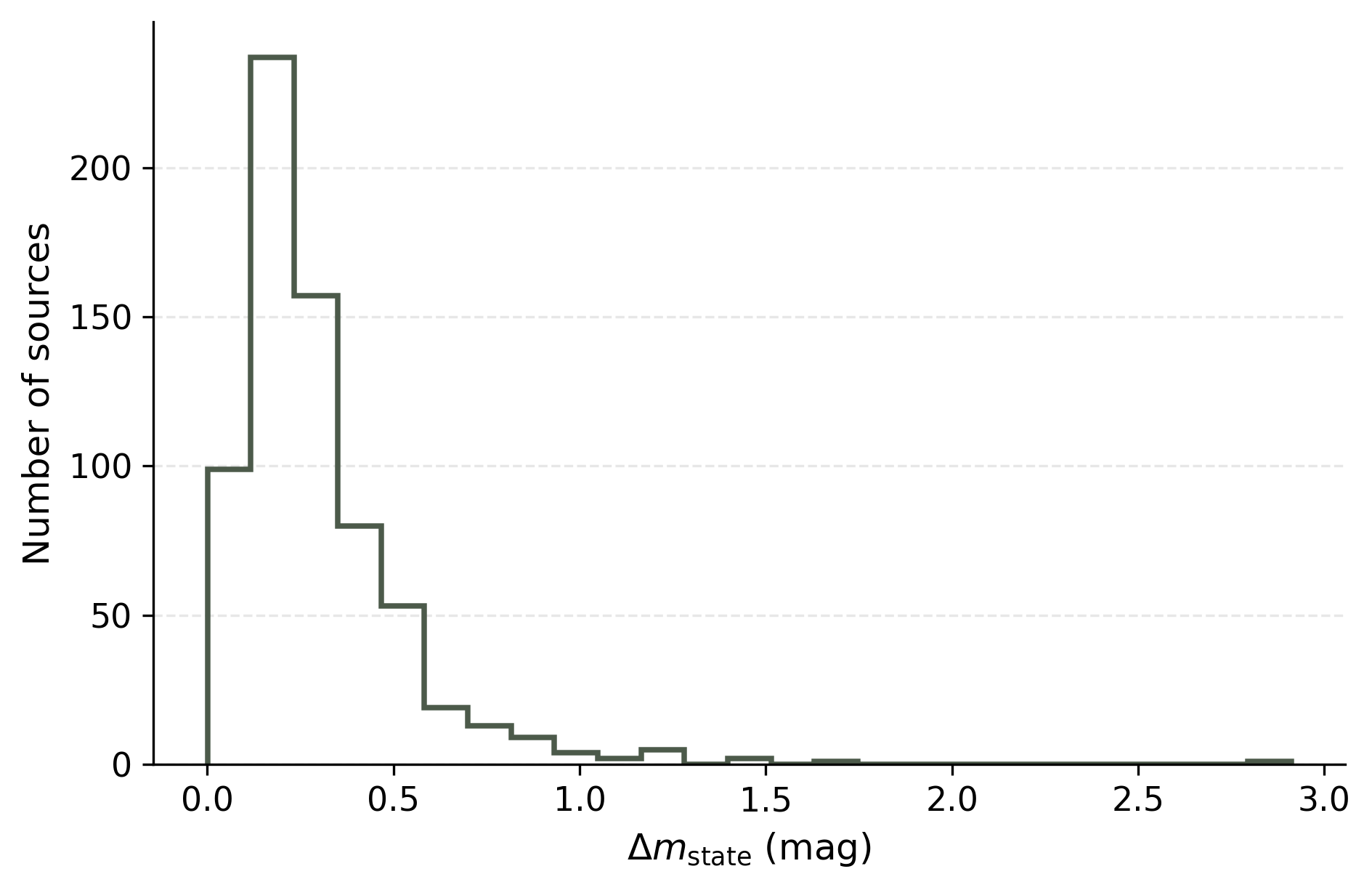}
    \caption{Distribution of the observed magnitude difference between the two photometric states for the 682 CSAGN candidates. The two states are assigned using the two-component GMM, and the state level is measured as the median observed Gaia \(G\)-band magnitude of each component. The distribution has a median of \(0.24\) mag and a mean of \(0.30\) mag, with most candidates showing state-to-state differences of only a few tenths of a magnitude.}
    \label{fig:state_delta_mag}
\end{figure}

Figure~\ref{fig:state_delta_mag} shows the distribution of \(\Delta m_{\rm state}\) for the 682 CSAGN candidates. The distribution is concentrated at relatively small values, with a median of \(0.24\) mag and a mean of \(0.30\) mag. Most candidates therefore show state-to-state differences of only a few tenths of a magnitude, while a smaller number extend to larger state separations, reaching a maximum of \(2.91\) mag.

Several previous photometric searches for changing-look AGNs have emphasized large-amplitude variability, often using thresholds such as \(|\Delta g|>1\) mag \citep{2019ApJ...874....8M, 2016MNRAS.457..389M, 2018ApJ...862..109Y}. However, changing-look behavior is not necessarily tied to a single large optical-amplitude cut. In particular, \citet{2020MNRAS.497..192H} showed that magnitude-based selection can become ineffective at lower redshifts, where host galaxy light can dilute the observed nuclear variability. More broadly, \citet{2026ApJ...997..100W} summarized several selection approaches used in the literature, including criteria based on infrared variability, color changes, spectral changes, and variability significance. They also argued that CLAGNs may occupy a critical state in which moderate accretion rate fluctuations can produce temporary spectral changes.

There are also individual changing-look systems with optical-continuum changes closer to the range found here. For example, NGC 3786 changed from type 1.8/1.9 to type 1 while showing only a moderate optical-continuum change of about \(0.2\)--\(0.3\) mag \citep{2026ApJ..1003....4S}. Similarly, the Mg~II changing-look AGN SDSS J152533.60+292012.1 showed a strong change in the broad Mg~II line while the optical continuum changed by only \(\Delta V_{\rm max-min}=0.17\pm0.05\) mag over the same long-term baseline \citep{2019ApJ...883L..44G}. These examples illustrate that relatively small optical-continuum changes can still be associated with spectroscopic state changes in some systems. \citet{2025ApJ...986..160D} also identified 31 changing-look candidates with pronounced broad-line variations but without significant photometric variability.

Our selection was not based on requiring a large magnitude change. Instead, it first identifies anomalous Gaia DR3 variable AGN light curves and then selects sources whose observed magnitudes show evidence for two preferred states. The candidates with \(\Delta m_{\rm state}\) of only a few tenths of a magnitude should therefore be interpreted as lower-amplitude photometric CSAGN candidates. Spectroscopic follow-up is needed to test whether these photometric states correspond to real changes in the broad-line emission.

\subsection{Why Most Candidates Are Not \(R\)-based Anomalies}

One of the main findings of this work is that most of the CSAGN candidates are selected through the Isolation Forest criterion rather than through the \(R\) reconstruction error criterion. Among the 682 candidates, 479 sources are IF-only anomalies, while 148 are \(R\)-only anomalies and 55 satisfy both criteria.

This result is reasonable in the context of the VRAE analysis. The reconstruction error measures how poorly the model reproduces the observed magnitudes. A light curve can have a clear transition between two brightness states and still be reconstructed reasonably well, especially if the transition is smooth or if the two states are well sampled. In such cases, the light curve may not appear as an extreme \(R\) outlier.

The latent-space IF criterion is sensitive to a different part of the model output. Instead of asking whether the reconstructed light curve is poor, it asks whether the compressed representation of the light curve is unusual compared with the rest of the sample. This is why IF can select sources with unusual overall variability patterns, even when their reconstruction errors are not among the largest. The dominance of IF-only candidates therefore suggests that, for many of the CSAGN candidates, the two-state behavior appears as an unusual variability pattern in latent space without producing an equally extreme reconstruction error. This also explains why using both anomaly criteria is useful. The \(R\) criterion recovers sources with large reconstruction residuals, while the IF criterion recovers sources with unusual latent space structure. The small overlap between the two groups shows that the two criteria are not redundant. The higher CSAGN scores reached by the \(R\)-selected sources may indicate that some of the strongest bimodal light curves are also those for which the VRAE reconstruction departs most clearly from the observed variability. This is consistent with the point raised by \citet{2021AJ....162..206S} that, for sources with large reconstruction errors, the latent space representation may no longer compress the true variability signal reliably.

\subsection{Comparison with Previous light curve based Searches}
\label{sec:comparison_variability_searches}

Our search is closely related to the anomaly detection approach of \citet{2021AJ....162..206S}, who used ZTF DR5 light curves and a VRAE-based anomaly-detection method to identify AGNs with unusual variability behavior. Their parent sample contained 230,451 AGN light curves, from which they selected 8,809 anomalies and then identified 75 promising CSAGN candidates after visual inspection. In comparison, our Gaia DR3 search starts from 872,228 AGN light curves and selects 682 photometrically selected CSAGN candidates using anomaly detection followed by an automated bimodality analysis. Therefore, our CSAGN candidate sample is roughly 9.1 times as large as the candidate sample presented by \citet{2021AJ....162..206S}.

The two searches are similar in spirit, but they differ in the final selection step. In \citet{2021AJ....162..206S}, the anomaly detection stage was followed by visual inspection of the anomalous light curves. In this work, the final selection is instead based on a reproducible GMM-based bimodality analysis. This difference is particularly useful for Gaia DR3, where the much larger parent sample makes a fully automated final selection more practical. The GMM step allows us to keep the selection tied to a specific photometric signature, which is the presence of two preferred brightness states in the Gaia G band light curves.

Our candidate sample is also substantially larger than some recent literature samples of known CLAGNs. For comparison, \citet{2026ApJ...997..100W} compiled a literature sample of 228 previously reported CLAGNs for their variability analysis. Our sample of 682 photometric candidates is about three times as large as that particular literature sample, although the two samples have very different selection and confirmation status. This comparison should not be interpreted as an increase in the number of confirmed CLAGNs, since our sources still require spectroscopic follow-up. Rather, it shows that Gaia DR3 provides a large independent sample of AGN light curves showing photometric behavior consistent with changing-state candidates.

This comparison also highlights the role of variability-based searches. Earlier photometric searches often focused on large-amplitude changes, while anomaly detection methods can select a wider range of unusual light curve behavior. In our case, the anomaly selection step identifies sources that are unusual either in reconstruction error or in VRAE latent space, and the GMM analysis then isolates the subset with bimodal magnitude distributions. This allows the method to recover candidates that may not be selected by a simple variability-amplitude threshold alone.

The CSAGN candidate sample expands the pool of plausible changing-state candidates in a reproducible way, using only Gaia DR3 photometry. Spectroscopic observations are still needed to determine which of these photometric candidates show the broad line changes required for confirmation as changing-state AGNs.

\subsection{Lower-luminosity candidates and accretion state}

There is evidence that CLAGNs are more prevalent among low accretion rate systems. \citet{2024ApJ...966...85Z} found that their changing-look sample has lower Eddington ratios than a redshift and luminosity-matched control sample. \citet{2025ApJS..278...28G} found a median Eddington ratio near $0.01$ in the dim states of their DESI sample. Individual systems also show why this regime may matter. For instance, in NGC~2992, broad H$\alpha$ has repeatedly disappeared near an estimated Eddington ratio of $1\%$ \citep{2021MNRAS.508..144G}. More recently, \citet{2026RAA....26i5001H} found that the UV--optical behavior of lower-luminosity changing-look AGNs is more consistent with a truncated disk than with a standard thin disk, which could reduce the ionizing radiation reaching the broad-line region.

Our CSAGN candidates are, on average, less luminous and have a higher fraction of galaxy-classified sources than the full Gaia variable AGN sample. Notably, the faintest and most galaxy-dominated sources cluster is in the faint, red region of the  CMD, which is underrepresented in the full Gaia sample. These properties may indicate a population of AGN accreting at lower rates, although spectroscopic follow-up is needed to confirm the CSAGN nature of these candidates and determine their accretion rates.

\section{Conclusions} \label{sec:conclusions}

We presented a systematic search for changing-state AGN candidates using Gaia DR3 G-band light curves. Starting from a parent sample of 872,228 Gaia DR3 AGN light curves, we used a pipeline based on machine learning to identify sources with unusual optical variability and then isolate the subset showing evidence for two preferred brightness states.

The first stage of the analysis used a VRAE model together with two anomaly criteria: the \(R\) reconstruction error and the Isolation Forest score in the VRAE latent space. This step reduced the full sample to 8,504 anomalous light curves. We then applied a bimodality analysis based on the bimodality coefficient and a comparison between one- and two-component GMM fits. This final step selected 682 photometric CSAGN candidates.

The CSAGN candidates represent only about \(0.08\%\) of the original Gaia DR3 AGN sample. However, the anomalous source population is strongly enriched in two-state light curves. Among the anomalous sources, \(8.02\%\) satisfy the bimodality criteria, compared with only \(0.85\%\) of the non-anomalous sources. This shows that the anomaly selection stage substantially increases the likelihood of finding sources with CSAGN-like photometric behavior.

The CSAGN candidate sample is dominated by candidates selected through the Isolation Forest criterion in the VRAE latent space, rather than by the \(R\) reconstruction error criterion alone. This indicates that many two-state candidates are not simply the sources with the largest reconstruction residuals, but instead have unusual light curve representations in the learned latent space. The measured magnitude differences between the two photometric states are typically modest, with a median \(\Delta m_{\rm state}=0.24\) mag and a mean \(\Delta m_{\rm state}=0.30\) mag.

Due to the limited Gaia DR3 time window of 34 months, the candidate light curves span several morphologies and may not show a complete transition between two long-lived plateau states. They can instead capture partial transitions, strongly unbalanced brightness modes, flare-like behavior, or a transition into or out of a relatively stable brightness state. The CSAGN score ranks the candidates after selection rather than defining the sample. IF-only selection criterion dominates the CSAGN candidate sample, although R-selected sources make up a larger fraction of the highest-ranked candidates.

The CSAGN candidates span much the same range of observed Gaia colors as the reference samples. However, the candidates do appear to have fainter absolute magnitudes on average and therefore lie at slightly lower redshifts than the broader Gaia variable AGN sample. About \(7\%\) of the candidates occupy the fainter and redder part of the CMD and this fraction is significantly higher than that of the non-anomalous AGN light curves. Galaxy classified sources are also in greater abundance in this region, which contains all eight CSAGN candidates classified as galaxies. We find the strongest candidates extend into the redder, lower-luminosity region of the CMD as well. The extension toward lower luminosities suggests that a subset of our CSAGN candidates may be accreting at lower rates than in typical quasars samples. Spectroscopic and multiwavelength follow-up will be needed to test this interpretation.

A key feature of this search is that the candidate selection is based only on the Gaia light curves. No spectroscopic information, multiwavelength data, or visual inspection was used to define the CSAGN candidate sample. This makes the method scalable to large time-domain surveys and provides a reproducible way to narrow a very large AGN sample to a smaller set of sources that are more likely to contain genuine changing-state systems. A detailed spectroscopic and multiwavelength analysis of the candidate sample will be presented in a forthcoming companion study.

This work addresses the growing need to work efficiently with very large photometric and spectroscopic data sets. This need will only become stronger with the enormous data volumes being produced or expected from facilities such as the Nancy Grace Roman Space Telescope, Euclid, the Vera C.\ Rubin Observatory, 4MOST, DESI, WEAVE, and SDSS-V. At this scale, visual inspection and other traditional selection approaches become increasingly difficult to use as the main discovery tool. Big data methods can instead become part of the everyday astronomical workflow, helping to narrow millions of sources to manageable samples for closer physical and spectroscopic study.

\begin{acknowledgments}

This work has made use of data from the European Space Agency (ESA) mission
{\it Gaia} (\url{https://www.cosmos.esa.int/gaia}), processed by the {\it Gaia}
Data Processing and Analysis Consortium (DPAC,
\url{https://www.cosmos.esa.int/web/gaia/dpac/consortium}). Funding for the DPAC
has been provided by national institutions, in particular the institutions
participating in the {\it Gaia} Multilateral Agreement.

K.M. is especially grateful to Dr.\ Lorenzo G.\ Rimoldini of the Department of Astronomy, University of Geneva, and the Gaia Data Processing and Analysis Consortium (DPAC) Coordination Unit 7 (CU7), for his generosity, continued support, and many helpful technical discussions throughout this work. K.M. also thanks Prof.\ Claudio Ricci of the Department of Astronomy, University of Geneva, for helpful advice related to this research, and Prof.\ Dr.\ Sergei Klioner of TU Dresden and the Gaia Science Team for his guidance on astrometric questions related to the Gaia mission.

\end{acknowledgments}

\section*{ORCID iDs}

Kasra Mokhtarpour \orcidlink{0009-0007-2139-1791}
\href{https://orcid.org/0009-0007-2139-1791}
{https://orcid.org/0009-0007-2139-1791}

Vicki Sarajedini \orcidlink{0000-0002-2304-0908}
\href{https://orcid.org/0000-0002-2304-0908}
{https://orcid.org/0000-0002-2304-0908}

\bibliography{References}{}

@ARTICLE{1978AnSta...6..461S,
       author = {{Schwarz}, Gideon},
        title = "{Estimating the Dimension of a Model}",
      journal = {Annals of Statistics},
         year = 1978,
        month = jul,
       volume = {6},
       number = {2},
        pages = {461-464},
       adsurl = {https://ui.adsabs.harvard.edu/abs/1978AnSta...6..461S}
}

@ARTICLE{1993ARA&A..31..473A,
       author = {{Antonucci}, Robert},
        title = "{Unified models for active galactic nuclei and quasars.}",
      journal = {\araa},
         year = 1993,
        month = jan,
       volume = {31},
        pages = {473-521},
          doi = {10.1146/annurev.aa.31.090193.002353},
       adsurl = {https://ui.adsabs.harvard.edu/abs/1993ARA&A..31..473A}
}

@ARTICLE{1995PASP..107..803U,
       author = {{Urry}, C. Megan and {Padovani}, Paolo},
        title = "{Unified Schemes for Radio-Loud Active Galactic Nuclei}",
      journal = {\pasp},
         year = 1995,
        month = sep,
       volume = {107},
        pages = {803},
          doi = {10.1086/133630},
archivePrefix = {arXiv},
       eprint = {astro-ph/9506063},
 primaryClass = {astro-ph},
       adsurl = {https://ui.adsabs.harvard.edu/abs/1995PASP..107..803U}
}

@ARTICLE{1997ARA&A..35..445U,
       author = {{Ulrich}, Marie-Helene and {Maraschi}, Laura and {Urry}, C. Megan},
        title = "{Variability of Active Galactic Nuclei}",
      journal = {\araa},
         year = 1997,
        month = jan,
       volume = {35},
        pages = {445-502},
          doi = {10.1146/annurev.astro.35.1.445},
       adsurl = {https://ui.adsabs.harvard.edu/abs/1997ARA&A..35..445U}
}

@ARTICLE{2003AJ....126.2687S,
       author = {{Soffel}, M. and {Klioner}, S.~A. and {Petit}, G. and {Wolf}, P. and {Kopeikin}, S.~M. and {Bretagnon}, P. and {Brumberg}, V.~A. and {Capitaine}, N. and {Damour}, T. and {Fukushima}, T. and {Guinot}, B. and {Huang}, T.-Y. and {Lindegren}, L. and {Ma}, C. and {Nordtvedt}, K. and {Ries}, J.~C. and {Seidelmann}, P.~K. and {Vokrouhlick{\'y}}, D. and {Will}, C.~M. and {Xu}, C.},
        title = "{The IAU 2000 Resolutions for Astrometry, Celestial Mechanics, and Metrology in the Relativistic Framework: Explanatory Supplement}",
      journal = {\aj},
         year = 2003,
        month = dec,
       volume = {126},
       number = {6},
        pages = {2687-2706},
          doi = {10.1086/378162},
archivePrefix = {arXiv},
       eprint = {astro-ph/0303376},
 primaryClass = {astro-ph},
       adsurl = {https://ui.adsabs.harvard.edu/abs/2003AJ....126.2687S}
}

@ARTICLE{2009ApJ...698..895K,
       author = {{Kelly}, Brandon C. and {Bechtold}, Jill and {Siemiginowska}, Aneta},
        title = "{Are the Variations in Quasar Optical Flux Driven by Thermal Fluctuations?}",
      journal = {\apj},
         year = 2009,
        month = jun,
       volume = {698},
       number = {1},
        pages = {895-910},
          doi = {10.1088/0004-637X/698/1/895},
archivePrefix = {arXiv},
       eprint = {0903.5315},
 primaryClass = {astro-ph.CO},
       adsurl = {https://ui.adsabs.harvard.edu/abs/2009ApJ...698..895K}
}

@ARTICLE{2010ApJ...721.1014M,
       author = {{MacLeod}, C.~L. and {Ivezi{\'c}}, {\v{Z}}. and {Kochanek}, C.~S. and {Koz{\l}owski}, S. and {Kelly}, B. and {Bullock}, E. and {Kimball}, A. and {Sesar}, B. and {Westman}, D. and {Brooks}, K. and {Gibson}, R. and {Becker}, A.~C. and {de Vries}, W.~H.},
        title = "{Modeling the Time Variability of SDSS Stripe 82 Quasars as a Damped Random Walk}",
      journal = {\apj},
         year = 2010,
        month = oct,
       volume = {721},
       number = {2},
        pages = {1014-1033},
          doi = {10.1088/0004-637X/721/2/1014},
archivePrefix = {arXiv},
       eprint = {1004.0276},
 primaryClass = {astro-ph.CO},
       adsurl = {https://ui.adsabs.harvard.edu/abs/2010ApJ...721.1014M}
}

@ARTICLE{2012ApJ...753..106M,
       author = {{MacLeod}, Chelsea L. and {Ivezi{\'c}}, {\v{Z}}eljko and {Sesar}, Branimir and {de Vries}, Wim and {Kochanek}, Christopher S. and {Kelly}, Brandon C. and {Becker}, Andrew C. and {Lupton}, Robert H. and {Hall}, Patrick B. and {Richards}, Gordon T. and {Anderson}, Scott F. and {Schneider}, Donald P.},
        title = "{A Description of Quasar Variability Measured Using Repeated SDSS and POSS Imaging}",
      journal = {\apj},
         year = 2012,
        month = jul,
       volume = {753},
       number = {2},
          eid = {106},
        pages = {106},
          doi = {10.1088/0004-637X/753/2/106},
archivePrefix = {arXiv},
       eprint = {1112.0679},
 primaryClass = {astro-ph.CO},
       adsurl = {https://ui.adsabs.harvard.edu/abs/2012ApJ...753..106M}
}

@ARTICLE{2013arXiv1312.6114K,
       author = {{Kingma}, Diederik P and {Welling}, Max},
        title = "{Auto-Encoding Variational Bayes}",
      journal = {arXiv e-prints},
         year = 2013,
        month = dec,
          eid = {arXiv:1312.6114},
        pages = {arXiv:1312.6114},
          doi = {10.48550/arXiv.1312.6114},
archivePrefix = {arXiv},
       eprint = {1312.6114},
 primaryClass = {stat.ML},
       adsurl = {https://ui.adsabs.harvard.edu/abs/2013arXiv1312.6114K}
}

@ARTICLE{10.3389/fpsyg.2013.00700,
    
AUTHOR={Pfister, Roland  and Schwarz, Katharina A. and Janczyk, Markus  and Dale, Rick  and Freeman, Jon },
           
TITLE={Good things peak in pairs: a note on the bimodality coefficient},
          
JOURNAL={Frontiers in Psychology},
          
VOLUME={Volume 4 - 2013},
  
YEAR={2013},
  
URL={https://www.frontiersin.org/journals/psychology/articles/10.3389/fpsyg.2013.00700},
  
DOI={10.3389/fpsyg.2013.00700},
  
ISSN={1664-1078},
  
}

@ARTICLE{2015ApJ...800..144L,
       author = {{LaMassa}, Stephanie M. and {Cales}, Sabrina and {Moran}, Edward C. and {Myers}, Adam D. and {Richards}, Gordon T. and {Eracleous}, Michael and {Heckman}, Timothy M. and {Gallo}, Luigi and {Urry}, C. Megan},
        title = "{The Discovery of the First {\textquotedblleft}Changing Look{\textquotedblright} Quasar: New Insights Into the Physics and Phenomenology of Active Galactic Nucleus}",
      journal = {\apj},
         year = 2015,
        month = feb,
       volume = {800},
       number = {2},
          eid = {144},
        pages = {144},
          doi = {10.1088/0004-637X/800/2/144},
archivePrefix = {arXiv},
       eprint = {1412.2136},
 primaryClass = {astro-ph.GA},
       adsurl = {https://ui.adsabs.harvard.edu/abs/2015ApJ...800..144L}
}

@ARTICLE{2016A&A...595A...1G,
       author = {{Gaia Collaboration} and {Prusti}, T. and {de Bruijne}, J.~H.~J. and {Brown}, A.~G.~A. and {Vallenari}, A. and {Babusiaux}, C. and {Bailer-Jones}, C.~A.~L. and {Bastian}, U. and {Biermann}, M. and {Evans}, D.~W. and {Eyer}, L. and {Jansen}, F. and {Jordi}, C. and {Klioner}, S.~A. and {Lammers}, U. and {Lindegren}, L. and {Luri}, X. and {Mignard}, F. and {Milligan}, D.~J. and {Panem}, C. and {Poinsignon}, V. and {Pourbaix}, D. and {Randich}, S. and {Sarri}, G. and {Sartoretti}, P. and {Siddiqui}, H.~I. and {Soubiran}, C. and {Valette}, V. and {van Leeuwen}, F. and {Walton}, N.~A. and {Aerts}, C. and {Arenou}, F. and {Cropper}, M. and {Drimmel}, R. and {H{\o}g}, E. and {Katz}, D. and {Lattanzi}, M.~G. and {O'Mullane}, W. and {Grebel}, E.~K. and {Holland}, A.~D. and {Huc}, C. and {Passot}, X. and {Bramante}, L. and {Cacciari}, C. and {Casta{\~n}eda}, J. and {Chaoul}, L. and {Cheek}, N. and {De Angeli}, F. and {Fabricius}, C. and {Guerra}, R. and {Hern{\'a}ndez}, J. and {Jean-Antoine-Piccolo}, A. and {Masana}, E. and {Messineo}, R. and {Mowlavi}, N. and {Nienartowicz}, K. and {Ord{\'o}{\~n}ez-Blanco}, D. and {Panuzzo}, P. and {Portell}, J. and {Richards}, P.~J. and {Riello}, M. and {Seabroke}, G.~M. and {Tanga}, P. and {Th{\'e}venin}, F. and {Torra}, J. and {Els}, S.~G. and {Gracia-Abril}, G. and {Comoretto}, G. and {Garcia-Reinaldos}, M. and {Lock}, T. and {Mercier}, E. and {Altmann}, M. and {Andrae}, R. and {Astraatmadja}, T.~L. and {Bellas-Velidis}, I. and {Benson}, K. and {Berthier}, J. and {Blomme}, R. and {Busso}, G. and {Carry}, B. and {Cellino}, A. and {Clementini}, G. and {Cowell}, S. and {Creevey}, O. and {Cuypers}, J. and {Davidson}, M. and {De Ridder}, J. and {de Torres}, A. and {Delchambre}, L. and {Dell'Oro}, A. and {Ducourant}, C. and {Fr{\'e}mat}, Y. and {Garc{\'\i}a-Torres}, M. and {Gosset}, E. and {Halbwachs}, J.-L. and {Hambly}, N.~C. and {Harrison}, D.~L. and {Hauser}, M. and {Hestroffer}, D. and {Hodgkin}, S.~T. and {Huckle}, H.~E. and {Hutton}, A. and {Jasniewicz}, G. and {Jordan}, S. and {Kontizas}, M. and {Korn}, A.~J. and {Lanzafame}, A.~C. and {Manteiga}, M. and {Moitinho}, A. and {Muinonen}, K. and {Osinde}, J. and {Pancino}, E. and {Pauwels}, T. and {Petit}, J.-M. and {Recio-Blanco}, A. and {Robin}, A.~C. and {Sarro}, L.~M. and {Siopis}, C. and {Smith}, M. and {Smith}, K.~W. and {Sozzetti}, A. and {Thuillot}, W. and {van Reeven}, W. and {Viala}, Y. and {Abbas}, U. and {Abreu Aramburu}, A. and {Accart}, S. and {Aguado}, J.~J. and {Allan}, P.~M. and {Allasia}, W. and {Altavilla}, G. and {{\'A}lvarez}, M.~A. and {Alves}, J. and {Anderson}, R.~I. and {Andrei}, A.~H. and {Anglada Varela}, E. and {Antiche}, E. and {Antoja}, T. and {Ant{\'o}n}, S. and {Arcay}, B. and {Atzei}, A. and {Ayache}, L. and {Bach}, N. and {Baker}, S.~G. and {Balaguer-N{\'u}{\~n}ez}, L. and {Barache}, C. and {Barata}, C. and {Barbier}, A. and {Barblan}, F. and {Baroni}, M. and {Barrado y Navascu{\'e}s}, D. and {Barros}, M. and {Barstow}, M.~A. and {Becciani}, U. and {Bellazzini}, M. and {Bellei}, G. and {Bello Garc{\'\i}a}, A. and {Belokurov}, V. and {Bendjoya}, P. and {Berihuete}, A. and {Bianchi}, L. and {Bienaym{\'e}}, O. and {Billebaud}, F. and {Blagorodnova}, N. and {Blanco-Cuaresma}, S. and {Boch}, T. and {Bombrun}, A. and {Borrachero}, R. and {Bouquillon}, S. and {Bourda}, G. and {Bouy}, H. and {Bragaglia}, A. and {Breddels}, M.~A. and {Brouillet}, N. and {Br{\"u}semeister}, T. and {Bucciarelli}, B. and {Budnik}, F. and {Burgess}, P. and {Burgon}, R. and {Burlacu}, A. and {Busonero}, D. and {Buzzi}, R. and {Caffau}, E. and {Cambras}, J. and {Campbell}, H. and {Cancelliere}, R. and {Cantat-Gaudin}, T. and {Carlucci}, T. and {Carrasco}, J.~M. and {Castellani}, M. and {Charlot}, P. and {Charnas}, J. and {Charvet}, P. and {Chassat}, F. and {Chiavassa}, A. and {Clotet}, M. and {Cocozza}, G. and {Collins}, R.~S. and {Collins}, P. and {Costigan}, G.},
        title = "{The Gaia mission}",
      journal = {\aap},
         year = 2016,
        month = nov,
       volume = {595},
          eid = {A1},
        pages = {A1},
          doi = {10.1051/0004-6361/201629272},
archivePrefix = {arXiv},
       eprint = {1609.04153},
 primaryClass = {astro-ph.IM},
       adsurl = {https://ui.adsabs.harvard.edu/abs/2016A&A...595A...1G}
}

@ARTICLE{2016MNRAS.457..389M,
       author = {{MacLeod}, Chelsea L. and {Ross}, Nicholas P. and {Lawrence}, Andy and {Goad}, Mike and {Horne}, Keith and {Burgett}, William and {Chambers}, Ken C. and {Flewelling}, Heather and {Hodapp}, Klaus and {Kaiser}, Nick and {Magnier}, Eugene and {Wainscoat}, Richard and {Waters}, Christopher},
        title = "{A systematic search for changing-look quasars in SDSS}",
      journal = {\mnras},
         year = 2016,
        month = mar,
       volume = {457},
       number = {1},
        pages = {389-404},
          doi = {10.1093/mnras/stv2997},
archivePrefix = {arXiv},
       eprint = {1509.08393},
 primaryClass = {astro-ph.GA},
       adsurl = {https://ui.adsabs.harvard.edu/abs/2016MNRAS.457..389M}
}

@ARTICLE{2018ApJ...862..109Y,
       author = {{Yang}, Qian and {Wu}, Xue-Bing and {Fan}, Xiaohui and {Jiang}, Linhua and {McGreer}, Ian and {Shangguan}, Jinyi and {Yao}, Su and {Wang}, Bingquan and {Joshi}, Ravi and {Green}, Richard and {Wang}, Feige and {Feng}, Xiaotong and {Fu}, Yuming and {Yang}, Jinyi and {Liu}, Yuanqi},
        title = "{Discovery of 21 New Changing-look AGNs in the Northern Sky}",
      journal = {\apj},
         year = 2018,
        month = aug,
       volume = {862},
       number = {2},
          eid = {109},
        pages = {109},
          doi = {10.3847/1538-4357/aaca3a},
archivePrefix = {arXiv},
       eprint = {1711.08122},
 primaryClass = {astro-ph.GA},
       adsurl = {https://ui.adsabs.harvard.edu/abs/2018ApJ...862..109Y}
}

@ARTICLE{2018ApJ...864...27S,
       author = {{Stern}, Daniel and {McKernan}, Barry and {Graham}, Matthew J. and {Ford}, K.~E.~S. and {Ross}, Nicholas P. and {Meisner}, Aaron M. and {Assef}, Roberto J. and {Balokovi{\'c}}, Mislav and {Brightman}, Murray and {Dey}, Arjun and {Drake}, Andrew and {Djorgovski}, S.~G. and {Eisenhardt}, Peter and {Jun}, Hyunsung D.},
        title = "{A Mid-IR Selected Changing-look Quasar and Physical Scenarios for Abrupt AGN Fading}",
      journal = {\apj},
         year = 2018,
        month = sep,
       volume = {864},
       number = {1},
          eid = {27},
        pages = {27},
          doi = {10.3847/1538-4357/aac726},
archivePrefix = {arXiv},
       eprint = {1805.06920},
 primaryClass = {astro-ph.GA},
       adsurl = {https://ui.adsabs.harvard.edu/abs/2018ApJ...864...27S}
}

@ARTICLE{2018MNRAS.480.4468R,
       author = {{Ross}, Nicholas P. and {Ford}, K.~E. Saavik and {Graham}, Matthew and {McKernan}, Barry and {Stern}, Daniel and {Meisner}, Aaron M. and {Assef}, Roberto J. and {Dey}, Arjun and {Drake}, Andrew J. and {Jun}, Hyunsung D. and {Lang}, Dustin},
        title = "{A new physical interpretation of optical and infrared variability in quasars}",
      journal = {\mnras},
         year = 2018,
        month = nov,
       volume = {480},
       number = {4},
        pages = {4468-4479},
          doi = {10.1093/mnras/sty2002},
archivePrefix = {arXiv},
       eprint = {1805.06921},
 primaryClass = {astro-ph.GA},
       adsurl = {https://ui.adsabs.harvard.edu/abs/2018MNRAS.480.4468R}
}

@ARTICLE{2019ApJ...874....8M,
       author = {{MacLeod}, Chelsea L. and {Green}, Paul J. and {Anderson}, Scott F. and {Bruce}, Alastair and {Eracleous}, Michael and {Graham}, Matthew and {Homan}, David and {Lawrence}, Andy and {LeBleu}, Amy and {Ross}, Nicholas P. and {Ruan}, John J. and {Runnoe}, Jessie and {Stern}, Daniel and {Burgett}, William and {Chambers}, Kenneth C. and {Kaiser}, Nick and {Magnier}, Eugene and {Metcalfe}, Nigel},
        title = "{Changing-look Quasar Candidates: First Results from Follow-up Spectroscopy of Highly Optically Variable Quasars}",
      journal = {\apj},
         year = 2019,
        month = mar,
       volume = {874},
       number = {1},
          eid = {8},
        pages = {8},
          doi = {10.3847/1538-4357/ab05e2},
archivePrefix = {arXiv},
       eprint = {1810.00087},
 primaryClass = {astro-ph.GA},
       adsurl = {https://ui.adsabs.harvard.edu/abs/2019ApJ...874....8M}
}

@ARTICLE{2019ApJ...883...94T,
       author = {{Trakhtenbrot}, Benny and {Arcavi}, Iair and {MacLeod}, Chelsea L. and {Ricci}, Claudio and {Kara}, Erin and {Graham}, Melissa L. and {Stern}, Daniel and {Harrison}, Fiona A. and {Burke}, Jamison and {Hiramatsu}, Daichi and {Hosseinzadeh}, Griffin and {Howell}, D. Andrew and {Smartt}, Stephen J. and {Rest}, Armin and {Prieto}, Jose L. and {Shappee}, Benjamin J. and {Holoien}, Thomas W.-S. and {Bersier}, David and {Filippenko}, Alexei V. and {Brink}, Thomas G. and {Zheng}, WeiKang and {Li}, Ruancun and {Remillard}, Ronald A. and {Loewenstein}, Michael},
        title = "{1ES 1927+654: An AGN Caught Changing Look on a Timescale of Months}",
      journal = {\apj},
         year = 2019,
        month = sep,
       volume = {883},
       number = {1},
          eid = {94},
        pages = {94},
          doi = {10.3847/1538-4357/ab39e4},
archivePrefix = {arXiv},
       eprint = {1903.11084},
 primaryClass = {astro-ph.GA},
       adsurl = {https://ui.adsabs.harvard.edu/abs/2019ApJ...883...94T}
}

@ARTICLE{2019ApJ...883L..44G,
       author = {{Guo}, Hengxiao and {Sun}, Mouyuan and {Liu}, Xin and {Wang}, Tinggui and {Kong}, Minzhi and {Wang}, Shu and {Sheng}, Zhenfeng and {He}, Zhicheng},
        title = "{Discovery of an Mg II Changing-look Active Galactic Nucleus and Its Implications for a Unification Sequence of Changing-look Active Galactic Nuclei}",
      journal = {\apjl},
         year = 2019,
        month = oct,
       volume = {883},
       number = {2},
          eid = {L44},
        pages = {L44},
          doi = {10.3847/2041-8213/ab4138},
archivePrefix = {arXiv},
       eprint = {1908.00072},
 primaryClass = {astro-ph.GA},
       adsurl = {https://ui.adsabs.harvard.edu/abs/2019ApJ...883L..44G}
}

@ARTICLE{2020ApJ...898L...1R,
       author = {{Ricci}, C. and {Kara}, E. and {Loewenstein}, M. and {Trakhtenbrot}, B. and {Arcavi}, I. and {Remillard}, R. and {Fabian}, A.~C. and {Gendreau}, K.~C. and {Arzoumanian}, Z. and {Li}, R. and {Ho}, L.~C. and {MacLeod}, C.~L. and {Cackett}, E. and {Altamirano}, D. and {Gandhi}, P. and {Kosec}, P. and {Pasham}, D. and {Steiner}, J. and {Chan}, C.-H.},
        title = "{The Destruction and Recreation of the X-Ray Corona in a Changing-look Active Galactic Nucleus}",
      journal = {\apjl},
         year = 2020,
        month = jul,
       volume = {898},
       number = {1},
          eid = {L1},
        pages = {L1},
          doi = {10.3847/2041-8213/ab91a1},
archivePrefix = {arXiv},
       eprint = {2007.07275},
 primaryClass = {astro-ph.HE},
       adsurl = {https://ui.adsabs.harvard.edu/abs/2020ApJ...898L...1R}
}

@ARTICLE{2020MNRAS.491.4925G,
       author = {{Graham}, Matthew J. and {Ross}, Nicholas P. and {Stern}, Daniel and {Drake}, Andrew J. and {McKernan}, Barry and {Ford}, K.~E. Saavik and {Djorgovski}, S.~G. and {Mahabal}, Ashish A. and {Glikman}, Eilat and {Larson}, Steve and {Christensen}, Eric},
        title = "{Understanding extreme quasar optical variability with CRTS - II. Changing-state quasars}",
      journal = {\mnras},
         year = 2020,
        month = feb,
       volume = {491},
       number = {4},
        pages = {4925-4948},
          doi = {10.1093/mnras/stz3244},
archivePrefix = {arXiv},
       eprint = {1905.02262},
 primaryClass = {astro-ph.GA},
       adsurl = {https://ui.adsabs.harvard.edu/abs/2020MNRAS.491.4925G}
}

@ARTICLE{2020ApJ...903...54T,
       author = {{Tachibana}, Yutaro and {Graham}, Matthew J. and {Kawai}, Nobuyuki and {Djorgovski}, S.~G. and {Drake}, Andrew J. and {Mahabal}, Ashish A. and {Stern}, Daniel},
        title = "{Deep Modeling of Quasar Variability}",
      journal = {\apj},
         year = 2020,
        month = nov,
       volume = {903},
       number = {1},
          eid = {54},
        pages = {54},
          doi = {10.3847/1538-4357/abb9a9},
archivePrefix = {arXiv},
       eprint = {2003.01241},
 primaryClass = {astro-ph.GA},
       adsurl = {https://ui.adsabs.harvard.edu/abs/2020ApJ...903...54T}
}

@ARTICLE{2020MNRAS.497..192H,
       author = {{Hon}, Wei Jeat and {Webster}, Rachel and {Wolf}, Christian},
        title = "{Changing look active galactic nuclei in the MaNGA survey}",
      journal = {\mnras},
         year = 2020,
        month = sep,
       volume = {497},
       number = {1},
        pages = {192-203},
          doi = {10.1093/mnras/staa1939},
       adsurl = {https://ui.adsabs.harvard.edu/abs/2020MNRAS.497..192H}
}

@ARTICLE{2020ApJS..250....8L,
       author = {{Lyke}, Brad W. and {Higley}, Alexandra N. and {McLane}, J.~N. and {Schurhammer}, Danielle P. and {Myers}, Adam D. and {Ross}, Ashley J. and {Dawson}, Kyle and {Chabanier}, Sol{\`e}ne and {Martini}, Paul and {Busca}, Nicol{\'a}s G. and {Mas des Bourboux}, H{\'e}lion du and {Salvato}, Mara and {Streblyanska}, Alina and {Zarrouk}, Pauline and {Burtin}, Etienne and {Anderson}, Scott F. and {Bautista}, Julian and {Bizyaev}, Dmitry and {Brandt}, W.~N. and {Brinkmann}, Jonathan and {Brownstein}, Joel R. and {Comparat}, Johan and {Green}, Paul and {de la Macorra}, Axel and {Mu{\~n}oz Guti{\'e}rrez}, Andrea and {Hou}, Jiamin and {Newman}, Jeffrey A. and {Palanque-Delabrouille}, Nathalie and {P{\^a}ris}, Isabelle and {Percival}, Will J. and {Petitjean}, Patrick and {Rich}, James and {Rossi}, Graziano and {Schneider}, Donald P. and {Smith}, Alexander and {Vivek}, M. and {Weaver}, Benjamin Alan},
        title = "{The Sloan Digital Sky Survey Quasar Catalog: Sixteenth Data Release}",
      journal = {\apjs},
         year = 2020,
        month = sep,
       volume = {250},
       number = {1},
          eid = {8},
        pages = {8},
          doi = {10.3847/1538-4365/aba623},
archivePrefix = {arXiv},
       eprint = {2007.09001},
 primaryClass = {astro-ph.GA},
       adsurl = {https://ui.adsabs.harvard.edu/abs/2020ApJS..250....8L}
}

@ARTICLE{2021ApJ...907...96S,
       author = {{Suberlak}, Krzysztof L. and {Ivezi{\'c}}, {\v{Z}}eljko and {MacLeod}, Chelsea},
        title = "{Improving Damped Random Walk Parameters for SDSS Stripe 82 Quasars with Pan-STARRS1}",
      journal = {\apj},
         year = 2021,
        month = feb,
       volume = {907},
       number = {2},
          eid = {96},
        pages = {96},
          doi = {10.3847/1538-4357/abc698},
archivePrefix = {arXiv},
       eprint = {2012.12907},
 primaryClass = {astro-ph.GA},
       adsurl = {https://ui.adsabs.harvard.edu/abs/2021ApJ...907...96S}
}

@ARTICLE{2021AJ....162..206S,
       author = {{S{\'a}nchez-S{\'a}ez}, P. and {Lira}, H. and {Mart{\'\i}}, L. and {S{\'a}nchez-Pi}, N. and {Arredondo}, J. and {Bauer}, F.~E. and {Bayo}, A. and {Cabrera-Vives}, G. and {Donoso-Oliva}, C. and {Est{\'e}vez}, P.~A. and {Eyheramendy}, S. and {F{\"o}rster}, F. and {Hern{\'a}ndez-Garc{\'\i}a}, L. and {Arancibia}, A.~M. Mu{\~n}oz and {P{\'e}rez-Carrasco}, M. and {Sep{\'u}lveda}, M. and {Vergara}, J.~R.},
        title = "{Searching for Changing-state AGNs in Massive Data Sets. I. Applying Deep Learning and Anomaly-detection Techniques to Find AGNs with Anomalous Variability Behaviors}",
      journal = {\aj},
         year = 2021,
        month = nov,
       volume = {162},
       number = {5},
          eid = {206},
        pages = {206},
          doi = {10.3847/1538-3881/ac1426},
archivePrefix = {arXiv},
       eprint = {2106.07660},
 primaryClass = {astro-ph.IM},
       adsurl = {https://ui.adsabs.harvard.edu/abs/2021AJ....162..206S}
}

@ARTICLE{2021MNRAS.508..144G,
       author = {{Guolo}, Muryel and {Ruschel-Dutra}, Daniel and {Grupe}, Dirk and {Peterson}, Bradley M. and {Storchi-Bergmann}, Thaisa and {Schimoia}, Jaderson and {Nemmen}, Rodrigo and {Robinson}, Andrew},
        title = "{The Eddington ratio-dependent 'changing look' events in NGC 2992}",
      journal = {\mnras},
         year = 2021,
        month = nov,
       volume = {508},
       number = {1},
        pages = {144-156},
          doi = {10.1093/mnras/stab2550},
archivePrefix = {arXiv},
       eprint = {2109.06914},
 primaryClass = {astro-ph.GA},
       adsurl = {https://ui.adsabs.harvard.edu/abs/2021MNRAS.508..144G}
}

@ARTICLE{2022MNRAS.512.3662D,
       author = {{Duncan}, Kenneth J.},
        title = "{All-purpose, all-sky photometric redshifts for the Legacy Imaging Surveys Data Release 8}",
      journal = {\mnras},
         year = 2022,
        month = may,
       volume = {512},
       number = {3},
        pages = {3662-3683},
          doi = {10.1093/mnras/stac608},
archivePrefix = {arXiv},
       eprint = {2203.01949},
 primaryClass = {astro-ph.GA},
       adsurl = {https://ui.adsabs.harvard.edu/abs/2022MNRAS.512.3662D}
}

@ARTICLE{2023NatAs...7.1282R,
       author = {{Ricci}, Claudio and {Trakhtenbrot}, Benny},
        title = "{Changing-look active galactic nuclei}",
      journal = {Nature Astronomy},
         year = 2023,
        month = nov,
       volume = {7},
        pages = {1282-1294},
          doi = {10.1038/s41550-023-02108-4},
archivePrefix = {arXiv},
       eprint = {2211.05132},
 primaryClass = {astro-ph.GA},
       adsurl = {https://ui.adsabs.harvard.edu/abs/2023NatAs...7.1282R}
}

@ARTICLE{2023A&A...674A..24C,
       author = {{Carnerero}, Maria I. and {Raiteri}, Claudia M. and {Rimoldini}, Lorenzo and {Busonero}, Deborah and {Licata}, Enrico and {Mowlavi}, Nami and {Lecoeur-Ta{\"\i}bi}, Isabelle and {Audard}, Marc and {Holl}, Berry and {Gavras}, Panagiotis and {Nienartowicz}, Krzysztof and {Jevardat de Fombelle}, Gr{\'e}gory and {Carballo}, Ruth and {Clementini}, Gisella and {Delchambre}, Ludovic and {Klioner}, Sergei and {Lattanzi}, Mario G. and {Eyer}, Laurent},
        title = "{Gaia Data Release 3. The first Gaia catalogue of variable AGN}",
      journal = {\aap},
         year = 2023,
        month = jun,
       volume = {674},
          eid = {A24},
        pages = {A24},
          doi = {10.1051/0004-6361/202244035},
archivePrefix = {arXiv},
       eprint = {2207.06849},
 primaryClass = {astro-ph.HE},
       adsurl = {https://ui.adsabs.harvard.edu/abs/2023A&A...674A..24C}
}

@ARTICLE{2023A&A...674A..25H,
       author = {{Holl}, B. and {Fabricius}, C. and {Portell}, J. and {Lindegren}, L. and {Panuzzo}, P. and {Bernet}, M. and {Casta{\~n}eda}, J. and {Jevardat de Fombelle}, G. and {Audard}, M. and {Ducourant}, C. and {Harrison}, D.~L. and {Evans}, D.~W. and {Busso}, G. and {Sozzetti}, A. and {Gosset}, E. and {Arenou}, F. and {De Angeli}, F. and {Riello}, M. and {Eyer}, L. and {Rimoldini}, L. and {Gavras}, P. and {Mowlavi}, N. and {Nienartowicz}, K. and {Lecoeur-Ta{\"\i}bi}, I. and {Garc{\'\i}a-Lario}, P. and {Pourbaix}, D.},
        title = "{Gaia Data Release 3. Gaia scan-angle-dependent signals and spurious periods}",
      journal = {\aap},
         year = 2023,
        month = jun,
       volume = {674},
          eid = {A25},
        pages = {A25},
          doi = {10.1051/0004-6361/202245353},
archivePrefix = {arXiv},
       eprint = {2212.11971},
 primaryClass = {astro-ph.IM},
       adsurl = {https://ui.adsabs.harvard.edu/abs/2023A&A...674A..25H}
}

@ARTICLE{2023A&A...674A..13E,
       author = {{Eyer}, L. and {Audard}, M. and {Holl}, B. and {Rimoldini}, L. and {Carnerero}, M.~I. and {Clementini}, G. and {De Ridder}, J. and {Distefano}, E. and {Evans}, D.~W. and {Gavras}, P. and {Gomel}, R. and {Lebzelter}, T. and {Marton}, G. and {Mowlavi}, N. and {Panahi}, A. and {Ripepi}, V. and {Wyrzykowski}, {\L}. and {Nienartowicz}, K. and {Jevardat de Fombelle}, G. and {Lecoeur-Taibi}, I. and {Rohrbasser}, L. and {Riello}, M. and {Garc{\'\i}a-Lario}, P. and {Lanzafame}, A.~C. and {Mazeh}, T. and {Raiteri}, C.~M. and {Zucker}, S. and {{\'A}brah{\'a}m}, P. and {Aerts}, C. and {Aguado}, J.~J. and {Anderson}, R.~I. and {Bashi}, D. and {Binnenfeld}, A. and {Faigler}, S. and {Garofalo}, A. and {Karbevska}, L. and {K{\'o}sp{\'a}l}, {\'A}. and {Kruszy{\'n}ska}, K. and {Kun}, M. and {Lanza}, A.~F. and {Leccia}, S. and {Marconi}, M. and {Messina}, S. and {Molinaro}, R. and {Moln{\'a}r}, L. and {Muraveva}, T. and {Musella}, I. and {Nagy}, Z. and {Pagano}, I. and {Palaversa}, L. and {Plachy}, E. and {Pr{\v{s}}a}, A. and {Rybicki}, K.~A. and {Shahaf}, S. and {Szabados}, L. and {Szegedi-Elek}, E. and {Trabucchi}, M. and {Barblan}, F. and {Grenon}, M. and {Roelens}, M. and {S{\"u}veges}, M.},
        title = "{Gaia Data Release 3. Summary of the variability processing and analysis}",
      journal = {\aap},
         year = 2023,
        month = jun,
       volume = {674},
          eid = {A13},
        pages = {A13},
          doi = {10.1051/0004-6361/202244242},
archivePrefix = {arXiv},
       eprint = {2206.06416},
 primaryClass = {astro-ph.SR},
       adsurl = {https://ui.adsabs.harvard.edu/abs/2023A&A...674A..13E}
}

@ARTICLE{2023A&A...674A...1G,
       author = {{Gaia Collaboration} and {Vallenari}, A. and {Brown}, A.~G.~A. and {Prusti}, T. and {de Bruijne}, J.~H.~J. and {Arenou}, F. and {Babusiaux}, C. and {Biermann}, M. and {Creevey}, O.~L. and {Ducourant}, C. and {Evans}, D.~W. and {Eyer}, L. and {Guerra}, R. and {Hutton}, A. and {Jordi}, C. and {Klioner}, S.~A. and {Lammers}, U.~L. and {Lindegren}, L. and {Luri}, X. and {Mignard}, F. and {Panem}, C. and {Pourbaix}, D. and {Randich}, S. and {Sartoretti}, P. and {Soubiran}, C. and {Tanga}, P. and {Walton}, N.~A. and {Bailer-Jones}, C.~A.~L. and {Bastian}, U. and {Drimmel}, R. and {Jansen}, F. and {Katz}, D. and {Lattanzi}, M.~G. and {van Leeuwen}, F. and {Bakker}, J. and {Cacciari}, C. and {Casta{\~n}eda}, J. and {De Angeli}, F. and {Fabricius}, C. and {Fouesneau}, M. and {Fr{\'e}mat}, Y. and {Galluccio}, L. and {Guerrier}, A. and {Heiter}, U. and {Masana}, E. and {Messineo}, R. and {Mowlavi}, N. and {Nicolas}, C. and {Nienartowicz}, K. and {Pailler}, F. and {Panuzzo}, P. and {Riclet}, F. and {Roux}, W. and {Seabroke}, G.~M. and {Sordo}, R. and {Th{\'e}venin}, F. and {Gracia-Abril}, G. and {Portell}, J. and {Teyssier}, D. and {Altmann}, M. and {Andrae}, R. and {Audard}, M. and {Bellas-Velidis}, I. and {Benson}, K. and {Berthier}, J. and {Blomme}, R. and {Burgess}, P.~W. and {Busonero}, D. and {Busso}, G. and {C{\'a}novas}, H. and {Carry}, B. and {Cellino}, A. and {Cheek}, N. and {Clementini}, G. and {Damerdji}, Y. and {Davidson}, M. and {de Teodoro}, P. and {Nu{\~n}ez Campos}, M. and {Delchambre}, L. and {Dell'Oro}, A. and {Esquej}, P. and {Fern{\'a}ndez-Hern{\'a}ndez}, J. and {Fraile}, E. and {Garabato}, D. and {Garc{\'\i}a-Lario}, P. and {Gosset}, E. and {Haigron}, R. and {Halbwachs}, J.-L. and {Hambly}, N.~C. and {Harrison}, D.~L. and {Hern{\'a}ndez}, J. and {Hestroffer}, D. and {Hodgkin}, S.~T. and {Holl}, B. and {Jan{\ss}en}, K. and {Jevardat de Fombelle}, G. and {Jordan}, S. and {Krone-Martins}, A. and {Lanzafame}, A.~C. and {L{\"o}ffler}, W. and {Marchal}, O. and {Marrese}, P.~M. and {Moitinho}, A. and {Muinonen}, K. and {Osborne}, P. and {Pancino}, E. and {Pauwels}, T. and {Recio-Blanco}, A. and {Reyl{\'e}}, C. and {Riello}, M. and {Rimoldini}, L. and {Roegiers}, T. and {Rybizki}, J. and {Sarro}, L.~M. and {Siopis}, C. and {Smith}, M. and {Sozzetti}, A. and {Utrilla}, E. and {van Leeuwen}, M. and {Abbas}, U. and {{\'A}brah{\'a}m}, P. and {Abreu Aramburu}, A. and {Aerts}, C. and {Aguado}, J.~J. and {Ajaj}, M. and {Aldea-Montero}, F. and {Altavilla}, G. and {{\'A}lvarez}, M.~A. and {Alves}, J. and {Anders}, F. and {Anderson}, R.~I. and {Anglada Varela}, E. and {Antoja}, T. and {Baines}, D. and {Baker}, S.~G. and {Balaguer-N{\'u}{\~n}ez}, L. and {Balbinot}, E. and {Balog}, Z. and {Barache}, C. and {Barbato}, D. and {Barros}, M. and {Barstow}, M.~A. and {Bartolom{\'e}}, S. and {Bassilana}, J.-L. and {Bauchet}, N. and {Becciani}, U. and {Bellazzini}, M. and {Berihuete}, A. and {Bernet}, M. and {Bertone}, S. and {Bianchi}, L. and {Binnenfeld}, A. and {Blanco-Cuaresma}, S. and {Blazere}, A. and {Boch}, T. and {Bombrun}, A. and {Bossini}, D. and {Bouquillon}, S. and {Bragaglia}, A. and {Bramante}, L. and {Breedt}, E. and {Bressan}, A. and {Brouillet}, N. and {Brugaletta}, E. and {Bucciarelli}, B. and {Burlacu}, A. and {Butkevich}, A.~G. and {Buzzi}, R. and {Caffau}, E. and {Cancelliere}, R. and {Cantat-Gaudin}, T. and {Carballo}, R. and {Carlucci}, T. and {Carnerero}, M.~I. and {Carrasco}, J.~M. and {Casamiquela}, L. and {Castellani}, M. and {Castro-Ginard}, A. and {Chaoul}, L. and {Charlot}, P. and {Chemin}, L. and {Chiaramida}, V. and {Chiavassa}, A. and {Chornay}, N. and {Comoretto}, G. and {Contursi}, G. and {Cooper}, W.~J. and {Cornez}, T. and {Cowell}, S. and {Crifo}, F. and {Cropper}, M. and {Crosta}, M. and {Crowley}, C. and {Dafonte}, C. and {Dapergolas}, A. and {David}, M. and {David}, P. and {de Laverny}, P. and {De Luise}, F. and {De March}, R.},
        title = "{Gaia Data Release 3. Summary of the content and survey properties}",
      journal = {\aap},
         year = 2023,
        month = jun,
       volume = {674},
          eid = {A1},
        pages = {A1},
          doi = {10.1051/0004-6361/202243940},
archivePrefix = {arXiv},
       eprint = {2208.00211},
 primaryClass = {astro-ph.GA},
       adsurl = {https://ui.adsabs.harvard.edu/abs/2023A&A...674A...1G}
}

@ARTICLE{2023A&A...674A..41G,
       author = {{Gaia Collaboration} and {Bailer-Jones}, C.~A.~L. and {Teyssier}, D. and {Delchambre}, L. and {Ducourant}, C. and {Garabato}, D. and {Hatzidimitriou}, D. and {Klioner}, S.~A. and {Rimoldini}, L. and {Bellas-Velidis}, I. and {Carballo}, R. and {Carnerero}, M.~I. and {Diener}, C. and {Fouesneau}, M. and {Galluccio}, L. and {Gavras}, P. and {Krone-Martins}, A. and {Raiteri}, C.~M. and {Teixeira}, R. and {Brown}, A.~G.~A. and {Vallenari}, A. and {Prusti}, T. and {de Bruijne}, J.~H.~J. and {Arenou}, F. and {Babusiaux}, C. and {Biermann}, M. and {Creevey}, O.~L. and {Evans}, D.~W. and {Eyer}, L. and {Guerra}, R. and {Hutton}, A. and {Jordi}, C. and {Lammers}, U.~L. and {Lindegren}, L. and {Luri}, X. and {Mignard}, F. and {Panem}, C. and {Pourbaix}, D. and {Randich}, S. and {Sartoretti}, P. and {Soubiran}, C. and {Tanga}, P. and {Walton}, N.~A. and {Bastian}, U. and {Drimmel}, R. and {Jansen}, F. and {Katz}, D. and {Lattanzi}, M.~G. and {van Leeuwen}, F. and {Bakker}, J. and {Cacciari}, C. and {Casta{\~n}eda}, J. and {De Angeli}, F. and {Fabricius}, C. and {Fr{\'e}mat}, Y. and {Guerrier}, A. and {Heiter}, U. and {Masana}, E. and {Messineo}, R. and {Mowlavi}, N. and {Nicolas}, C. and {Nienartowicz}, K. and {Pailler}, F. and {Panuzzo}, P. and {Riclet}, F. and {Roux}, W. and {Seabroke}, G.~M. and {Sordo}, R. and {Th{\'e}venin}, F. and {Gracia-Abril}, G. and {Portell}, J. and {Altmann}, M. and {Andrae}, R. and {Audard}, M. and {Benson}, K. and {Berthier}, J. and {Blomme}, R. and {Burgess}, P.~W. and {Busonero}, D. and {Busso}, G. and {C{\'a}novas}, H. and {Carry}, B. and {Cellino}, A. and {Cheek}, N. and {Clementini}, G. and {Damerdji}, Y. and {Davidson}, M. and {de Teodoro}, P. and {Nu{\~n}ez Campos}, M. and {Dell'Oro}, A. and {Esquej}, P. and {Fern{\'a}ndez-Hern{\'a}ndez}, J. and {Fraile}, E. and {Garc{\'\i}a-Lario}, P. and {Gosset}, E. and {Haigron}, R. and {Halbwachs}, J.-L. and {Hambly}, N.~C. and {Harrison}, D.~L. and {Hern{\'a}ndez}, J. and {Hestroffer}, D. and {Hodgkin}, S.~T. and {Holl}, B. and {Jan{\ss}en}, K. and {Jevardat de Fombelle}, G. and {Jordan}, S. and {Lanzafame}, A.~C. and {L{\"o}ffler}, W. and {Marchal}, O. and {Marrese}, P.~M. and {Moitinho}, A. and {Muinonen}, K. and {Osborne}, P. and {Pancino}, E. and {Pauwels}, T. and {Recio-Blanco}, A. and {Reyl{\'e}}, C. and {Riello}, M. and {Roegiers}, T. and {Rybizki}, J. and {Sarro}, L.~M. and {Siopis}, C. and {Smith}, M. and {Sozzetti}, A. and {Utrilla}, E. and {van Leeuwen}, M. and {Abbas}, U. and {{\'A}brah{\'a}m}, P. and {Abreu Aramburu}, A. and {Aerts}, C. and {Aguado}, J.~J. and {Ajaj}, M. and {Aldea-Montero}, F. and {Altavilla}, G. and {{\'A}lvarez}, M.~A. and {Alves}, J. and {Anderson}, R.~I. and {Anglada Varela}, E. and {Antoja}, T. and {Baines}, D. and {Baker}, S.~G. and {Balaguer-N{\'u}{\~n}ez}, L. and {Balbinot}, E. and {Balog}, Z. and {Barache}, C. and {Barbato}, D. and {Barros}, M. and {Barstow}, M.~A. and {Bartolom{\'e}}, S. and {Bassilana}, J.-L. and {Bauchet}, N. and {Becciani}, U. and {Bellazzini}, M. and {Berihuete}, A. and {Bernet}, M. and {Bertone}, S. and {Bianchi}, L. and {Binnenfeld}, A. and {Blanco-Cuaresma}, S. and {Boch}, T. and {Bombrun}, A. and {Bossini}, D. and {Bouquillon}, S. and {Bragaglia}, A. and {Bramante}, L. and {Breedt}, E. and {Bressan}, A. and {Brouillet}, N. and {Brugaletta}, E. and {Bucciarelli}, B. and {Burlacu}, A. and {Butkevich}, A.~G. and {Buzzi}, R. and {Caffau}, E. and {Cancelliere}, R. and {Cantat-Gaudin}, T. and {Carlucci}, T. and {Carrasco}, J.~M. and {Casamiquela}, L. and {Castellani}, M. and {Castro-Ginard}, A. and {Chaoul}, L. and {Charlot}, P. and {Chemin}, L. and {Chiaramida}, V. and {Chiavassa}, A. and {Chornay}, N. and {Comoretto}, G. and {Contursi}, G. and {Cooper}, W.~J. and {Cornez}, T. and {Cowell}, S. and {Crifo}, F. and {Cropper}, M. and {Crosta}, M. and {Crowley}, C. and {Dafonte}, C. and {Dapergolas}, A. and {David}, P. and {de Laverny}, P.},
        title = "{Gaia Data Release 3. The extragalactic content}",
      journal = {\aap},
         year = 2023,
        month = jun,
       volume = {674},
          eid = {A41},
        pages = {A41},
          doi = {10.1051/0004-6361/202243232},
archivePrefix = {arXiv},
       eprint = {2206.05681},
 primaryClass = {astro-ph.GA},
       adsurl = {https://ui.adsabs.harvard.edu/abs/2023A&A...674A..41G}
}

@ARTICLE{2023A&A...674A..31D,
       author = {{Delchambre}, L. and {Bailer-Jones}, C.~A.~L. and {Bellas-Velidis}, I. and {Drimmel}, R. and {Garabato}, D. and {Carballo}, R. and {Hatzidimitriou}, D. and {Marshall}, D.~J. and {Andrae}, R. and {Dafonte}, C. and {Livanou}, E. and {Fouesneau}, M. and {Licata}, E.~L. and {Lindstr{\o}m}, H.~E.~P. and {Manteiga}, M. and {Robin}, C. and {Silvelo}, A. and {Abreu Aramburu}, A. and {{\'A}lvarez}, M.~A. and {Bakker}, J. and {Bijaoui}, A. and {Brouillet}, N. and {Brugaletta}, E. and {Burlacu}, A. and {Casamiquela}, L. and {Chaoul}, L. and {Chiavassa}, A. and {Contursi}, G. and {Cooper}, W.~J. and {Creevey}, O.~L. and {Dapergolas}, A. and {de Laverny}, P. and {Demouchy}, C. and {Dharmawardena}, T.~E. and {Edvardsson}, B. and {Fr{\'e}mat}, Y. and {Garc{\'\i}a-Lario}, P. and {Garc{\'\i}a-Torres}, M. and {Gavel}, A. and {Gomez}, A. and {Gonz{\'a}lez-Santamar{\'\i}a}, I. and {Heiter}, U. and {Jean-Antoine Piccolo}, A. and {Kontizas}, M. and {Kordopatis}, G. and {Korn}, A.~J. and {Lanzafame}, A.~C. and {Lebreton}, Y. and {Lobel}, A. and {Lorca}, A. and {Magdaleno Romeo}, A. and {Marocco}, F. and {Mary}, N. and {Nicolas}, C. and {Ordenovic}, C. and {Pailler}, F. and {Palicio}, P.~A. and {Pallas-Quintela}, L. and {Panem}, C. and {Pichon}, B. and {Poggio}, E. and {Recio-Blanco}, A. and {Riclet}, F. and {Rybizki}, J. and {Santove{\~n}a}, R. and {Sarro}, L.~M. and {Schultheis}, M.~S. and {Segol}, M. and {Slezak}, I. and {Smart}, R.~L. and {Sordo}, R. and {Soubiran}, C. and {S{\"u}veges}, M. and {Th{\'e}venin}, F. and {Torralba Elipe}, G. and {Ulla}, A. and {Utrilla}, E. and {Vallenari}, A. and {van Dillen}, E. and {Zhao}, H. and {Zorec}, J.},
        title = "{Gaia Data Release 3. Apsis. III. Non-stellar content and source classification}",
      journal = {\aap},
         year = 2023,
        month = jun,
       volume = {674},
          eid = {A31},
        pages = {A31},
          doi = {10.1051/0004-6361/202243423},
archivePrefix = {arXiv},
       eprint = {2206.06710},
 primaryClass = {astro-ph.GA},
       adsurl = {https://ui.adsabs.harvard.edu/abs/2023A&A...674A..31D}
}

@ARTICLE{2023A&A...674A..14R,
       author = {{Rimoldini}, Lorenzo and {Holl}, Berry and {Gavras}, Panagiotis and {Audard}, Marc and {De Ridder}, Joris and {Mowlavi}, Nami and {Nienartowicz}, Krzysztof and {Jevardat de Fombelle}, Gr{\'e}gory and {Lecoeur-Ta{\"\i}bi}, Isabelle and {Karbevska}, Lea and {Evans}, Dafydd W. and {{\'A}brah{\'a}m}, P{\'e}ter and {Carnerero}, Maria I. and {Clementini}, Gisella and {Distefano}, Elisa and {Garofalo}, Alessia and {Garc{\'\i}a-Lario}, Pedro and {Gomel}, Roy and {Klioner}, Sergei A. and {Kruszy{\'n}ska}, Katarzyna and {Lanzafame}, Alessandro C. and {Lebzelter}, Thomas and {Marton}, G{\'a}bor and {Mazeh}, Tsevi and {Molinaro}, Roberto and {Panahi}, Aviad and {Raiteri}, Claudia M. and {Ripepi}, Vincenzo and {Szabados}, L{\'a}szl{\'o} and {Teyssier}, David and {Trabucchi}, Michele and {Wyrzykowski}, {\L}ukasz and {Zucker}, Shay and {Eyer}, Laurent},
        title = "{Gaia Data Release 3. All-sky classification of 12.4 million variable sources into 25 classes}",
      journal = {\aap},
         year = 2023,
        month = jun,
       volume = {674},
          eid = {A14},
        pages = {A14},
          doi = {10.1051/0004-6361/202245591},
archivePrefix = {arXiv},
       eprint = {2211.17238},
 primaryClass = {astro-ph.GA},
       adsurl = {https://ui.adsabs.harvard.edu/abs/2023A&A...674A..14R}
}

@ARTICLE{2023A&A...674A...2D,
       author = {{De Angeli}, F. and {Weiler}, M. and {Montegriffo}, P. and {Evans}, D.~W. and {Riello}, M. and {Andrae}, R. and {Carrasco}, J.~M. and {Busso}, G. and {Burgess}, P.~W. and {Cacciari}, C. and {Davidson}, M. and {Harrison}, D.~L. and {Hodgkin}, S.~T. and {Jordi}, C. and {Osborne}, P.~J. and {Pancino}, E. and {Altavilla}, G. and {Barstow}, M.~A. and {Bailer-Jones}, C.~A.~L. and {Bellazzini}, M. and {Brown}, A.~G.~A. and {Castellani}, M. and {Cowell}, S. and {Delchambre}, L. and {De Luise}, F. and {Diener}, C. and {Fabricius}, C. and {Fouesneau}, M. and {Fr{\'e}mat}, Y. and {Gilmore}, G. and {Giuffrida}, G. and {Hambly}, N.~C. and {Hidalgo}, S. and {Holland}, G. and {Kostrzewa-Rutkowska}, Z. and {van Leeuwen}, F. and {Lobel}, A. and {Marinoni}, S. and {Miller}, N. and {Pagani}, C. and {Palaversa}, L. and {Piersimoni}, A.~M. and {Pulone}, L. and {Ragaini}, S. and {Rainer}, M. and {Richards}, P.~J. and {Rixon}, G.~T. and {Ruz-Mieres}, D. and {Sanna}, N. and {Sarro}, L.~M. and {Rowell}, N. and {Sordo}, R. and {Walton}, N.~A. and {Yoldas}, A.},
        title = "{Gaia Data Release 3. Processing and validation of BP/RP low-resolution spectral data}",
      journal = {\aap},
         year = 2023,
        month = jun,
       volume = {674},
          eid = {A2},
        pages = {A2},
          doi = {10.1051/0004-6361/202243680},
archivePrefix = {arXiv},
       eprint = {2206.06143},
 primaryClass = {astro-ph.IM},
       adsurl = {https://ui.adsabs.harvard.edu/abs/2023A&A...674A...2D}
}

@ARTICLE{2023ApJS..265...25J,
       author = {{Jin}, Jun-Jie and {Wu}, Xue-Bing and {Fu}, Yuming and {Yao}, Su and {Ai}, Yan-Li and {Feng}, Xiao-Tong and {He}, Zi-Qi and {Ma}, Qin-Chun and {Pang}, Yu-Xuan and {Zhu}, Rui and {Zhang}, Yan-xia and {Yuan}, Hai-long and {Huo}, Zhi-Ying},
        title = "{The Large Sky Area Multi-Object Fiber Spectroscopic Telescope (LAMOST) Quasar Survey: Quasar Properties from Data Releases 6 to 9}",
      journal = {\apjs},
         year = 2023,
        month = mar,
       volume = {265},
       number = {1},
          eid = {25},
        pages = {25},
          doi = {10.3847/1538-4365/acaf89},
archivePrefix = {arXiv},
       eprint = {2212.12876},
 primaryClass = {astro-ph.GA},
       adsurl = {https://ui.adsabs.harvard.edu/abs/2023ApJS..265...25J}
}

@ARTICLE{2023OJAp....6E..49F,
       author = {{Flesch}, Eric Wim},
        title = "{The Million Quasars (Milliquas) Catalogue, v8}",
      journal = {The Open Journal of Astrophysics},
         year = 2023,
        month = dec,
       volume = {6},
          eid = {49},
        pages = {49},
          doi = {10.21105/astro.2308.01505},
archivePrefix = {arXiv},
       eprint = {2308.01505},
 primaryClass = {astro-ph.GA},
       adsurl = {https://ui.adsabs.harvard.edu/abs/2023OJAp....6E..49F}
}

@ARTICLE{2024ApJ...966...85Z,
       author = {{Zeltyn}, Grisha and {Trakhtenbrot}, Benny and {Eracleous}, Michael and {Yang}, Qian and {Green}, Paul and {Anderson}, Scott F. and {LaMassa}, Stephanie and {Runnoe}, Jessie and {Assef}, Roberto J. and {Bauer}, Franz E. and {Brandt}, W.~N. and {Davis}, Megan C. and {Frederick}, Sara E. and {Fries}, Logan B. and {Graham}, Matthew J. and {Grogin}, Norman A. and {Guolo}, Muryel and {Hern{\'a}ndez-Garc{\'\i}a}, Lorena and {Koekemoer}, Anton M. and {Krumpe}, Mirko and {Liu}, Xin and {Mart{\'\i}nez-Aldama}, Mary Loli and {Ricci}, Claudio and {Schneider}, Donald P. and {Shen}, Yue and {{\'S}niegowska}, Marzena and {Temple}, Matthew J. and {Trump}, Jonathan R. and {Xue}, Yongquan and {Brownstein}, Joel R. and {Dwelly}, Tom and {Morrison}, Sean and {Bizyaev}, Dmitry and {Pan}, Kaike and {Kollmeier}, Juna A.},
        title = "{Exploring Changing-look Active Galactic Nuclei with the Sloan Digital Sky Survey V: First Year Results}",
      journal = {\apj},
         year = 2024,
        month = may,
       volume = {966},
       number = {1},
          eid = {85},
        pages = {85},
          doi = {10.3847/1538-4357/ad2f30},
archivePrefix = {arXiv},
       eprint = {2401.01933},
 primaryClass = {astro-ph.GA},
       adsurl = {https://ui.adsabs.harvard.edu/abs/2024ApJ...966...85Z}
}

@ARTICLE{2024ApJ...966..128W,
       author = {{Wang}, Shu and {Woo}, Jong-Hak and {Gallo}, Elena and {Guo}, Hengxiao and {Son}, Donghoon and {Kong}, Minzhi and {Mandal}, Amit Kumar and {Cho}, Hojin and {Kim}, Changseok and {Shin}, Jaejin},
        title = "{Identifying Changing-look AGNs Using Variability Characteristics}",
      journal = {\apj},
         year = 2024,
        month = may,
       volume = {966},
       number = {1},
          eid = {128},
        pages = {128},
          doi = {10.3847/1538-4357/ad3049},
archivePrefix = {arXiv},
       eprint = {2402.18131},
 primaryClass = {astro-ph.GA},
       adsurl = {https://ui.adsabs.harvard.edu/abs/2024ApJ...966..128W}
}

@ARTICLE{2025A&A...693A..35J,
       author = {{Jana}, A. and {Ricci}, C. and {Temple}, M.~J. and {Chang}, H.-K. and {Shablovinskaya}, E. and {Trakhtenbrot}, B. and {Diaz}, Y. and {Ilic}, D. and {Nandi}, P. and {Koss}, M.},
        title = "{Investigating changing-look active galactic nuclei with long-term optical and X-ray observations}",
      journal = {\aap},
         year = 2025,
        month = jan,
       volume = {693},
          eid = {A35},
        pages = {A35},
          doi = {10.1051/0004-6361/202451058},
archivePrefix = {arXiv},
       eprint = {2411.08676},
 primaryClass = {astro-ph.GA},
       adsurl = {https://ui.adsabs.harvard.edu/abs/2025A&A...693A..35J}
}

@ARTICLE{2025ApJS..278...28G,
       author = {{Guo}, Wei-Jian and {Zou}, Hu and {Greenwell}, Claire L. and {Alexander}, David M. and {Fawcett}, Victoria A. and {Pan}, Zhiwei and {Siudek}, Ma{\l}gorzata and {Aguilar}, Jessica Nicole and {Ahlen}, Steven and {Brooks}, David and {Claybaugh}, Todd and {Dawson}, Kyle and {de la Macorra}, Axel and {Doel}, Peter and {Font-Ribera}, Andreu and {Gazta{\~n}aga}, Enrique and {Gontcho A Gontcho}, Satya and {Gutierrez}, Gaston and {Kehoe}, Robert and {Kisner}, Theodore and {Landriau}, Martin and {Le Guillou}, Laurent and {Manera}, Marc and {Meisner}, Aaron and {Miquel}, Ramon and {Moustakas}, John and {Prada}, Francisco and {Rossi}, Graziano and {Sanchez}, Eusebio and {Schubnell}, Michael and {Sprayberry}, David and {Sui}, Jipeng and {Tarl{\'e}}, Gregory and {Weaver}, Benjamin Alan and {Xiao}, Yun-Ao and {Zou}, Siwei},
        title = "{Changing-look Active Galactic Nuclei from the Dark Energy Spectroscopic Instrument. II. Statistical Properties from the First Data Release}",
      journal = {\apjs},
         year = 2025,
        month = may,
       volume = {278},
       number = {1},
          eid = {28},
        pages = {28},
          doi = {10.3847/1538-4365/adc124},
archivePrefix = {arXiv},
       eprint = {2408.00402},
 primaryClass = {astro-ph.GA},
       adsurl = {https://ui.adsabs.harvard.edu/abs/2025ApJS..278...28G}
}

@ARTICLE{2025ApJ...986..160D,
       author = {{Dong}, Qian and {Zhang}, Zhi-Xiang and {Gu}, Wei-Min and {Sun}, Mouyuan and {Zheng}, Yong-Gang},
        title = "{Newly Discovered Changing-look Active Galactic Nuclei from SDSS and LAMOST Survey}",
      journal = {\apj},
         year = 2025,
        month = jun,
       volume = {986},
       number = {2},
          eid = {160},
        pages = {160},
          doi = {10.3847/1538-4357/add331},
archivePrefix = {arXiv},
       eprint = {2408.07335},
 primaryClass = {astro-ph.GA},
       adsurl = {https://ui.adsabs.harvard.edu/abs/2025ApJ...986..160D}
}

@ARTICLE{2026ApJ...997..100W,
       author = {{Wang}, Huimei and {Wu}, Xue-Bing and {Yao}, Nanyu and {Lyu}, Bing and {Pang}, Yuxuan and {Fu}, Yuming and {Zhu}, Rui and {Yang}, Qian},
        title = "{Systematic Analysis of Changing-Look Active Galactic Nucleus Variability Using ZTF Light Curves}",
      journal = {\apj},
         year = 2026,
        month = jan,
       volume = {997},
       number = {1},
          eid = {100},
        pages = {100},
          doi = {10.3847/1538-4357/ae1fdc},
archivePrefix = {arXiv},
       eprint = {2511.10217},
 primaryClass = {astro-ph.GA},
       adsurl = {https://ui.adsabs.harvard.edu/abs/2026ApJ...997..100W}
}

@ARTICLE{2026Ap&SS.371...12D,
       author = {{Dennefeld}, M. and {Pursimo}, T. and {Carvalho}, C. and {Biancalani}, E. and {D{\'\i}az Teodori}, M.~A. and {D{\"u}rfeldt Pedros}, O. and {Fetzner Keniger}, M.~A. and {Kasikov}, A. and {Koivisto}, N. and {Martikainen}, J. and {Matilainen}, K. and {Sinkbaek Thomsen}, J. and {Terwel}, J. and {Viitanen}, A.},
        title = "{AGN in Gaia alerts: from flares to changing look quasars}",
      journal = {\apss},
         year = 2026,
        month = jan,
       volume = {371},
       number = {1},
          eid = {12},
        pages = {12},
          doi = {10.1007/s10509-026-04534-y},
archivePrefix = {arXiv},
       eprint = {2601.14932},
 primaryClass = {astro-ph.HE},
       adsurl = {https://ui.adsabs.harvard.edu/abs/2026Ap&SS.371...12D}
}

@ARTICLE{2026ApJ..1004..121V,
       author = {{Verrico}, Margaret E. and {French}, K. Decker and {Baldassare}, Vivienne F. and {Burke}, Colin J. and {Duffy}, Laura and {Earl}, Nicholas and {Harrison}, Megan and {Hinkle}, Jason T. and {Messick}, Alexander and {Mondal}, Samaresh and {Moon}, Yashasvi and {Shepherd}, Margaret and {Stone}, Zachary},
        title = "{Identifying Transitions in Changing-look Active Galactic Nuclei in Light-curve Data with the Zwicky Transient Facility}",
      journal = {\apj},
         year = 2026,
        month = jun,
       volume = {1004},
       number = {1},
          eid = {121},
        pages = {121},
          doi = {10.3847/1538-4357/ae6b77},
archivePrefix = {arXiv},
       eprint = {2604.12109},
 primaryClass = {astro-ph.GA},
       adsurl = {https://ui.adsabs.harvard.edu/abs/2026ApJ..1004..121V}
}

@ARTICLE{2026ApJ..1003....4S,
       author = {{Son}, Suyeon and {Kim}, Minjin and {Ho}, Luis C. and {Kim}, Dohyeong and {Kim}, Taehyun},
        title = "{The Changing-look Phenomenon Accompanied by an Accretion Mode Transition in NGC 3786}",
      journal = {\apj},
         year = 2026,
        month = may,
       volume = {1003},
       number = {1},
          eid = {4},
        pages = {4},
          doi = {10.3847/1538-4357/ae6106},
archivePrefix = {arXiv},
       eprint = {2604.17173},
 primaryClass = {astro-ph.GA},
       adsurl = {https://ui.adsabs.harvard.edu/abs/2026ApJ..1003....4S}
}

@ARTICLE{2026AJ....171..285D,
       author = {{DESI Collaboration} and {Abdul Karim}, M. and {Adame}, A.~G. and {Aguado}, D. and {Aguilar}, J. and {Ahlen}, S. and {Alam}, S. and {Aldering}, G. and {Alexander}, D.~M. and {Alfarsy}, R. and {Allen}, L. and {Allende Prieto}, C. and {Alves}, O. and {Anand}, A. and {Andrade}, U. and {Armengaud}, E. and {Avila}, S. and {Aviles}, A. and {Awan}, H. and {Bailey}, S. and {Baleato Lizancos}, A. and {Ballester}, O. and {Bault}, A. and {Bautista}, J. and {Bean}, R. and {Behera}, J. and {BenZvi}, S. and {Beraldo e Silva}, L. and {Bermejo-Climent}, J.~R. and {Beutler}, F. and {Bianchi}, D. and {Blake}, C. and {Blum}, R. and {Bolton}, A.~S. and {Bonici}, M. and {Brieden}, S. and {Brodzeller}, A. and {Brooks}, D. and {Buckley-Geer}, E. and {Burtin}, E. and {Bystr{\"o}m}, A. and {Canning}, R. and {Carnero Rosell}, A. and {Carr}, A. and {Carrilho}, P. and {Casas}, L. and {Castander}, F.~J. and {Cereskaite}, R. and {Cervantes-Cota}, J.~L. and {Chaussidon}, E. and {Chaves-Montero}, J. and {Chen}, S. and {Chen}, X. and {Circosta}, C. and {Claybaugh}, T. and {Cole}, S. and {Cooper}, A.~P. and {Cousinou}, M.-C. and {Cuceu}, A. and {Davis}, T.~M. and {Dawson}, K.~S. and {de Belsunce}, R. and {de la Cruz}, R. and {de la Macorra}, A. and {de Mattia}, A. and {Deiosso}, N. and {Della Costa}, J. and {Demina}, R. and {Demirbozan}, U. and {DeRose}, J. and {Dey}, A. and {Dey}, B. and {Ding}, J. and {Ding}, Z. and {Doel}, P. and {Douglass}, K. and {Dowicz}, M. and {Ebina}, H. and {Edelstein}, J. and {Eisenstein}, D.~J. and {Elbers}, W. and {Emas}, N. and {Escoffier}, S. and {Fagrelius}, P. and {Fan}, X. and {Fanning}, K. and {Favole}, G. and {Fawcett}, V.~A. and {Fern{\'a}ndez-Garc{\'\i}a}, E. and {Ferraro}, S. and {Findlay}, N. and {Font-Ribera}, A. and {Forero-Romero}, J.~E. and {Forero-S{\'a}nchez}, D. and {Frenk}, C.~S. and {G{\"a}nsicke}, B.~T. and {Galbany}, L. and {Garc{\'\i}a-Bellido}, J. and {Garcia-Quintero}, C. and {Garrison}, L.~H. and {Gazta{\~n}aga}, E. and {Gil-Mar{\'\i}n}, H. and {Gloudemans}, A. and {Gnedin}, O.~Y. and {Gontcho A Gontcho}, S. and {Gonzalez}, D. and {Gonzalez-Morales}, A.~X. and {Gonzalez-Perez}, V. and {Gordon}, C. and {Graur}, O. and {Green}, D. and {Gruen}, D. and {Gsponer}, R. and {Guandalin}, C. and {Gutierrez}, G. and {Guy}, J. and {Hahn}, C. and {Han}, J.~J. and {Han}, J. and {He}, S. and {Herrera-Alcantar}, H.~K. and {Heydenreich}, S. and {Honscheid}, K. and {Hou}, J. and {Howlett}, C. and {Huterer}, D. and {Ir{\v{s}}i{\v{c}}}, V. and {Ishak}, M. and {Jacques}, A. and {Jiang}, L. and {Jimenez}, J. and {Jing}, Y.~P. and {Joachimi}, B. and {Joudaki}, S. and {Joyce}, R. and {Jullo}, E. and {Juneau}, S. and {Kara{\c{c}}ayl{\i}}, N.~G. and {Karim}, T. and {Kehoe}, R. and {Kent}, S. and {Khederlarian}, A. and {Kirkby}, D. and {Kisner}, T. and {Kitaura}, F.-S. and {Kizhuprakkat}, N. and {Kong}, H. and {Koposov}, S.~E. and {Kremin}, A. and {Krolewski}, A. and {Lahav}, O. and {Lai}, Y. and {Lamman}, C. and {Lan}, T.-W. and {Landriau}, M. and {Lang}, D. and {Lange}, J.~U. and {Lasker}, J. and {Le Goff}, J.~M. and {Le Guillou}, L. and {Leauthaud}, A. and {Levi}, M.~E. and {Li}, S. and {Li}, T.~S. and {Liu}, W. and {Lodha}, K. and {Lokken}, M. and {Luo}, Y. and {Magneville}, C. and {Manera}, M. and {Manser}, C.~J. and {Margala}, D. and {Martini}, P. and {Maus}, M. and {McCullough}, J. and {McDonald}, P. and {Medina}, G.~E. and {Medina-Varela}, L. and {Meisner}, A. and {Mena-Fern{\'a}ndez}, J. and {Menegas}, A. and {Meneses-Rizo}, J. and {Mezcua}, M. and {Miquel}, R. and {Montero-Camacho}, P. and {Moon}, J. and {Moustakas}, J. and {Mu{\~n}oz-Guti{\'e}rrez}, A. and {Mu noz-Santos}, D. and {Myers}, A.~D. and {Myles}, J. and {Nadathur}, S. and {Najita}, J. and {Napolitano}, L. and {Newman}, J.~A. and {Nikakhtar}, F. and {Nikutta}, R. and {Niz}, G. and {Noriega}, H.~E. and {Nugent}, P.},
        title = "{Data Release 1 of the Dark Energy Spectroscopic Instrument}",
      journal = {\aj},
         year = 2026,
        month = may,
       volume = {171},
       number = {5},
          eid = {285},
        pages = {285},
          doi = {10.3847/1538-3881/ae4c43},
archivePrefix = {arXiv},
       eprint = {2503.14745},
 primaryClass = {astro-ph.CO},
       adsurl = {https://ui.adsabs.harvard.edu/abs/2026AJ....171..285D}
}

@ARTICLE{2026RAA....26i5001H,
       author = {{Han}, Pengyuan and {Lu}, Huaiyuan and {Lyu}, Bing and {Wu}, Jiancheng and {Wu}, Qingwen},
        title = "{Accretion-mode Transition: The Driver Behind Spectral Changes in Changing-look AGNs}",
      journal = {Research in Astronomy and Astrophysics},
         year = 2026,
        month = sep,
       volume = {26},
       number = {9},
          eid = {095001},
        pages = {095001},
          doi = {10.1088/1674-4527/ae5f71},
archivePrefix = {arXiv},
       eprint = {2604.12427},
 primaryClass = {astro-ph.HE},
       adsurl = {https://ui.adsabs.harvard.edu/abs/2026RAA....26i5001H}
}
\bibliographystyle{aasjournalv7}



\end{document}